\documentclass[a4paper,12pt]{article}
\usepackage{soul}
\usepackage[usenames,dvipsnames]{color}
\usepackage{jheppub}
\usepackage{esvect}
\usepackage{multirow}
\usepackage{comment}
\usepackage{amsmath, amssymb, slashed, epsf, color, graphicx, latexsym, bm}
\usepackage{mathrsfs}
\usepackage{tensor}
\usepackage{physics}
\usepackage{epsfig}
\usepackage{graphics}
\usepackage{mathtools}
\usepackage{caption}
\usepackage{float}
\usepackage{bbm}
\usepackage{subcaption}
\usepackage{enumitem}
\usepackage{verbatim}
\usepackage{blindtext}
\usepackage[mathlines]{lineno}

\begin{document}
\title{Soft and Collinear Limits in $\mathcal{N}=8$ Supergravity  using Double Copy Formalism}
\author[a]{Nabamita Banerjee,}
\author[a]{Tabasum Rahnuma,}
\author[b]{and Ranveer Kumar Singh.}
\affiliation[a]{Department of Physics, Indian Institute of Science Education and Research Bhopal,
	Bhopal Bypass, Bhopal 462066, India.}
\affiliation[b]{NHETC and Department of physics and Astronomy, Rutgers University, 126
Frelinghuysen Rd., Piscataway NJ 08855, USA}
\emailAdd{nabamita@iiserb.ac.in}
\emailAdd{tabasum19@iiserb.ac.in}
\emailAdd{ranveer.singh@rutgers.edu}

\abstract{It is known that $\mathcal{N}=8$ supergravity  is dual to $\mathcal{N}=4$ super Yang-Mills (SYM) via the double copy relation. Using the explicit relation between scattering amplitudes in the two theories, we calculate the soft and collinear limits in $\mathcal{N}=8$ supergravity from know results in $\mathcal{N}=4$ SYM.  In our application of double copy, a particular self-duality condition is chosen for scalars that allows us  to constrain and determine the R-symmetry indices of the supergravity states in the collinear limit.}
\maketitle
\section{Introduction}
Symmetries in a quantum field theory are their most important features. The difficulty of solving a theory, which means to compute the scattering amplitudes in terms of the correlation functions, is often dictated by the amount of symmetries the theory possesses. This is because symmetries in the theory are reflected in the amplitudes via the Ward identities and one can use these identities to constrain the correlation functions. This is called bootstrapping and has been a very effective tool in conformal field theory \cite{RevModPhys.91.015002,Poland:2016chs,BELAVIN1984333}. This also goes in the reverse direction, that is knowledge about the nature of amplitudes can help us discover non-trivial symmetries of the theory. Important examples include soft and collinear limits of amplitudes. Soft limit of an amplitude is defined by taking the momenta of an external particle to be zero.\footnote{Of course the particle has to be massless for such a limit to make sense.} In soft limit, the amplitude factorises into a universal soft factor which contains the divergence of the amplitude times the amplitude without the soft particle. The soft factor is universal in the sense that it does not depend on the intricate detailes of the theory, but only the helicity of the external particles. This universal factorisation can be extended to subleading order in electromagnetism and to sub-subleading order in gravity \cite{Sahoo:2020ryf, Saha:2019tub, PhysRev.135.B1049, PhysRev.140.B516, PhysRev.166.1287, PhysRev.168.1623, He:2014laa,Bern:2014vva, Campiglia:2014yka, Laddha:2017ygw,Klose:2015xoa,  Cachazo:2014fwa, Casali:2014xpa}. Soft limits of amplitudes at tree level provided important new insights about the symmetries of certain theories when they were interpreted as Ward identities of certain non-trivial symmetries \cite{PhysRevD.97.106019,PhysRevLett.120.201601,AtulBhatkar:2019vcb,Campiglia:2019wxe,Hijano:2020szl}. For example, soft gluon theorem in Yang-Mills theory is related to large gauge transformations \cite{Catani:2000pi,Mao:2017tey,Klose:2015xoa,Casali:2014xpa,Strominger:2017zoo} and soft graviton theorem in Einstein's gravity is related to the so called Bondi-Metzner-Sachs (BMS) symmetries \cite{Bondi:1962px, Sachs:1962wk, Sachs:1962zza,Strominger:2017zoo}. Thus studying soft limits of amplitudes even at tree level teach us more about the symmetries of the theory. \par Another important limit in which one can study the amplitudes is the collinear limit, in which the momenta of two external particles is taken to be collinear. Again the amplitude factorises into a collinear factor containing the divergence times the amplitude with the collinear particles replaced by another particle (see Section \ref{symco} and Section \ref{sec:sugrafac} for precise details). Collinear limits have played important role in flat space holography \cite{Taylor:2017sph,Fan:2019emx}. The collinear limit of amplitude turns into an operator product expansion (OPE) of conformal operators of the celestial conformal field theory (CCFT) on the celestial sphere\footnote{In CCFT, one describes the four dimensional physics in terms of the conformal correlators of two dimensional CFT on the celestial sphere living at the null infinities of the Minkowski flat spacetime. The map from amplitudes in the bulk to conformal correlators on the boundary is the Mellin transform.} on the boundary \cite{Schreiber:2017jsr,Fan:2019emx,Fan:2022vbz,Mizera:2022sln,Pasterski:2021raf,Pasterski:2021rjz}. These OPEs can be used to calculate the non-trivial asymptotic symmetries of the theory. The usual method of calculating asymptotic symmetries is by finding conformal Killing vectors and spinors becomes intractable in the presence of other fields in the theory. That is where CCFT becomes important. A recent proposal of Taylor et. al. asserts that one can calculate the asymptotic symmetries of gravity theories using soft and collinear limit of amplitude in the framework of CCFT. This has been confirmed to give consistent results in the few cases it has been implemented \cite{Fotopoulos:2020bqj,Banerjee:2021uxe,Fotopoulos:2019vac}. Hence the study of soft and collinear limits in gravity theories becomes important from this perspective. 

$\mathcal{N}=4$ supersymmetric Yang-Mills and $\mathcal{N}=8$ supergravity are maximally supersymmetric theories and are rich in symmetries.  Due to enormous symmetries, one can compute higher and higher loop amplitudes and show that they are finite \cite{Bern:2006kd}. In fact people argue that these are one of the simplest quantum field theories \cite{Arkani-Hamed:2008owk}. One can then study the soft and collinear limits of amplitudes in these theories to learn more about their symmetries. The study of soft and collinear limits in $\mathcal{N}=4$ SYM has already been done \cite{Golden:2012hi,Bourjaily:2011hi,Nandan:2012rk} and the corresponding CCFT was studied in \cite{Jiang:2021xzy}. On the other hand, recent investigations into gravity and gauge theory amplitudes have resulted in non-trivial relationships between the two \cite{Bern:2002kj}. Gravity tree level amplitudes can be expressed in terms of sums of products of gauge theory tree level amplitudes. This can be described by different double copy formalisms  \cite{Kawai:1985xq,Bern:2008qj,Cachazo:2013gna, Cachazo:2013hca}. One can then naturally ask if it is possible to relate the soft and collinear limits in $\mathcal{N}=4$ SYM to soft and collinear limits in $\mathcal{N}=8$ supergravity. Indeed this can be done \cite{Liu:2014vva,Bianchi:2008pu,Bern:1998sv,Bern:1998xc}. The relevant double copy formalism are reviewed in \cite{Adamo:2022dcm} which was originally formulated as a relation between open and closed string amplitudes \cite{Kawai:1985xq}. The corresponding relation in the low energy effective theory gives a relation between gauge theory and gravity amplitudes. Thus one can explicitly calculate soft and collinear limits of amplitudes in gravity using the corresponding results in gauge theory.

In this paper we explicitly calculate the soft and collinear limits of all possible helicity combination in $\mathcal{N}=8$ supergravity using the known double copy relation to $\mathcal{N}=4$ SYM. General formulas for the double copy of amplitudes exists in literature \cite{Bern:1998xc,Bern:1998sv} but to our knowledge, they have not been worked out explicitly. These relations are explicitly derived and stated in this paper. \par 
The paper is organised as follows. In Section \ref{notation}, we set up the notations that we follow throughout the paper. In Section \ref{symco}, we briefly review soft and collinear limits in $\mathcal{N}=4$ SYM which we use later in the paper. Double copy formalism and the relevant formula relating the amplitudes are reviewed in Section \ref{sec:dc}. In Section \ref{sec:sugrafac} we recall some basic facts about $\mathcal{N}=8$ supergravity and state our conventions for its factorisation into a pair of $\mathcal{N}=4$ SYM theories. Finally in Sections \ref{collinear} and \ref{sec:softlimsugra} we record the explicit soft and collinear limits of supergravity amplitudes. In the main body of the paper we have tabulated the collinear and soft limits of the amplitudes with the appropriate R-symmetry indices and the detailed calculations have been postponed to the appendices for reference. The appendices also include spinor-helicity formalism and a list of computational results.
\section{Notations}\label{notation}
The Minkowski space can be parameterized using the Bondi coordinates $(u,r,z,\bar{z})$ where $(z,\bar{z})$ parameterises the celestial sphere $\mathcal{CS}^2$ at null infinity. The Lorentz group $\mathrm{SL}(2,\mathbb{C})$  acts on  $\mathcal{CS}^2$ as follows:
\[
(z,\bar{z})\longmapsto\left(\frac{az+b}{cz+d},\frac{\bar{a}\bar{z}+\bar{b}}{\bar{c}\bar{z}+\bar{d}}\right),\quad \begin{pmatrix}
a&b\\c&d
\end{pmatrix}\in\mathrm{SL}(2,\mathbb{C}).
\]
A general null momentum vector $p^{\mu}$ can be parametrized as 
\[
p^{\mu}=\omega q^{\mu},\quad q^{\mu}=\frac{1}{2}\left(1+|z|^{2}, z+\bar{z},-i(z-\bar{z}), 1-|z|^{2}\right),
\]
where $q^{\mu}$ is a null vector, $\omega$ is identified with the light cone energy and all the particles momenta are taken to be outgoing. Under the Lorentz group the four momentum transforms as a Lorentz vector $p^{\mu}\mapsto \Lambda^{\mu}_{~\nu}p^{\nu}$. This induces the following transformation of $\omega$ and $q^{\mu}$ as
$$
\omega \mapsto (c z+d)(\bar{c} \bar{z}+\bar{d}) \omega, \quad q^{\mu} \mapsto q^{\prime \mu}=(c z+d)^{-1}(\bar{c} \bar{z}+\bar{d})^{-1} \Lambda_{~\nu}^{\mu} q^{\nu}.
$$ 
It is useful to introduce the bispinor notation at this stage. We can write the basic null momentum vector $q^{\mu}$ as 
\begin{equation}
    q^{\alpha \dot{\alpha}}=\sigma_{\mu}^{\alpha \dot{\alpha}} q^{\mu}=\left(\begin{array}{cc}
1 &\;  \bar{z} \\
z &\; z \bar{z}
\end{array}\right)=\left(\begin{array}{l}
1 \\
z
\end{array}\right)\left(\begin{array}{ll}
1 & \bar{z}
\end{array}\right)
\end{equation}
Here $\sigma_{\mu}^{\alpha \dot{\alpha}}=(1, \sigma_x,\sigma_y,\sigma_z)$.We can thus introduce the familiar angle and square bracket spinor notation (see Appendix \ref{app:spinorhelicity} for a brief review of spinor-helicity formalism) for the left and right-handed momentum spinors:
\begin{equation}
h^{\alpha} \equiv\langle p|^{\alpha}=\sqrt{ \omega}\left(\begin{array}{l}
1 \\
z
\end{array}\right)=\sqrt{\omega}\langle q|^{\alpha}, \quad \tilde{h}^{\dot{\alpha}} \equiv| p]^{\dot{\alpha}}=\sqrt{\omega}\left(\begin{array}{l}
1 \\
\bar{z}
\end{array}\right)=\sqrt{\omega} |q]^{\dot{\alpha}},
\end{equation}
where we write
\begin{equation}
\langle q |^{\alpha}=\left(\begin{array}{l}
1 \\
z
\end{array}\right), \quad| q]^{\dot{\alpha}}=\left(\begin{array}{c}
1 \\
\bar{z}
\end{array}\right).    
\end{equation}
To shorten the notation, we denote the spinors for momenta $p_i$ by $\langle i|^{\alpha}$ and $|i]^{\dot{\alpha}}$ respectively. The inner product of momentas $p_i$ and $p_j$ can then be written in terms of the angle and square brackets of the corresponding spinors which are now given by
\begin{equation}
 \langle i j\rangle=-\sqrt{\omega_{i} \omega_{j}} z_{i j}, \quad[i j]= \sqrt{\omega_{i} \omega_{j}} \bar{z}_{i j}.  
 \label{eq:paramominncs2}
\end{equation}
where $z_{ij}=z_i-z_j$ and similarly $\bar{z}_{ij}=\bar{z}_i-\bar{z}_j$.
\section{Soft and Collinear Limits in $\mathcal{N}=4$ SYM} \label{symco}
As detailed in the introduction, in this paper we shall be studying the interesting limits of supergravity amplitudes using double copy relations. For this purpose we use $\mathcal{N}=4$ SYM as a machinary to find our desired results for gravity. Let us briefly recall some of the prime properties of $\mathcal{N}=4$ SYM. 
There are 16 different fields in $\mathcal{N}=4$ SYM, all of which can be packaged in a single superfield. Let $\{\eta_a\}_{a=1}^4$ be the Grassmann odd coordinates on the superspace. Then the superfield for $\mathcal{N}=4$ SYM can be written as 
\begin{equation}
\begin{aligned}
\Psi(p, \eta)=& G^{+}(p)+\eta_{a} \Gamma^{a}_+(p)+\frac{1}{2 !} \eta_{a} \eta_{b} \Phi^{a b}(p) \\
&+\frac{1}{3 !} \epsilon^{abcd} \eta_{a} \eta_{b} \eta_{c} \Gamma_{d}^-(p)+\frac{1}{4 !} \epsilon^{abcd} \eta_{a} \eta_{b} \eta_{c} \eta_{d} G^{-}(p)
\end{aligned}   
\end{equation}
where $G^{\pm}(p)$ denote positive and negative helicity gluons, $\Gamma^a_+,\Gamma_a^-$ denote positive and negative helicity gluinos respectively and $\Phi^{ab}$ denotes the scalars. The superamplitude of $n$ such superfields is then given by the $n$-point correlation function
\begin{equation}
    \mathcal{A}_n(\{p_1,\eta^1\},\dots\{p_n,\eta^n\})\equiv \langle\Psi_{1}(p_1,\eta^1)\dots\Psi_{n}(p_n,\eta^n)\rangle.
\end{equation}
We sometimes suppress the momenta $p_i$ and superspace Grassmann coordinates $\eta^i$ and simply write $\mathcal{A}_n(1,2,\dots,n)$.  Expanding both sides in $\eta$ and comparing, one gets the scattering amplitude of all
the component fields. Next we find the soft and collinear limits of the superamplitude. We begin with the soft theorem following \cite{He:2014bga}:
\begin{equation}
    \mathcal{A}_{n}\left(\cdots, a, s, b, \cdots\right) \stackrel{p_{s} \rightarrow 0}{\longrightarrow} \operatorname{Soft}^{\text{SYM}}\left(a, s, b\right) \mathcal{A}_{n-1}(\cdots, a, b, \cdots),
\end{equation}
where $p_s$ is the momenta of the soft superfield and $a,b$ are the adjacent superfields. The soft factor $\operatorname{Soft}^{\text{SYM}}\left(a, s, b\right)$ is given by \cite{Liu:2014vva} 
\begin{equation}
    \text {Soft}_{\mathrm{hol}}^{\mathrm{SYM}}(a, s, b)=\frac{1}{\varepsilon^2}\text {Soft(0)}_{\mathrm{hol}}^{\mathrm{SYM}}(a, s, b)+\frac{1}{\varepsilon}\text {Soft(1)}_{\mathrm{hol}}^{\mathrm{SYM}}(a, s, b).
\label{eq:}
\end{equation}
where $(0)$ and $(1)$ indicate the leading and subleading terms. Let us explain the above notation. We associate a pair of spinors
$\left(h_{s}, \tilde{h}_{s}\right) $ with every soft momenta $p_s$.
The limit $\left(h_{s}, \tilde{h}_{s}, \eta^{s}\right) \rightarrow\left(\varepsilon h_{s}, \tilde{h}_{s}, \eta^{s}\right)$ with $\varepsilon\to 0$ and $h_s$ some fixed spinor (namely $h_{s} \rightarrow 0$) is known as the \textit{holomorphic soft limit}. The holomorphic soft factor is then given by \cite{Liu:2014vva}
\begin{equation}
\begin{split}
\text {Soft($k$)}_{\mathrm{hol}}^{\mathrm{SYM}}(a, s, b)=\frac{1}{k!}\frac{\langle a b\rangle}{\langle a s\rangle\langle s b\rangle}\bigg[\frac{\langle sa\rangle}{\langle  ba\rangle}\bigg(\tilde{h}_{s}^{\dot{\alpha}} \frac{\partial}{\partial \tilde{h}_{b}^{\dot{\alpha}}}&+(\eta^{s})_{c} \frac{\partial}{\partial (\eta^{b})_{c}}\bigg)\\&+\frac{\langle sb\rangle}{\langle a b\rangle}\left(\tilde{h}_{s}^{\dot{\alpha}} \frac{\partial}{\partial \tilde{h}_{a}^{\dot{\alpha}}}+(\eta^{s})_{c} \frac{\partial}{\partial (\eta^{a})_{c}}\right)\bigg]^k.
\end{split}
\label{eq:holsoftsym}
\end{equation}
Similarly the limit $\left(h_{s}, \tilde{h}_{s}, \eta^{s}\right) \rightarrow\left(h_{s}, \varepsilon \tilde{h}_{s}, \eta^{s}\right)$ with $\tilde{h}_s$ a fixed spinor (namely $\tilde{h}_{s} \rightarrow 0$) is known as the \textit{anti-holomorphic} soft limit. The anti-holomorphic soft factor is given by
\begin{equation}
\text {Soft($k$)}_{\text {anti-hol }}^{\mathrm{SYM}}(a, s, b)=\frac{1}{k!}\frac{[a b]}{[a s][s b]} \delta^{4}\left(\eta^{s}+\frac{[a s]}{[a b]} \eta^{b}+\frac{[s b]}{[a b]} \eta^{a}\right)\left[\frac{[s b]}{[a b]} h_{s}^{\alpha} \frac{\partial}{\partial h_{a}^{\alpha}}+\frac{[a s]}{[a b]} h_{s}^{\alpha} \frac{\partial}{\partial h_{b}^{\alpha}}\right]^k
\label{eq:antiholsoftsym}
\end{equation}
The physical soft limit $p_{s} \rightarrow 0$ is equivalent to considering both $h_{s}, \tilde{h}_{s} \rightarrow 0$ simultaneously.  Thus in the physical soft lomit, the soft factor splits as the sum of holomorphic as well as the anti-holomorphic soft factors. We use these results in Section \ref{sec:softlimsugra} to compute soft limits in supergravity. 

Next we discuss the collinear limits. In collinear limit, we take the momenta of two adjacent superfields $p_1$ and $p_2$ to be collinear. Under  this limit, the two supefields can fuse to give another supefield with momentum $p_{12}=p_1+p_2$. We parametrize the momenta of the collinear superfields as 
\begin{equation}
    p_1=zp_{12},\quad p_2=(1-z)p_{12},
\end{equation}
where $z$ corresponds to the combined momentum $p_{12}$ on the celestial sphere $\mathcal{CS}^2$. Since $p_1+p_2=p_{12}$, we see that, for massless fields, the collinear limit $p_1||p_2$ implies $p_{1}\cdot p_2\propto p_1^2=0$ which is equivalent to the condition $p_{12}^2\to 0$. Now the collinear limit in $\mathcal{N}=4$ SYM is given by \cite{Jiang:2021xzy,Ferro:2020lgp} 
\begin{equation}
    \mathcal{A}_{n}(1,2,3, \cdots, n) \stackrel{p_{12}^{2} \rightarrow 0}{\longrightarrow} \sum_{l=1}^{2} \int d^{4} \eta^{p_{12}} \operatorname{Split}_{1-l}\left(1,2, p_{12}\right) \mathcal{A}_{n-1}\left(p_{12}, 3, \cdots, n\right).
    \label{eq:collimsymsupfd}
\end{equation}
The $l=1,2$ terms in the collinear limits are called the \textit{helicity-preserving} and \textit{helicity-decreasing} processes. The collinear singularity is contained in the split factors. The split factor of helicity preserving process is given by  \cite{Ferro:2020lgp} 
\begin{equation}
    \operatorname{Split}_{0}\left(z ; \eta^{1}, \eta^{2}, \eta^{p_{12}}\right)=\frac{1}{\sqrt{z(1-z)}} \frac{1}{\langle 12\rangle} \prod_{a=1}^{4}\left(\eta^{p_{12}}_{a}-\sqrt{z} \eta^{1}_{a}-\sqrt{1-z} \eta^{2}_{a}\right).
\end{equation}
Whereas for helicity-decreasing process, the split factor is given by \cite{Ferro:2020lgp}
\begin{equation}
    \text {Split}_{-1}\left(z; \eta^{1}, \eta^{2}, \eta^{p_{12}}\right)=\frac{1}{\sqrt{z(1-z)}} \frac{1}{[12]} \prod_{a=1}^{4}\left(\eta^{1}_{a} \eta^{2}_{a}-\sqrt{1-z} \eta^{1}_{a} \eta^{p_{12}}_{a}+\sqrt{z} \eta^{2}_{a} \eta^{p_{12}}_{a}\right).
\end{equation}
The integral over $\eta^{p_{12}}$ can be performed using general results of Grassmann integration.\footnote{As an example for any function $f(\eta)$ we have \cite{Jiang:2021xzy}
\[
    \int d^{4} \eta^{p}\;  \delta^{(4)}\left(\eta^{p}-\eta \right) f\left(\eta^{p}\right)=\int d^{4} \eta^{p} \prod_{a=1}^{4}\left(\eta^{p}_{a}-\eta_{a}\right) f\left(\eta^{p}\right)=f(\eta).
\]
where, $\delta^{(4)}(\eta^{p}-\eta)=\prod_{a=1}^{4}\left(\eta^{p}_{a}-\eta_{a}\right).$} 
Here
\begin{equation}
\begin{split}
\int d \eta^{p_{12}}_{a}\prod_{a=1}^4\big(\eta^{1}_{a} \eta^{2}_{a}-\sqrt{1-z}\eta^{1}_{a} \eta^{p_{12}}_{a}&+\sqrt{z}\eta^{2}_{a} \eta^{p_{12}}_{a}\big) f\left(\eta^{p_{12}}_{a}\right)\\&=\delta^{(4)}\left(\sqrt{1-z}\eta^{1}_{a}-\sqrt{z}\eta^{2}_{a} \right) f\left(\frac{\eta^{2}_{a}}{\sqrt{1-z}}\right),    
\end{split}
\end{equation}
Using these, we express the collinear limit \eqref{eq:collimsymsupfd} as 
\begin{equation}
    \begin{split}
        \mathcal{A}_{n}(1,2,3, \cdots, n)& \stackrel{p_{12}^{2} \rightarrow 0}{\longrightarrow} \frac{1}{\sqrt{z(1-z)}} \frac{1}{[12]}\delta^{(4)}\left(\sqrt{1-z}\eta^{1}_{a}-\sqrt{z}\eta^{2}_{a} \right)\mathcal{A}_{n-1}\left(\{p_{12},\frac{\eta^{2}_a}{\sqrt{1-z}}\},3, \cdots, n\right)\\&+\frac{1}{\sqrt{z(1-z)}} \frac{1}{\langle 12\rangle} \mathcal{A}_{n-1}(\{p_{12},\sqrt{z} \eta^{1}_{a}+\sqrt{1-z} \eta^{2}_{a}\},3, \cdots, n).
    \end{split}
\label{eq:collimkinfacsymsupfd}
\end{equation}
Expanding both sides in $\eta^1$ and $\eta^2$, we can get collinear limits of the component fields. For collinear limit of component fields, we use the following notation:
\begin{equation}
    A_n(1^{h_1},2^{h_2},\dots,n)\stackrel{1||2}{\longrightarrow}\sum_{h}\text{Split}_{-h}^{\text{SYM}}(z,1^{h_1},2^{h_2})A_{n-1}(p^{h},\dots,n),
\end{equation}
where $A_n$ is the amplitude of $n$ different fields in the theory and the sum is over all helicities in the theory. Note that the split factor is trivial for helicities $h$ which does not have corresponding interaction with $h_1$ and $h_2$. The split also satisfies the conjugation relation \cite{Ferro:2020lgp}
\begin{equation}\label{helflip}
\begin{split}
  \text {Split}_{-h}\left(z ; a^{h_{1}}, b^{h_{2}}\right)=\text {Split}_{+h}\left(z ; a^{-h_{1}}, b^{-h_{2}}\right)|_{[ab] \leftrightarrow\langle ab\rangle}  
  \end{split}
\end{equation}
The split factor of component fields in SYM has two parts, the kinematic part and the index structure part. Kinematic part only depends on the momenta of the collinear particles while the index structure consists of the $\mathrm{SU}(4)$ R-symmetry indices of the component fields. 
As indicated earlier, one can compute the kinematic part of the split factors for various combinations of helicities by expanding both sides of \eqref{eq:collimkinfacsymsupfd} in $\eta^1,\eta^2$ and then comparing the coefficients. This has been done using mathematica. The non-trivial split factor for collinear gluons are: 
\begin{equation}
\begin{split}
& \text{Split}_{+1}^{\text {SYM}}\left(z, a^{+1}, b^{+1}\right)=0,\qquad  \text{Split}_{-1}^{\text {SYM}}\left(z, a^{+1}, b^{+1}\right)=\frac{1}{\sqrt{z(1-z)}}\frac{1}{\langle a b\rangle},\\
&\text{Split}_{+1}^{\text {SYM}}\left(z, a^{-1}, b^{+1}\right)=\sqrt{\frac{z^{3}}{1-z}} \frac{1}{\langle a b\rangle},\qquad \text{Split}_{+1}^{\text {SYM}}\left(z, a^{+1}, b^{-1}\right)=\frac{(1-z)^2}{\sqrt{z(1-z)}} \frac{1}{\langle a b\rangle}.
    \end{split}
\label{eq:}
\end{equation}
The split factor for collinear gluinos and scalars are:
\begin{equation}\label{list}
\begin{split}
&\text{Split}_{0}^{\text {SYM}}\left(z, a^{+\frac{1}{2}}, b^{+\frac{1}{2}}\right)= \frac{1}{\langle ab\rangle}, \qquad \text{Split}_{+1}^{\text {SYM}}\left(z, a^{+\frac{1}{2}}, b^{-\frac{1}{2}}\right)= \frac{(1-z)}{\langle ab\rangle}, \\
&\text{Split}_{+1}^{\text {SYM}}\left(z, a^{-\frac{1}{2}}, b^{+\frac{1}{2}}\right)=\frac{z}{\langle ab \rangle}, \qquad \text{Split}_{1}^{\text {SYM}}\left(z, a^{0}, b^{0}\right)=\sqrt{z(1-z)} \frac{1}{\langle ab \rangle}. 
\end{split}
\end{equation}
Finally the split factor for mixed helicities are 
\begin{equation}
\begin{split}
&\text{Split}_{+\frac{1}{2}}^{\text {SYM}}\left(z, a^{-\frac{1}{2}}, b^{+1}\right)=\frac{z}{\sqrt{(1-z)}} \frac{1}{\langle ab \rangle}, \quad \text{Split}_{+\frac{1}{2}}^{\text {SYM}}\left(z, a^{+1}, b^{-\frac{1}{2}}\right)=\frac{1-z}{\sqrt{z}} \frac{1}{\langle ab \rangle},\\
&\text{Split}_{-\frac{1}{2}}^{\text {SYM}}\left(z, a^{+\frac{1}{2}}, b^{+1}\right)=\frac{1}{\sqrt{(1-z)}} \frac{1}{\langle ab \rangle}, \quad \text{Split}_{-\frac{1}{2}}^{\text {SYM}}\left(z, a^{+1}, b^{+\frac{1}{2}}\right)=\frac{1}{\sqrt{z}} \frac{1}{\langle ab \rangle}, \\
& \text{Split}_{0}^{\text {SYM}}\left(z, a^{0}, b^{+1}\right)=\sqrt{\frac{z}{1-z}} \frac{1}{\langle ab\rangle},\quad \text{Split}_{0}^{\text {SYM}}\left(z, a^{+1}, b^{0}\right)=\sqrt{\frac{1-z}{z}} \frac{1}{\langle ab\rangle},\\
&\text{Split}_{\frac{1}{2}}^{\text {SYM}}\left(z, a^{0}, b^{+\frac{1}{2}}\right)= \sqrt{z} \frac{1}{\langle ab \rangle},\quad \text{Split}_{\frac{1}{2}}^{\text {SYM}}\left(z, a^{+\frac{1}{2}}, b^{0}\right)= \sqrt{(1-z)} \frac{1}{\langle ab \rangle}.
\end{split}
\end{equation}
All the other split factors can easily be obtained from Eq. \eqref{helflip}.
We now list the index structure part of the split factors for various component fields. We obtain it by expanding both sides of \eqref{eq:collimkinfacsymsupfd} in $\eta^1$ and $\eta^2$ and comparing the coefficients. Some of the index structures have been worked out in \cite{Jiang:2021xzy}. We complete the list here. Note the index structure in the collinear limit of a gluon with any other component field is trivially determined, hence we omit them from Table \ref{tab:indstrsym}.
\begin{table}[H]
    \centering
    \begin{tabular}{|c|c|}
\hline    Collinear fields     & Resulting index structure \\\hline
    $\Gamma^a_+,\Gamma^b_+$     & $\Phi^{ab}$\\$\Gamma_a^-,\Gamma_b^-$     & $\frac{1}{2!}\epsilon_{abcd}\Phi^{cd}$\\$\Gamma^a_+,\Gamma_b^-$ & $\delta^a_bG^{\pm}$\\$\Gamma^a_+,\Phi^{bc}$&$\epsilon^{abcd}\Gamma_d^-$\\$\Gamma_a^-,\Phi^{bc}$&2!$\delta^{[b}_{a}\Gamma^{c]}_+$\\$\Phi^{ab},\Phi^{cd}$&$\epsilon^{abcd}G^{\pm}$\\\hline
    \end{tabular}
    \caption{Index structure in collinear limit in $\mathcal{N}=4$ SYM}
    \label{tab:indstrsym}
\end{table}

\section{Double Copy : a brief review}\label{sec:dc}
Let us briefly review double copy (DC) technique that plays a crucial role in our analysis.  It is a multiplicative bilinear operation to compute the amplitudes in one theory using amplitudes from other simpler theories. This is a method to express gravity tree level amplitudes in terms of sums of products of gauge theory tree level amplitudes. There are three different double copy formalisms for tree level amplitudes: KLT (named after Kawai, Lewellen, and Tye) \cite{Kawai:1985xq}, BCJ (named after  Bern, Carrasco, and Johansson) \cite{Bern:2008qj} and  CHY (named after Cachazo, He, and Yuan) \cite{Cachazo:2013gna, Cachazo:2013hca} formalism. We refer to \cite{Adamo:2022dcm} for detailed review of these formalisms. Here we restrict our discussion to the application of double copy to soft and collinear limit of gravity amplitudes in terms of soft and collinear limits of gauge theory amplitudes. 
\subsection{Double copy and collinear limit}
The KLT double copy was originally discovered in string theory as a relation between open and closed string amplitudes. Once the large string tension limit (also called the field theory limit) is taken, the KLT relation turns into a relation between gravity and gauge theory tree level amplitudes \cite{Bern:1998sv}. The general KLT relation for a general gravity tree level amplitude $M_n^{\text{tree}}(1,2,\dots,n)$ with $n$ external legs (we have assumed $n$ to be even below but the odd case can also be written in a similar way with appropriate modifications) with color-ordered gauge theory tree level amplitude $A_n^{\text{tree}}(1,2,\dots,n)$ is given by \cite{Bern:1998sv}.
\begin{equation}
\begin{split}
M_{n}^{\text {tree }}(1,2, \ldots, n)&=i(-1)^{n+1}A_{n}^{\text {tree }}(1,2, \ldots, n) \\&\times\sum_{\substack{\sigma\in S_{n/2-1}\\\tau\in S_{n/2-2}}} f\left(\sigma(1), \ldots, \sigma(n/2-1)\right)\bar{f}\left(\tau(n/2+1), \ldots, \tau(n-2)\right)\\&\times
A_{n}^{\text {tree }}\left(\sigma(1), \ldots, \sigma(n/2-1), 1, n-1, \tau(n/2+1), \ldots,\tau(n-2), n\right) \\
&+\text{Permutations of $(2, \ldots, n-2)$}.
\end{split}
\label{eq:genkltrel}
\end{equation}
The functions $f$ and $\bar{f}$ are defined as 
\begin{equation}
    \begin{aligned}
&f\left(i_{1}, \ldots, i_{j}\right)=s\left(1, i_{j}\right) \prod_{m=1}^{j-1}\left(s\left(1, i_{m}\right)+\sum_{k=m+1}^{j} g\left(i_{m}, i_{k}\right)\right) \\
&\bar{f}\left(l_{1}, \ldots, l_{j^{\prime}}\right)=s\left(l_{1}, n-1\right) \prod_{m=2}^{j^{\prime}}\left(s\left(l_{m}, n-1\right)+\sum_{k=1}^{m-1} g\left(l_{k}, l_{m}\right)\right)
\end{aligned}
\end{equation}
where 
\begin{equation}
    g(i,j)=\begin{cases}
    s(i,j):=s_{ij}:=\langle ij\rangle[ji],& i>j\\0,&\text{otherwise}.
    \end{cases}
\end{equation}
Thus every gravity state $j$ on the LHS can be interpreted as the tensor product of the two gauge theory state on the RHS. For example, $\mathcal{N}=8$ supergravity amplitude can be related to the amplitudes in $\mathcal{N}=4$ super Yang-Mills in this way and this leads to the relation  
\[
\mathcal{N}=8 ~~\text{Supergravity}\sim (\mathcal{N}=4~~\text{Super Yang-Mills})~\otimes~(\mathcal{N}=4~~\text{Super Yang-Mills}).
\]
Note that the doubling of supersymmetry in this double copy relation can be understood by counting the degrees of freedom on the two sides. Indeed $\mathcal{N}=8$ supergravity has 264 states which is twice the 128 states in $\mathcal{N}=4$ SYM. One can take collinear limit on both sides of the KLT relation \eqref{eq:genkltrel} to obtain a relation between the split factor for collinear states in gravity to the split factors in gauge theory. We describe this relation below. The collinear limit in gravity is written as \cite{Bern:1998xc}
\begin{equation}
M_{n}^{\text {tree }}(1^{h_1},2^{h_2},\dots,n) \stackrel{1 \| 2}{\longrightarrow} \sum_{h=\pm} \text { Split }_{-h}^{\text {gravity }}(z,1^{h_1},2^{h_2}) \times M_{n-1}^{\text {tree }}\left(P^{h}, 3, \ldots, n\right).
\end{equation}
Using the KLT relation, the gravity split factor can be related to the ``square" of gauge split factors as \cite{Bern:1998sv},
\begin{equation}
\begin{split}
\text { Split }_{-(h+\tilde{h})}^{\text {gravity }}\left(z, 1^{h_{1}+\tilde{h}_{1}}, 2^{h_{2}+\tilde{h}_{2}}\right)&=-s_{12} \times \text { Split }_{-h}^{\text {gauge}}\left(z, 1^{h_{1}}, 2^{h_{2}}\right)\\
&\hspace{15ex}\times \text { Split }_{-\tilde{h}}^{\text {gauge}}\left(z, 2^{\tilde{h}_{2}}, 1^{\tilde{h}_{1}}\right) .
\end{split}
\end{equation} 
Here a state $h+ \tilde h $ in gravity theory is written as product of states $h,\tilde{h}$ in the two gauge theories and $s_{12} = \left<12\right>[21]$. We will explain the explicit factorisation of states for the case of $\mathcal{N}=8$ supergravity into $\mathcal{N}=4$ super Yang-Mills states in Section \ref{sec:sugrafac}.
\subsection{Double copy and soft limit}
Similarly, one can take the soft limit of the double copy relation to relate the soft factors in gravity and gauge theories. Let us start with the universal soft behaviour of the tree level $n$-gluon amplitude. The soft factor when the $i$-th particle is taken to be soft, for either helicities, is given by,
\begin{equation}
\begin{split}
\mathcal{A}_{n}^{\text {tree}}\left(\ldots, a, \varepsilon i^{\pm}, b, \ldots\right) \stackrel{\varepsilon \rightarrow 0}{\longrightarrow} \Bigg(\frac{1}{\varepsilon^2} \mathcal{S}_{\mathrm{Gauge}}^{(0)}(i,a,b)+\frac{1}{\varepsilon} \mathcal{S}_{\mathrm{Gauge}}^{(1)}(i,a,b) + \mathcal{O}\left(1\right)\Bigg)\\\times \mathcal{A}_{n-1}^{\text {tree }}(\ldots, a, b, \ldots)
\end{split}
\end{equation}
Here the soft limit is parameterized by a factor $\varepsilon \to 0$, as described in the last section.  The factors $\mathcal{S}_{\mathrm{Gauge}}^{(0)}$ and  $\mathcal{S}_{\mathrm{Gauge}}^{(1)}$ contains the soft divergences to leading and subleading order in the gauge theory. Similarly, the gravity amplitude also has this universal soft behaviour with $i$-th particle going soft and is given by,
\begin{equation}
\begin{split}
 \mathcal{M}_{n}^{\text {tree }}\left(\ldots, a,\varepsilon i^{\pm}, b, \ldots\right) \stackrel{\varepsilon \rightarrow 0}{\longrightarrow} \Bigg(\frac{1}{\varepsilon^3}\mathcal{S}^{(0)}_{\mathrm{Gravity}}(i,a,b)+\frac{1}{\varepsilon^2}\mathcal{S}^{(1)}_{\mathrm{Gravity}}(i,a,b)\\+\frac{1}{\varepsilon}\mathcal{S}_{\mathrm{Gravity}}^{(2)}(i,a,b)+\mathcal{O}\left(1\right) \Bigg)\mathcal{M}_{n-1}^{\text {tree}}(\ldots, a, b, \ldots)
\end{split}
\end{equation}
where $\mathcal{S}^{(0)}_{\mathrm{Gravity}}$, $\mathcal{S}^{(1)}_{\mathrm{Gravity}}$ and $\mathcal{S}^{(2)}_{\mathrm{Gravity}}$ are leading, subleading and subsubleading soft factors in the gravity theory. Double copy relates these soft factors as follows \cite{He:2014bga, Bern:2008qj} 
\begin{equation}
\begin{split}
   \frac{1}{\varepsilon^{3}} \mathcal{S}_{\mathrm{Gravity}}^{(0)}(s,n,1)+\frac{1}{\varepsilon^{2}} \mathcal{S}_{\mathrm{Gravity}}^{(1)}(s,n,1)+\frac{1}{\varepsilon} \mathcal{S}_{\mathrm{Gravity}}^{(2)}(s,n,1)\\=\sum_{j=1}^{n} K_{sj}^{2}\left(\frac{1}{\varepsilon^{2}} \mathcal{S}_{\mathrm{Gauge}}^{(0)}(j, s, n)+\frac{1}{2 \varepsilon} \mathcal{S}_{\mathrm{Gauge}}^{(1)}(j, s, n)\right)^{2} 
\end{split}
\end{equation}
where $K_{sj}^{2}=\varepsilon\langle sj\rangle[sj]$.

This completes a brief review of double copy relations that we shall be using in the present work.

\section{$\mathcal{N}$=8 Supergravity}\label{sec:sugrafac}
In this section, we briefly review the field contents and basic properties of the theory and also establish notations that we follow in the remainder of the paper.

Let $\{\eta_A\}_{A=1}^8$ be the Grassmann coordinates on the $\mathcal{N}=8$ superspace. The degrees of $\mathcal{N}=8$ supergravity for an on-shell superfield is defined as
\begin{equation}
\begin{split}
\Psi(p, \eta)&=H^{+}(p)+\eta_{A} \psi^{A}_+(p)+\eta_{AB} G^{AB}_+(p)+\eta_{ABC} \chi^{ABC}_+(p)\\
&+\eta_{ABCD} \Phi^{ABCD}(p)+\tilde{\eta}^{ABC} \chi_{ABC}^-(p)+\tilde{\eta}^{AB} G_{AB}^-(p)+\tilde{\eta}^{A} \psi_{A}^-(p)+\tilde{\eta} H^{-}(p),
\end{split}
\end{equation}
where we have introduced the notation 
\begin{equation}
\begin{split}
& \eta_{A_{1} \ldots A_{n}} \equiv \frac{1}{n !} \eta_{A_{1}} \ldots \eta_{A_{2}}\\&\tilde{\eta}^{A_{1} \ldots A_{n}} \equiv \epsilon^{A_{1} \ldots A_{n} B_{1} \ldots B_{8-n}} \eta_{B^{1} \ldots B^{8-n}}\\&\tilde{\eta} \equiv \prod_{A=1}^8 \eta_A.   
\end{split}    
\end{equation} 
The fields $H^{\pm}$ represent graviton, $G^{AB}_+$ and $G_{AB}^-$ represent gluons, $\psi^A_+$ and $\psi_A^-$ represent gravitinos, $\chi^{ABC}_+$ and $\chi_{ABC}^-$ represent gluinos and finally $\Phi^{ABCD}$ represent the real scalars. The (sub)super scripts ${\pm}$ denote positive and negative helicity of various fields. The superamplitude is then defined by 
\begin{equation}
  \mathcal{M}_n(\{p_1,\eta^1\},\dots\{p_n,\eta^n\})=\langle\Psi_{1}(p_1,\eta^1)\dots\Psi_{n}(p_n,\eta^n)\rangle.  
\end{equation}
 We now explain the factorisation of states in supergravity into tensor product of states in super Yang-Mills.   
We begin by counting the degrees of freedom in the two theories. It is summarised in Table \ref{tab:facSGtoSYM} below. 
\begin{table}[H]
\renewcommand{\arraystretch}{1.5}
\centering
    \begin{tabular}{|c|c|}
    \hline
        $\mathcal{N}=8$ Supergravity & ($\mathcal{N}=4$ SYM) $\otimes$ ($\mathcal{N}=4$ SYM)\\\hline
        70 Scalars & $36 (0\otimes 0);1 (-1\otimes +1);1(+1\otimes -1);16(+\frac{1}{2}\otimes-\frac{1}{2});16(-\frac{1}{2}\otimes+\frac{1}{2})$\\\hline
        112 Gravi-photinos ($\pm$) & $48(\pm\frac{1}{2}\otimes 0);48(0\otimes \pm\frac{1}{2});8(\pm\frac{1}{2}\otimes\mp 1);8(\pm 1\otimes\mp\frac{1}{2})$\\\hline 56 Graviphotons $(\pm)$ & $12(\pm 1\otimes 0);16(+\frac{1}{2}\otimes +\frac{1}{2});16(-\frac{1}{2}\otimes-\frac{1}{2})$ \\\hline16 Gravitinos $(\pm)$&$8(\pm\frac{1}{2}\otimes\pm 1);8(\pm 1\otimes\pm \frac{1}{2})$ \\\hline 2 Gravitons $(\pm)$&$2(\pm 1\otimes\pm 1)$\\
    \hline
    \end{tabular}
    \caption{Factorisation of $\mathcal{N}=8$ supergravity states into $\mathcal{N}=4$ super Yang-Mills states}
    \label{tab:facSGtoSYM}
\end{table}
\noindent The precise factorisation of fields and operators are given in \cite{Bianchi:2008pu}. We summarise the factorisation in Table \ref{tab:statefac} below. The second factor of $\mathcal{N}=4$ SYM is written with a tilde to emphasize that the factors of the two gauge theories are not identical. The following notation is used in the table below  and in the rest of the paper: uppercase indices $A,B,C,D,...\in\{1,\dots,8\}$ will denote indices in $\mathcal{N}=8$ supergravity, lower case indices $a,b,c,d\in\{1,2,3,4\}$ correspond to first SYM factor and $r,s,t,u\in\{5,6,7,8\}$ correspond to second SYM factor. In particular, in equations where both upper and lower case indices have been used, we will assume $A=a$ and $A=r$ and so on when $1\leq A\leq 4$ and $5\leq A\leq 8$ respectively.  
\begin{table}[H]
\centering
\begin{tabular}{|c|c|}
\hline $H^{+} =G^{+} \tilde{G}^{+}$ & $H^{-}  =G^{-} \tilde{G}^{-}$ \\
\hline $S_{+}^{a} =\Gamma_{+}^{a} \tilde{G}^{+}$ & $f_{a}^{-} =\Gamma_{a}^{-} \tilde{G}^{-}$ \\
$S_{+}^{r} =G^{+} \tilde{\Gamma}_{+}^{r}$ & $f_{r}^{-} =G^{-} \tilde{\Gamma}_{r}^{-}$ \\
\hline $G^{+}_{ab} =\Phi^{ab} \tilde{G}^{+}$ & $G_{ab}^{-} =\Phi_{ab} \tilde{G}^{-}$ \\
$G^{+}_{ar} =\Gamma_{+}^{a} \tilde{\Gamma}_{+}^{r}$ & $G_{a r}^{-}  =-\Gamma_{a}^{-} \tilde{\Gamma}_{r}^{-}$ \\
$G^{+}_{rs} =G^{+} \tilde{\Phi}^{r s}$ & $\bar{G}_{rs}^{-}  =G^{-} \tilde{\Phi}_{r s}$ \\
\hline $\chi_{+}^{a b c}=\alpha_4 \epsilon^{a b c d} \Gamma_{d}^{-} \tilde{G}^{+}$ & $\chi_{a b c}^{-} =-\alpha_4 \epsilon_{a b c d} \Gamma_{+}^{d} \tilde{G}^{-}$ \\
$\chi_{+}^{a b r} =\Phi^{ab} \tilde{\Gamma}_{+}^{r}$ & $\chi_{a b r}^{-} =\Phi_{ab} \tilde{\Gamma}_{r}^{-}$ \\
$\chi_{+}^{a r s} =\Gamma_{+}^{a} \tilde{\Phi}^{r s}$ &$ \chi_{a r s}^{-} =\Gamma_{a}^{-} \tilde{\Phi}_{r s}$ \\
$\chi_{+}^{r s t}=\tilde{\alpha_4} \epsilon^{r s t u} G^{+} \tilde{\Gamma}_{u}^{-}$ & $\chi_{r s t}^{-}=  -\tilde{\alpha_4}\epsilon_{r s t u} G^{-} \tilde{\Gamma}_{+}^{u} $\\
\hline $\Phi^{a b c d}  =\alpha_4\epsilon^{a b c d} G^{-} \tilde{G}^{+}$ & $\Phi_{a b c d} =\alpha_4\epsilon_{a b c d} G^{+} \tilde{G}^{-}$ \\
$\Phi^{a b c r}  =\alpha_4\epsilon^{a b c d} \Gamma_{d}^{-} \tilde{\Gamma}_{+}^{r}$ & $\Phi_{a b c r} =\alpha_4 \epsilon_{a b c d} \Gamma_{+}^{d} \tilde{\Gamma}_{r}^{-}$ \\
$\Phi^{a b r s} =\Phi^{ab} \tilde{\Phi}^{r s}$ & $\Phi_{a b r s}  =\Phi_{ab} \tilde{\Phi}_{r s}$ \\
$\Phi^{a r s t} =\tilde{\alpha_4}\epsilon^{r s t u} \Gamma_{+}^{a} \tilde{\Gamma}_{u}^{-}$ & $\Phi_{a r s t}  =\tilde{\alpha_4} \epsilon_{r s t u} \Gamma_{a}^{-} \tilde{\Gamma}_{+}^{u}$ \\
$\Phi^{r s t u}  =\tilde{\alpha_4}\epsilon^{r s t u} G^{+} \tilde{G}^{-}$ & $\Phi_{r s t u}  = \tilde{\alpha_4}\epsilon_{r s t u} G^{-} \tilde{G}^{+}$ \\
\hline
    \end{tabular}
    \caption{Factorisation of states in supergravity into states in super Yang-Mills}
    \label{tab:statefac}
\end{table}
\noindent Further note that the scalars in supergravity and super Yang-Mills satisfy the self duality relation, \cite{Bianchi:2008pu} 
\begin{equation}
\begin{split}
    &\Phi_{ABCD}=\frac{1}{4!}\alpha_8\epsilon_{ABCDEFGH}\Phi^{EFGH}\\&\Phi_{ab}=\frac{1}{2!}\alpha_4\epsilon_{abcd}\Phi^{cd}\\&\tilde{\Phi}_{rs}=\frac{1}{2!}\tilde{\alpha}_4\epsilon_{rstu}\Phi^{tu}
\end{split}
\label{eq:selfdualcond}
\end{equation}
with $\alpha_4,\tilde{\alpha}_4,\alpha_8\in\{\pm 1\}$ along with the consistency condition \cite[Eq. 2.12]{Bianchi:2008pu}, 
\begin{equation}
    \alpha_4\tilde{\alpha}_4=\alpha_8.
\end{equation}
and $\epsilon_{AB...H}$ is the Levi-Civita tensor in 8 dimensions and $\epsilon_{abcd},\epsilon_{rstu}$ are Levi-Civita tensor in 4 dimensions. Note that since $5\leq r,s,t,u\leq 8$, $\epsilon_{rstu}$ is defined using permutations of $5,6,7,8$.
Using this factorisation, we can find the collinear limit of any two states in $\mathcal{N}=8$ supergravity. The possible choices of the self-duality factors $(\alpha_4,\tilde{\alpha}_4,\alpha_8)$ are $(1,1,1),(1,-1,-1),(-1,1,-1),(-1,-1,1)$. Based on possible choices of the self duality factors we have four different ways of getting the supergravity amplitudes via double copy.

\section{Collinear limits in $\mathcal{N}=8$ supergravity}\label{collinear}
In this section, we compute the collinear limits using the component field formalism. The double copy relation of collinear limits in component formalism is given by 
\begin{equation}\label{col}
    M_n(1^{h_1},2^{h_2},\dots,n)\stackrel{1||2}{\longrightarrow}\sum_{h}\text{Split}_{-h}^{\text{SG}}(z,1^{h_1},2^{h_2})M_{n-1}(p^{h},\dots,n),
\end{equation}
where the split factor $\text{Split}_{-h}^{\text{SG}}(z,1^{h_1},2^{h_2})$ is given in terms of the split factors in $\mathcal{N}=4$ super Yang-Mills theory as follows:
\begin{equation} \label{fact}
\begin{split}
    \text{Split}_{-(h+\tilde{h})}^{\text {SG}}\left(z, 1^{h_{1}+\tilde{h}_{1}}, 2^{h_{2}+\tilde{h}_{2}}\right)=-s_{12} & \times \text{Split}_{-h}^{\text {SYM}}\left(z, 1^{h_{1}}, 2^{h_{2}}\right) \\
& \times \text{Split}_{-\tilde{h}}^{\text {SYM}}\left(z, 2^{\tilde{h}_{2}}, 1^{\tilde{h}_{1}}\right)
\end{split} 
\end{equation}
where $(h+\tilde{h})$ is the factorisation of $\mathcal{N}=8$ supergravity state with total spin $h+\tilde{h}$ in terms of two copies of $\mathcal{N}=4$ super Yang-Mills states with spins $h,\Tilde{h}$ respectively according to Table \ref{tab:statefac} and $s_{12} = \left<12\right>[21]$. 
The sum over all $\mathcal{N}=8$ supergravity states is interpreted as a double sum over a tensor product of $\mathcal{N}=4$ SYM states \cite{Bern:1998sv}. The calculation of collinear limit then involves two steps:
\begin{enumerate}
    \item Calculate the split factors for all possible factorisation channels, that is, for all possible values of spin and helicity states $h$ in $\mathcal{N}=8$ supergravity. This can be done in such that the factorisation $h=h_1+h_2$ into different spin and helicity states in $\mathcal{N}=4$ SYM from Table \ref{tab:statefac} gives nontrivial split factors. In general one only needs to calculate half of all possible combinations of helicities.
    The remaining split factors can be calculated using
\begin{equation} \label{flip}
\text {Split}_{-}\left(z ; a^{h_{1}}, b^{h_{2}}\right)=\text {Split}_{+}\left(z ; a^{-h_{1}}, b^{-h_{2}}\right)|_{[ab] \leftrightarrow\langle ab\rangle}
\end{equation}
    \item Write the collinear limit of amplitudes by consistently matching the R-symmetry factors using Table \ref{tab:indstrsym} which is non-trivial in  case of $\mathcal{N} > 1$ theories.
\end{enumerate}
\subsection{Collinear limits of like spins}\label{likespin}
Here we compute the collinear amplitudes from the splits for states of same spin. We will show the computation for some cases and summarise the results for the rests in tabular form  and refer the reader to Appendix \ref{like} for all the details of the computations. Moreover we only summarise the collinear limits for independent cases not related by Eq.\eqref{flip}. \\\\
\textbf{\textit{Gravitons}:}\\
When both collinear gravitons are of same helicity (positive or negative), then from Table \ref{tab:statefac}, we see that 
\[
\begin{split}
    \text{Split}_{-(h+\tilde{h})}^{\text {SG}}\left(z, 1^{\pm 2}, 2^{\pm 2}\right)=-s_{12} & \times \text{Split}_{-h}^{\text {SYM}}\left(z, 1^{\pm 1}, 2^{\pm 1}\right) \\
& \times \text{Split}_{-\tilde{h}}^{\text {SYM}}\left(z, 2^{\pm 1}, 1^{\pm 1}\right).
\end{split} 
\]
A similar factorisation is true for opposite helicities.
Thus split factors in $\mathcal{N}=8$ supergravity for two collinear gravitons is 
\begin{equation}
    \begin{split}
        & \text{Split}_{+2}^{\text {SG}}\left(z, a^{+2}, b^{+2}\right)=0=\text{Split}_{-2}^{\text {SG}}\left(z, a^{-2}, b^{-2}\right),\\
&\text{Split}_{-2}^{\text {SG}}\left(z, a^{+2}, b^{+2}\right)=-\frac{1}{z(1-z)} \frac{[a b]}{\langle a b\rangle},\quad \text{Split}_{+2}^{\text {SG}}\left(z, a^{-2}, b^{-2}\right)=-\frac{1}{z(1-z)} \frac{\langle a b\rangle}{[a b]}\\
&\text{Split}_{+2}^{\text {SG}}\left(z, a^{-2}, b^{+2}\right)=-\frac{z^3}{(1-z)} \frac{[a b]}{\langle a b\rangle},\quad \text{Split}_{-2}^{\text {SG}}\left(z, a^{-2}, b^{+2}\right)=-\frac{(1-z)^3}{z}\frac{\langle a b\rangle}{[a b]}.
\end{split}
\end{equation}
Writing the momenta of the collinear particles as $p_i=\omega_iq_i,~i=1,2$, the momenta along the collinear channel is $p=p_{1}+p_{2}=\omega_{p} q_{p}$ with $\omega_{p}=\omega_{1}+\omega_{2}$ and we can write \begin{equation}\label{colpa}
    p_1=zp,\quad p_2=(1-z)p.
\end{equation}
Note that $q_{p}=q_{1}=q_{2}$ and hence
\[
\begin{aligned}
z=\frac{\omega_{1}}{\omega_{p}},\quad \left(1-z\right)=\frac{\omega_{2}}{\omega_{p}}
\end{aligned}\]
With this parametrization, the collinear limits can be tabulated as,
\begin{table}[H]
\renewcommand{\arraystretch}{1.5}
\centering
    \begin{tabular}{|c|c|}
    \hline
     $  M_n\left(1^{+2}, 2^{+2}, \cdots, n\right) $ & $\frac{\omega_{p}^{2}}{\omega_{1} \omega_{2}} \frac{\bar{z}_{12}}{z_{12}} M_{n-1}\left(p^{+2} , \ldots, n\right)$\\\hline
        $M_n\left(1^{-2}, 2^{-2}, \cdots, n\right) $ & $ \frac{\omega_{p}^{2}}{\omega_{1} \omega_{2}} \frac{z_{12}}{\bar{z}_{12}} M_{n-1}\left(p^{-2}, \ldots, n\right)$\\\hline
       $ M_n\left(1^{+2}, 2^{-2},\ldots, n\right)$  & $ \frac{\omega_{1}^{3}}{\omega_{p}^{2} \omega_{2}} \frac{\bar{z}_{12}}{z_{12}} M_{n-1}\left(p^{-2}, 3, \ldots, n\right)+\frac{\omega_{2}^{3}}{\omega_{p}^{2} \omega_{1}} \frac{z_{12}}{\bar{z}_{12}} M_{n-1}\left(p^{+2}, 3, \ldots, n\right)$\\
    \hline
    \end{tabular}
    \caption{Amplitude corresponding to two collinear gravitons}
    \label{tab:gravitoncol}
\end{table}
Here in LHS $1,2,\dots,n$ refers to external particles with momenta $p_1,p_2,\dots,p_n$ and $p_1$ is taken collinear to $p_2$ according to the parametrization in Eq. \eqref{colpa}. We will carry this notation throughout the paper.\\

Note that the collinear limit of two negative helicity gravitons from the collinear limit of two positive helicity gravitons by flipping the helicities throughout and $z_{12}\leftrightarrow \Bar{z}_{12}$. This is reminiscent of Eq.\eqref{flip}.\\\\
\noindent\textbf{\textit{Gravitinos}:}\\
The non-trivial split factors in $\mathcal{N}=8$ Supergravity for two collinear gravitinos are given by 
\begin{equation}
\begin{split}
&\text{Split}_{-1}^{\text {SG}}\left(z, 1^{\frac{1}{2}+1}, 2^{\frac{1}{2}+1}\right)=-\frac{1}{\sqrt{z(1-z)}}\frac{[12]}{\langle 12 \rangle }, \quad \text{Split}_{-1}^{\text {SG}}\left(z, 1^{\frac{1}{2}+1}, 2^{1+\frac{1}{2}}\right)=-\frac{1}{\sqrt{z(1-z)}}\frac{[12]}{\langle 12 \rangle }\\
&\text{Split}_{-2}^{\text {SG}}\left(z, 1^{\frac{1}{2}+1}, 2^{-\frac{1}{2}-1}\right)=-\sqrt{\frac{z^5}{(1-z)}}\frac{\langle 12 \rangle}{[12]}, \quad \text{Split}_{+2}^{\text {SG}}\left(z, 1^{\frac{1}{2}+1}, 2^{-\frac{1}{2}-1}\right)=-\sqrt{\frac{(1-z)^5}{z}} \frac{[12]}{\langle 12 \rangle }
\end{split} 
\end{equation}
We can calculate other split factors using Eq.\eqref{flip}. The factorisation of R-symmetry indices has the form 
\[
\begin{cases}\left(a;\frac{3}{2}\right)=\left(a;\frac{1}{2}\right)\otimes 1 \\
\left(r;\frac{3}{2}\right)=1\otimes\left(r;\frac{1}{2}\right)
\end{cases}
\]
Corresponding to the above two factorization the amplitudes following Eq.\eqref{col} and Eq.\eqref{fact}, can be combined and written in the table below. For details we refer the readers to section \ref{like} in the Appendix.
\begin{table}[H]
\renewcommand{\arraystretch}{1.5}
\centering
    \begin{tabular}{|c|c|}
    \hline
     $ M_n\left(1^{A;+\frac{3}{2}}, 2^{B;+\frac{3}{2}}, \cdots, n\right)$ & $\frac{\omega_p}{\sqrt{\omega_1 \omega_2}} \frac{\bar{z}_{12}}{z_{12}} \; M_{n-1}\left(p^{AB;+1},\cdots, n\right)$\\\hline
        $M_n\left(1^{A;+\frac{3}{2}}, 2_B^{-\frac{3}{2}}, \cdots, n\right)$& $ \delta^{A}_B\frac{\omega_2^{\frac{5}{2}}}{\omega_1^{\frac{1}{2}}\omega_p^2}\frac{\bar{z}_{12}}{z_{12}}M_{n-1}\left(p^{-2},\cdots, n\right)+\delta^{A}_B\frac{\omega_1^{\frac{5}{2}}}{\omega_2^{\frac{1}{2}}\omega_p^2}\frac{z_{12}}{\bar{z}_{12}}M_{n-1}\left(p^{+2},\cdots, n\right)$\\
    \hline
    \end{tabular}
    \caption{Amplitude corresponding to two collinear gravitinos}
    \label{tab:gravitinocol}
\end{table}
\noindent\textbf{\textit{Gravi-photons}:}\\
Using the factorisation in Eq.\eqref{fact} the non-trivial split factors for two collinear Graviphotons is given in Appendix \ref{A1}. The calculation of collinear limits is done in Appendix \ref{like}. The result is recorded in the table below.\par 
The R-symmetry index factorizes as follows: 
\[
\begin{cases}\left(ab;1\right)=\left(ab;0\right)\otimes 1 \\
\left(ar;1\right)=\left(a,\frac{1}{2}\right)\otimes\left(r;\frac{1}{2}\right)\\
\left(rs;1\right)=1\otimes\left(rs;0\right)
\end{cases}
\]
From the above factorisation we can combine all of the non-trivial amplitudes for $1\leq A,B\leq 8$ as,
\begin{table}[H]
\renewcommand{\arraystretch}{1.5}
\centering
    \begin{tabular}{|c|c|}
    \hline
     $  M_n\left(1^{AB; +1}, 2^{CD; +1}, \cdots \right)$ & $\frac{\bar{z}_{12}}{z_{12}} \times M_{n-1}\left(p^{ABCD;0},\cdots \right)$\\\hline
        $ M_n\left(1^{AB;+1}, 2_{CD}^{-1}, \cdots \right)$& $-\delta^{AB}_{CD}\Big[\frac{\omega_2^2 }{\omega_p^2}\frac{\bar{z}_{12}}{z_{12}} \times M_{n-1}\left(p^{-2},\cdots \right)+ \frac{\omega_1^2 }{\omega_p^2}\frac{z_{12}}{\bar{z}_{12}} \times M_{n-1}\left(p^{+2},\cdots\right)\Big]$\\
    \hline
    \end{tabular}
    \caption{Amplitude corresponding to two collinear graviphotons}
    \label{tab:graviphotoncol}
\end{table}
In writing the collinear limit of opposite helicity graviphotons, we made a choice of self-duality factors $\alpha_4=\Tilde{\alpha}_4=-1,\alpha_8=1$. This choice is unique and motivated by our aim to make the R-symmetry indices consistent in both sides of the amplitude calculations. See Appendix \ref{details} for details.\\\\
\noindent\textbf{\textit{Graviphotinos}:}\\
Following the factorisation in Eq. \eqref{fact}, the non-trivial split factors for this channel in $\mathcal{N}=8$ supergravity are given in Appendix \ref{A2}.

The factorisation of the R-symmetry indices is as follows: 
\[
\begin{cases} 
\left(abr;\frac{1}{2}\right)= \left( ab; 0 \right)\otimes\left(r;\frac{1}{2}\right)&\\
\left(ars;\frac{1}{2}\right)=\left(a;\frac{1}{2}\right)\otimes \left( rs; 0 \right)
\end{cases}
\]\[
\begin{cases}
\left(rst;\frac{1}{2}\right)=-\epsilon^{rstu}(1\otimes (u;-\frac{1}{2})) &\\
\left(abc;\frac{1}{2}\right)=-\epsilon^{abcd}((d;-\frac{1}{2})\otimes 1)
\end{cases} \quad (\text{sum over}~ u,d)
\]
The amplitudes corresponding to the above factorisation channels are summarised as follows
\begin{table}[H]
\renewcommand{\arraystretch}{1.5}
\centering
    \begin{tabular}{|c|c|}
    \hline
 $ M_n\left(1^{ars;+\frac{1}{2}}, 2^{btu; +\frac{1}{2}}, \cdots\right)$ & $\epsilon^{rstu} \epsilon^{abcd}\; \frac{\sqrt{\omega_1 \omega_2}}{\omega_p}\frac{\bar{z}_{12}}{z_{12}}\times M_{n-1}\left(p_{cd}^{-1},\cdots\right)$\\\hline
 $ M_n\left(1^{ars;+\frac{1}{2}}, 2^{bct; +\frac{1}{2}}, \cdots\right)$& $\epsilon^{abcd}\epsilon^{rstu} \; \frac{\sqrt{\omega_1 \omega_2}}{\omega_p}\frac{\bar{z}_{12}}{z_{12}}\times M_{n-1}\left(p^{-1}_{du},\cdots\right)$\\ \hline
 $M_n\left(1^{rst;+\frac{1}{2}}, 2^{abc; +\frac{1}{2}}, \cdots\right)$ & $\epsilon^{rstu}\epsilon^{abcd} \; \frac{\sqrt{\omega_1 \omega_2}}{\omega_p}\frac{\bar{z}_{12}}{z_{12}}\times M_{n-1}\left(p_{ud}^{-1},\cdots\right) $\\ \hline
$ M_n\left(1^{ars;+\frac{1}{2}}, 2_{btu}^{-\frac{1}{2}}, \cdots\right)$ & $\epsilon_{tuvw}\epsilon^{rsvw} \delta^{a}_{b}\Big[ \frac{\omega_1^{\frac{3}{2}} \omega_2^{\frac{1}{2}}}{\omega_p^2}\frac{z_{12}}{\bar{z}_{12}} M_{n-1}\left(p^{+2},\cdots\right)+ \frac{\omega_2^{\frac{3}{2}} \omega_1^{\frac{1}{2}}}{\omega_p^2}\frac{\bar{z}_{12}}{z_{12}} M_{n-1}\left(p^{-2},\cdots\right) \Big]$\\ 
    \hline
    \end{tabular}
    \caption{Amplitude corresponding to two collinear graviphotinos}
    \label{tab:graviphotinocol}
\end{table}
\noindent\textbf{\textit{Scalars}:}\\
The three possible channels are $0= 0 \otimes 0$, $0 = \pm 1 \otimes \mp 1$ and $0 = \pm\frac{1}{2} \otimes \mp\frac{1}{2}$. We have the non-trivial splits given in Appendix \ref{A3}.

The factorizations of R-symmetry indices are given by
\[
\left(abrs;0\right)=\left(ab;0\right)\otimes (rs; 0)
\]
\[
\begin{cases}\left(abcd;0\right)=-\epsilon^{abcd} (-1\otimes 1) \\ \left(rstu;0\right)=-\epsilon^{rstu} (1\otimes -1 )
\end{cases}
\]
\[
\begin{cases}\left(abcr;0\right)=-\epsilon^{abcd}\left(d;-\frac{1}{2}\right)\otimes (r;\frac{1}{2})\\ \left(arst;0\right)=-\epsilon^{rstu}(a;\frac{1}{2})\otimes\left(u;-\frac{1}{2}\right) 
\end{cases}
\]
The factorised amplitudes are,
\begin{table}[H]
\renewcommand{\arraystretch}{1.5}
\centering
    \begin{tabular}{|c|c|}
    \hline
 $ M_n\left(1^{abrs;0}, 2^{cdtu; 0}, \cdots \right)$ & $\epsilon^{abcd} \epsilon^{rstu}\Big[\frac{\omega_1 \omega_2}{\omega_p^2}\frac{z_{12}}{\bar{z}_{12}}\times M_{n-1}\left(p^{+2},\cdots \right) + \frac{\omega_1 \omega_2}{\omega_p^2}\frac{\bar{z}_{12}}{z_{12}}\times M_{n-1}\left(p^{-2},\cdots \right)\Big]$\\\hline
 $ M_n\left(1^{abcd;0}, 2^{rstu; 0}, \cdots \right)$& $\epsilon^{abcd}\epsilon^{rstu}\Big[ \frac{\omega_2\omega_1}{\omega_p^2 }\frac{z_{12}}{\bar{z}_{12}}\times M_{n-1}\left(p^{+2},\cdots \right) \frac{\omega_1\omega_2}{\omega_p^2}\frac{\bar{z}_{12}}{z_{12}}\times M_{n-1}\left(p^{-2},\cdots \right)\Big]$\\ \hline
 $M_n\left(1^{abcu;0}, 2^{drst; 0}, \cdots \right)$ & $\epsilon^{abcd}\epsilon^{rstu}\Big[ \frac{\omega_2\omega_1}{\omega_p^2}\frac{z_{12}}{\bar{z}_{12}}\times M_{n-1}\left(p^{+2},\cdots \right)+ \frac{\omega_1\omega_2}{\omega_p^2}\frac{\bar{z}_{12}}{z_{12}}\times M_{n-1}\left(p^{-2},\cdots \right)\Big]$\\ \hline
$  M_n\left(1^{arst;0}, 2^{bcdu; 0}, \cdots \right)$ & $\epsilon^{rstu}\epsilon^{abcd}\Big[ \frac{\omega_2\omega_1}{\omega_p^2}\frac{z_{12}}{\bar{z}_{12}}\times M_{n-1}\left(p^{+2},\cdots \right)+ \frac{\omega_1\omega_2}{\omega_p^2}\frac{\bar{z}_{12}}{z_{12}}\times M_{n-1}\left(p^{-2},\cdots \right)\Big]$\\ 
    \hline
    \end{tabular}
    \caption{Amplitude corresponding to two collinear scalars}
    \label{tab:scalarcol}
\end{table}
\subsection{Collinear limits of Mixed Spins}
In this section, we list the collinear limit of states with different spins. The non-trivial split factors are listed in Appendix \ref{split} and the detailed calculation is done in Appendix \ref{unlike}.\\\\
\noindent\textbf{\textit{Graviton-Gravitino}:}\\
The non-trivial split factors for this collinear pair are given in Appendix \ref{A4}. \\
Using different factorisation channels of the Gravitinos we have,
\begin{table}[H]
\renewcommand{\arraystretch}{1.5}
\centering
\begin{tabular}{|c|c|}
\hline
 $M_n\left(1^{+2}, 2^{r; +\frac{3}{2}}, \cdots, n\right)$ & $\frac{\omega_p^{\frac{3}{2}}}{\omega_2^{\frac{1}{2}} \omega_1} \frac{\bar{z}_{12}}{z_{12}}\times M_{n-1}\left(p^{r; +\frac{3}{2}},\cdots, n\right)$\\\hline
 $M_n\left(1^{+2}, 2_{r}^{ -\frac{3}{2}}, \cdots, n\right)$& $\frac{\omega_2^{\frac{5}{2}}}{\omega_p^{\frac{3}{2}}\omega_1} \frac{\bar{z}_{12}}{z_{12}}\times M_{n-1}\left(p_{r}^{ -\frac{3}{2}},\cdots, n\right)$\\\hline
 $M_n\left(1^{+2}, 2^{a; +\frac{3}{2}}, \cdots, n\right) $ & $\frac{\omega_p^{\frac{3}{2}}}{\omega_2^{\frac{1}{2}} \omega_1} \frac{\bar{z}_{12}}{z_{12}}\times M_{n-1}\left(p^{a; +\frac{3}{2}},\cdots, n\right)$\\\hline
$  M_n\left(1^{+2}, 2_{a}^{ -\frac{3}{2}}, \cdots, n\right)$ & $\frac{\omega_2^{\frac{5}{2}}}{\omega_p^{\frac{3}{2}}\omega_1} \frac{\bar{z}_{12}}{z_{12}} \times M_{n-1}\left(p_{a}^{ -\frac{3}{2}},\cdots, n\right)$\\ 
\hline
\end{tabular}
\caption{Amplitude corresponding to collinear graviton and gravitino}    \label{tab:gravitongravitinocol}
\end{table}
\noindent\textbf{\textit{Graviton-Graviphoton}:}\\
The split factors are given in Appendix \ref{A5}. For $1\leq A,B\leq 8$ all the amplitudes corresponding to different factorisation channels are summarised as
\begin{equation*}
    \boxed{ M_n\left(1^{+2}, 2_{AB}^{-1}, \cdots, n\right)=\frac{\omega_2^2}{\omega_1 \omega_p} \frac{\bar{z}_{12}}{z_{12}}\times M_{n-1}\left(p_{AB}^{-1},\cdots, n\right)}
\end{equation*}
\textbf{\textit{Graviton-Graviphotino}:}\\
Non-trivial split factors are given in Appendix \ref{A6}.
\begin{table}[H]
\renewcommand{\arraystretch}{1.5}
\centering
\begin{tabular}{|c|c|}
\hline
 $M_n\left(1^{+2}, 2^{abr; +\frac{1}{2}}, \cdots, n\right)$ & $\frac{\sqrt{\omega_2 \omega_p}}{\omega_1} \frac{\bar{z}_{12}}{z_{12}}\times M_{n-1}\left(p^{abr; +\frac{1}{2}},\cdots, n\right)$\\\hline
 $M_n\left(1^{+2}, 2^{ars; +\frac{1}{2}}, \cdots, n\right)$& $\frac{\sqrt{\omega_2 \omega_p}}{\omega_1} \frac{\bar{z}_{12}}{z_{12}}\times M_{n-1}\left(p^{ars; +\frac{1}{2}},\cdots, n\right)$\\\hline
 $M_n\left(1^{+2}, 2^{abc; +\frac{1}{2}}, \cdots, n\right)$ & $\frac{\sqrt{\omega_2 \omega_p}}{\omega_1} \frac{\bar{z}_{12}}{z_{12}}\times M_{n-1}\left(p^{abc; +\frac{1}{2}},\cdots, n\right)$\\\hline
$ M_n\left(1^{+2}, 2^{rst; +\frac{1}{2}}, \cdots, n\right)$ &$\frac{\sqrt{\omega_2 \omega_p}}{\omega_1} \frac{\bar{z}_{12}}{z_{12}}\times M_{n-1}\left(p^{rst; +\frac{1}{2}},\cdots, n\right)$\\ \hline 
$M_n\left(1^{+2}, 2_{abr}^{ -\frac{1}{2}}, \cdots, n\right)$ & $\frac{\omega_2^{\frac{3}{2}}}{\omega_p^{\frac{1}{2}}\omega_1} \frac{\bar{z}_{12}}{z_{12}}\times M_{n-1}\left(p_{abr}^{ -\frac{1}{2}},\cdots, n\right)$\\ \hline
$M_n\left(1^{+2}, 2_{ars}^{ -\frac{1}{2}}, \cdots, n\right) $& $\frac{\omega_2^{\frac{3}{2}}}{\omega_p^{\frac{1}{2}}\omega_1} \frac{\bar{z}_{12}}{z_{12}}\times M_{n-1}\left(p_{ars}^{ -\frac{1}{2}},\cdots, n\right)$\\ \hline
$M_n\left(1^{+2}, 2_{abc}^{ -\frac{1}{2}}, \cdots, n\right)  $& $-\frac{\omega_2^{\frac{3}{2}}}{\omega_p^{\frac{1}{2}}\omega_1} \frac{\bar{z}_{12}}{z_{12}}\times M_{n-1}\left(p_{abc}^{ -\frac{1}{2}},\cdots, n\right)$\\ \hline
$M_n\left(1^{+2}, 2_{rst}^{ -\frac{1}{2}}, \cdots, n\right) $& $-\frac{\omega_2^{\frac{3}{2}}}{\omega_p^{\frac{1}{2}}\omega_1} \frac{\bar{z}_{12}}{z_{12}}\times M_{n-1}\left(p_{rst}^{ -\frac{1}{2}},\cdots, n\right)$ \\ \hline
\end{tabular}
\caption{Amplitude corresponding to collinear graviton and graviphotino}    \label{tab:gravitongravitinocol}
\end{table}
\noindent\textbf{\textit{Graviton-Scalar}:}\\
The non-trivial split factors are given in Appendix \ref{A7}. For all the factorisation channel for the Scalars in $\mathcal{N}=8$ the split factors will remain the same. Hence
\begin{equation*}
    \boxed{M_n\left(1^{+2}, 2^{ABCD; 0}, \cdots, n\right)=\frac{\omega_2}{\omega_1} \frac{\bar{z}_{12}}{z_{12}}\times M_{n-1}\left(p^{ABCD; 0},\cdots, n\right)}
\end{equation*}
\textbf{\textit{Gravitino-Graviphoton}:}\\
The split factors for this collinear pair are given in Appendix \ref{A8}. For any $1\leq A,B\leq 8$ we have
\begin{equation*}
    \boxed{M_n\left(1^{A;+\frac{3}{2}}, 2_{BC}^{-1}, \cdots, n\right) =\frac{\omega_2^2}{\omega_p^{\frac{3}{2}} \omega_1^{\frac{1}{2}}} \frac{\bar{z}_{12}}{z_{12}} 2!\delta^A_{[B}\times M_{n-1}\left(p_{C]}^{ -\frac{3}{2}},\cdots, n\right)}
\end{equation*}
Here $[...]$ indicates antisymmetrized indices defined by 
\begin{equation}
p_{[A_1\dots A_n]}:=\frac{1}{n!}\sum_{\sigma\in S_n}\text{sign}(\sigma)\;p_{A_{\sigma(1)}\dots A_{\sigma(n)}}.
\end{equation}
\textbf{\textit{Gravitino-Graviphotino}:}\\
The split factors are given in Appendix \ref{A9}.
\begin{table}[H]
\renewcommand{\arraystretch}{1.5}
\centering
\begin{tabular}{|c|c|}
\hline
 $M_n\left(1^{A;+\frac{3}{2}}, 2^{BCD; +\frac{1}{2}}, \cdots \right)$ & $\sqrt{\frac{\omega_2}{\omega_1}} \frac{\bar{z}_{12}}{z_{12}}\times M_{n-1}\left(p^{ABCD; 0},\cdots \right) $\\\hline
 $M_n\left(1^{A;+\frac{3}{2}}, 2_{BCD}^{ -\frac{1}{2}}, \cdots \right)$& $-\frac{\omega_2^{\frac{3}{2}}}{\omega_p\omega_1^{\frac{1}{2}}}  \frac{\bar{z}_{12}}{z_{12}}\times 3\delta^A_{(B} M_{n-1}\left(p^{-1}_{CD)},\cdots \right)$\\\hline
\end{tabular}
\caption{Amplitude corresponding to collinear gravitino and graviphotino}    \label{tab:gravitinograviphotinocol}
\end{table}
Here $(...)$ indicates symmetrized indices defined by
\begin{equation}
p_{(A_1\dots A_n)}:=\frac{1}{n!}\sum_{\sigma\in S_n}p_{A_{\sigma(1)}\dots A_{\sigma(n)}}.
\end{equation}
\noindent\textbf{\textit{Gravitino-scalar}:}\\
The splits are given in Appendix \ref{A10}.
\begin{equation*}
    \boxed{M_n\left(1^{A;+\frac{3}{2}}, 2^{BCDE; 0}, \cdots, n\right)=-\frac{1}{3!}\epsilon^{ABCDEFGH}\frac{\omega_2}{\sqrt{\omega_1 \omega_p}} \frac{\bar{z}_{12}}{z_{12}}\times M_{n-1}\left(p_{FGH}^{-\frac{1}{2}},\cdots, n\right) }
\end{equation*}
\textbf{\textit{Graviphoton-Graviphotino}:}\\
The splits are in Appendix \ref{A11}.
\begin{table}[H]
\renewcommand{\arraystretch}{1.5}
\centering
\begin{tabular}{|c|c|}
\hline
 $M_n\left(1^{ab;+1}, 2^{cdr; +\frac{1}{2}}, \cdots, n\right)$ & $\sqrt{\frac{\omega_2}{\omega_1}} \frac{\bar{z}_{12}}{z_{12}}\times M_{n-1}\left(p^{ABCD; 0},\cdots \right) $\\\hline
 $M_n\left(1^{ab;+1}, 2_{cdr; \; -\frac{1}{2}}, \cdots, n\right)$& $-\delta^{ab}_{cd} \; \frac{\omega_2^{\frac{3}{2}}}{\omega_p^{\frac{3}{2}}} \frac{\bar{z}_{12}}{z_{12}}\times M_{n-1}\left(p_{r}^{ -\frac{3}{2}},\cdots, n\right)$\\\hline
\end{tabular}
\caption{Amplitude corresponding to collinear graviphoton and graviphotino}    \label{tab:graphphcol}
\end{table}
\noindent\textbf{\textit{Graviphoton-scalar}:}\\
The splits for this collinear pair are in Appendix \ref{A12}.
\begin{table}[H]
\renewcommand{\arraystretch}{1.5}
\centering
\begin{tabular}{|c|c|}
\hline
 $M_n\left(1^{ab;+1}, 2^{cdrs; 0}, \cdots, n\right)$ & $\epsilon^{abcd} \epsilon^{rstu}\;\frac{\omega_2}{\omega_p}\frac{\bar{z}_{12}}{z_{12}}\times M_{n-1}\left(p_{tu}^{-1},\cdots, n\right) $\\\hline
 $M_n\left(1^{rs;+1}, 2^{abtu; 0}, \cdots, n\right)$& $\epsilon^{rstu} \epsilon^{abcd} \;\frac{\omega_2}{\omega_p}\frac{\bar{z}_{12}}{z_{12}}\times M_{n-1}\left(p_{cd}^{-1},\cdots, n\right)$\\\hline
 $M_n\left(1^{ar;+1}, 2^{bcst; 0}, \cdots, n\right)$ & $\epsilon^{abcd} \epsilon^{rstu} \;\frac{\omega_2}{\omega_p}\frac{\bar{z}_{12}}{z_{12}}\times M_{n-1}\left(p_{du}^{-1},\cdots, n\right)$ \\\hline
 $M_n\left(1^{ab;+1}, 2^{cdef; 0}, \cdots, n\right)$ & $-\epsilon^{cdef}  \epsilon^{abgh} \;\frac{\omega_2}{\omega_p}\frac{\bar{z}_{12}}{z_{12}}\times M_{n-1}\left(p_{gh}^{-1},\cdots, n\right)$ \\\hline
 $M_n\left(1^{rs;+1}, 2^{cdef; 0}, \cdots, n\right)$ & $-\epsilon^{cdef} \epsilon^{rstu}\;\frac{\omega_2}{\omega_p}\frac{\bar{z}_{12}}{z_{12}}\times M_{n-1}\left(p_{tu}^{-1},\cdots, n\right)$ \\\hline
 $M_n\left(1^{ar;+1}, 2^{bcds; 0}, \cdots, n\right)$ & $-\epsilon^{abcd} \epsilon^{rstu} \;\frac{\omega_2}{\omega_p}\frac{\bar{z}_{12}}{z_{12}}\times M_{n-1}\left(p_{tu}^{-1},\cdots, n\right)$ \\\hline
 $M_n\left(1_{ab}^{-1}, 2^{cdef; 0}, \cdots, n\right)$ & $-\epsilon^{cdef}  \epsilon_{abgh} \;\frac{\omega_2}{\omega_p}\frac{z_{12}}{\bar{z}_{12}}\times M_{n-1}\left(p^{gh;+1},\cdots, n\right)$ \\ \hline
 $M_n\left(1_{ar}^{-1}, 2^{bcds; 0}, \cdots, n\right)$ & $-\epsilon^{bcde} \delta^{s}_{r}\epsilon_{aefg}\;\frac{\omega_1}{\omega_p}\frac{z_{12}}{\bar{z}_{12}}\times M_{n-1}\left(p^{fg;+1},\cdots, n\right)$ \\ \hline
 $M_n\left(1_{ar}^{-1}, 2^{bcst; 0}, \cdots, n\right)$ & $-\frac{\omega_2}{\omega_p}\frac{z_{12}}{\bar{z}_{12}}4!\delta^{[b}_{a} \delta^{s}_{r}\;  M_{n-1}\left(p^{tc];+1},\cdots, n\right)$\\ \hline
\end{tabular}
\caption{Amplitude corresponding to collinear graviphoton and scalar}    \label{tab:graphphsccol}
\end{table}
\noindent\textbf{\textit{Graviphotino-scalar}:}\\
The splits are given in Appendix \ref{A13}.
\begin{table}[H]
\renewcommand{\arraystretch}{1.5}
\centering
\begin{tabular}{|c|c|}
\hline
 $M_n\left(1^{abr;+\frac{1}{2}}, 2^{cdst; 0}, \cdots, n\right)$ & $\epsilon^{abcd}\epsilon^{rstu} \;\frac{\omega_1^{\frac{1}{2}}  \omega_2}{\omega_p^{\frac{3}{2}}}\frac{\bar{z}_{12}}{z_{12}}\times M_{n-1}\left(p_{u}^{-\frac{3}{2}},\cdots, n\right) $\\\hline
 $M_n\left(1^{abr;+\frac{1}{2}}, 2^{cstu; 0}, \cdots, n\right)$& $-\epsilon^{abcd}\epsilon^{rstu} \;\frac{\omega_1^{\frac{1}{2}}  \omega_2}{\omega_p^{\frac{3}{2}}}\frac{\bar{z}_{12}}{z_{12}}\times M_{n-1}\left(p_{d}^{-\frac{3}{2}},\cdots, n\right)$\\\hline
 $M_n\left(1^{ars;+\frac{1}{2}}, 2^{bctu; 0}, \cdots, n\right)$ & $\epsilon^{abcd}\epsilon^{rstu} \;\frac{\omega_1^{\frac{1}{2}}  \omega_2}{\omega_p^{\frac{3}{2}}}\frac{\bar{z}_{12}}{z_{12}}\times M_{n-1}\left(p_{d}^{-\frac{3}{2}},\cdots, n\right)$ \\\hline
 $M_n\left(1_{ars}^{-\frac{1}{2}}, 2^{bctu;0}, \cdots, n\right)$ & $-2!\delta^{tu}_{rs}\frac{\omega_1^{\frac{1}{2}}  \omega_2}{\omega_p^{\frac{3}{2}}}\frac{z_{12}}{\bar{z}_{12}}\delta^{[b}_{a} \times M_{n-1}\left(p^{c];+\frac{3}{2}},\cdots, n\right)$ \\\hline
 $M_n\left(1_{ars}^{-\frac{1}{2}}, 2^{btuv~0}, \cdots, n\right)$ & $-\delta_{a}^{b}\epsilon^{tuvw} \epsilon_{wrsx}\;\frac{\omega_1^{\frac{1}{2}}  \omega_2}{\omega_p^{\frac{3}{2}}}\frac{z_{12}}{\bar{z}_{12}}\times M_{n-1}\left(p^{x;+\frac{3}{2}},\cdots, n\right)$ \\\hline
 $M_n\left(1_{rst}^{-\frac{1}{2}}, 2^{avwx; 0}, \cdots, n\right)$ & $-\epsilon_{rstu} \epsilon^{vwxu} \;\frac{\omega_1^{\frac{1}{2}}  \omega_2}{\omega_p^{\frac{3}{2}}}\frac{z_{12}}{\bar{z}_{12}}\times M_{n-1}\left(p^{a;+\frac{3}{2}},\cdots, n\right)$ \\\hline
 $M_n\left(1_{rst}^{-\frac{1}{2}}, 2^{uvwx; 0}, \cdots, n\right)$ & $-\epsilon_{rsty} \epsilon^{uvwx} \;\frac{\omega_1^{\frac{1}{2}}  \omega_2}{\omega_p^{\frac{3}{2}}}\frac{z_{12}}{\bar{z}_{12}}\times M_{n-1}\left(p^{y;+\frac{3}{2}},\cdots, n\right)$ \\ \hline
\end{tabular}
\caption{Amplitude corresponding to collinear graviphotino and scalar}    \label{tab:graphtphsccol}
\end{table}
\section{Soft Limits in $\mathcal{N}$=8 supergravity}\label{sec:softlimsugra}
To complete our study, we now move on to study the soft limit of supergravity amplitudes. In particular, in this section, we will compute the soft limits of graviton and gravitinos up to sub-subleaing order. As explained earlier,
for both holomorphic and antiholomorphic soft limits for supergravity amplitudes we have,
\begin{equation}
\begin{split}
\mathcal{M}_{n+1}\left(\Psi_s,\Psi_1, \ldots, \Psi_n\right)&\stackrel{\varepsilon \to 0}{=}\sum_{k=0}^2 \frac{1}{\varepsilon^{3-k}} \mathcal{S}^{(k)} \mathcal{M}_n\left(\Psi_1, \ldots, \Psi_n\right)\quad\text{(holomorphic soft limit)}\\
\mathcal{M}_{n+1}\left(\Psi_s,\Psi_1, \ldots, \Psi_n\right)&\stackrel{\varepsilon \to 0}{=}\sum_{k=0}^2 \frac{1}{\varepsilon^{3-k}} \overline{\mathcal{S}}^{(k)} \mathcal{M}_n\left(\Psi_1, \ldots, \Psi_n\right) \quad\text{(antiholomorphic soft limit)} 
\end{split}
\end{equation}
where in both the cases the holomorphic and antiholomorphic soft limits are  parametrised by $\varepsilon \to 0$ for the soft superfield $\Psi_s$ and $\mathcal{S}^{(k)}$ and $\overline{\mathcal{S}}^{(k)}$ are soft operators corresponding to these limits.

\subsection{Graviton soft limit}\label{subsec:softgraviton}
Recall that in the physical soft limit $p_{s} \rightarrow 0$ or equivalently $h_{s}, \tilde{h}_{s} \rightarrow 0$, the leading soft factor in SYM is given by the sum of leading soft factors in holomorphic and anti-holomorphic soft limit:
\begin{equation}
\text {Soft}_{\text {leading }}^{\mathrm{SYM}}(a, s, b)=\frac{\langle a b\rangle}{\langle a s\rangle\langle s b\rangle}+\frac{[a b]}{[a s][s b]} \delta^{4}\left(\eta^{s}\right)
\end{equation}
As described in Section \ref{sec:dc}, we will use the double copy relation \cite[Eq. 2.15]{He:2014bga}
\begin{equation}
\begin{split}
    \frac{1}{\varepsilon^3}\text {Soft($0$)}^{\mathrm{SG}}&+\frac{1}{\varepsilon^2}\text {Soft($1$)}^{\mathrm{SG}}+\frac{1}{\varepsilon}\text {Soft($2$)}^{\mathrm{SG}}\\&=\sum_{i=1}^n\varepsilon\langle si\rangle[is]\left(\frac{1}{\varepsilon^2}\text {Soft($0$)}^{\mathrm{SYM}}(i, s, a)+\frac{1}{2\varepsilon}\text {Soft($1$)}^{\mathrm{SYM}}(i, s, a)\right)^2.
    \end{split}
\end{equation}
Comparing the coefficients of $\varepsilon$ powers, we get 
\begin{equation}
    \begin{split}
       &\text {Soft($0$)}^{\mathrm{SG}}=\sum_{i=1}^n\langle si\rangle[is]\left[\text {Soft($0$)}^{\mathrm{SYM}}(i, s, a)\right]^2 \\&\text {Soft($1$)}^{\mathrm{SG}}=\sum_{i=1}^n\langle si\rangle[is]\left[\text {Soft($0$)}^{\mathrm{SYM}}(i, s, a)\times\text {Soft($1$)}^{\mathrm{SYM}}(i, s, a)\right]\\&\text {Soft($2$)}^{\mathrm{SG}}=\frac{1}{4}\sum_{i=1}^n\langle si\rangle[is]\left[\text {Soft($1$)}^{\mathrm{SYM}}(i, s, a)\right]^2
    \end{split}
\end{equation}
Thus the double copy relation gives the sum of leading, subleading and subsubleading soft factors in supergravity in terms of the leading and subleading soft factors in SYM. It is clear that the leading and subleading soft factors in supergravity are given by 
\begin{equation}
    \text {Soft}_{\text {leading }}^{\mathrm{SG}}=\sum_{i=1}^n\langle si\rangle[is]\left([\text {Soft(0)}_{\mathrm{hol}}^{\mathrm{SYM}}(i, s, a)]^2+[\text {Soft(0)}_{\mathrm{anti-hol}}^{\mathrm{SYM}}(i, s, a)]^2\right)
    \label{eq:leadsoftfacsugra}
\end{equation}
and 
\begin{equation}
\begin{split}
    \text {Soft}_{\text {subleading }}^{\mathrm{SG}}=\sum_{i=1}^n\langle si\rangle[is]\bigg[&\text {Soft(0)}_{\mathrm{hol}}^{\mathrm{SYM}}(i, s, a)\times\text {Soft(1)}_{\mathrm{hol}}^{\mathrm{SYM}}(i, s, a)\\&+\text {Soft(0)}_{\mathrm{anti-hol}}^{\mathrm{SYM}}(i, s, a)\times\text {Soft(1)}_{\mathrm{anti-hol}}^{\mathrm{SYM}}(i, s, a)\bigg]
\end{split}
\label{eq:subleadsoft}
\end{equation}
We now substitute \eqref{eq:holsoftsym} and \eqref{eq:antiholsoftsym} into \eqref{eq:leadsoftfacsugra} and \eqref{eq:subleadsoft} to get the leading and subleading soft factors in supergravity. Note that the nonholomorphic soft factor in SYM includes the Grassmann delta function $\delta^4(\eta)$. So while squaring the nonholomorphic soft factor of SYM, the square of this delta function in double copy is interpreted as the Grassmann delta function on $\mathcal{N}=8$ superspace: 
\begin{equation}
    \left(\delta^4(\eta_a)\right)^2=\delta^8(\eta_A)
\end{equation}
where the indices have the usual meanings with $a$ running from 1 to 4 and $A$ running from 1 to 8. The leading soft factor is then given by
\begin{equation}
    \text {Soft}_{\text {leading }}^{\mathrm{SG}}(a,s,b)=\sum_{i=1}^n\left(\frac{[si]}{\langle si\rangle}\frac{\langle ai\rangle^2}{\langle as\rangle^2}+\frac{\langle si\rangle}{[si]}\frac{[ai]^2}{[as]^2}\delta^8(\eta_A)\right)
\end{equation}
We now evaluate the subleading soft limit. From \eqref{eq:antiholsoftsym} and \eqref{eq:subleadsoft} we have,
\begin{equation*}
\begin{split}
    &\text {Soft}_{\text {subleading }}^{\mathrm{SG}}\\
    &= \sum_{i=1}^n\langle si\rangle[is] \Bigg[ \frac{\langle ia \rangle }{\langle is \rangle \langle sa \rangle} \Bigg\{ \frac{1}{\langle sa \rangle} \left(\tilde{h}_{s}^{\dot{\alpha}} \frac{\partial}{\partial \tilde{h}_{a}^{\dot{\alpha}}}+\eta^{s}_{A} \frac{\partial}{\partial \eta^{a}_{A}}\right) + \frac{1}{\langle is \rangle}\left(\tilde{h}_{s}^{\dot{\alpha}} \frac{\partial}{\partial \tilde{h}_{i}^{\dot{\alpha}}}+\eta_{s}^{A} \frac{\partial}{\partial \eta_{i}^{A}}\right) \Bigg\}\\
    &+ \frac{[ia]}{[is][sa]}\delta^{8}\left(\eta^{s}+\frac{[a s]}{[a b]} \eta^{b}+\frac{[s b]}{[a b]} \eta^{a}\right) \Bigg(\frac{1}{[is]} h_{s}^{\alpha} \frac{\partial}{\partial h_{i}^{\alpha}}+\frac{1}{[sa]} h_{s}^{\alpha} \frac{\partial}{\partial h_{a}^{\alpha}}\Bigg)\Bigg]\\&=\sum_{i=1}^n\langle si\rangle[is] \Bigg[ \frac{\langle ia \rangle }{\langle is \rangle^2 \langle sa \rangle} \left(\tilde{h}_{s}^{\dot{\alpha}} \frac{\partial}{\partial \tilde{h}_{i}^{\dot{\alpha}}}+\eta^{s}_{A} \frac{\partial}{\partial \eta^{i}_{A}}\right)+ \frac{[ia]}{[is]^2[sa]}\delta^{8}\left(\eta^{s}+\frac{[a s]}{[a b]} \eta^{b}+\frac{[s b]}{[a b]} \eta^{a}\right) \Bigg( h_{s}^{\alpha} \frac{\partial}{\partial h_{i}^{\alpha}}\Bigg)\Bigg]
 \end{split}
\end{equation*}
where we used the momentum conservation 
\[
\sum_i\langle si\rangle[ia]=\sum_i[si]\langle ia\rangle =0.
\]
Note that in the soft superfield, $\eta_s\to 0$ gives the positive helicity soft graviton and $\delta^8(\eta_s)$ gives the negative helicity soft graviton. Thus we only retain these terms in the soft factor.  
Thus we get 
\[
\text {Soft}_{\text {subleading }}^{\mathrm{SG}}(a,s,b)=\sum_{i=1}^n\frac{[is]\langle ia\rangle}{\langle si\rangle\langle sa\rangle}\tilde{h}_{s}^{\dot{\alpha}} \frac{\partial}{\partial \tilde{h}_{i}^{\dot{\alpha}}}+\frac{\langle si\rangle [ia]}{[is][sa]}\delta^8(\eta^s)h_{s}^{\alpha} \frac{\partial}{\partial h_{i}^{\alpha}}
\]
which is the sum of soft factor for positive and negative helicity soft graviton in pure gravity \cite[Eq. 2.9]{He:2014bga}. Note that in the above formula, the momenta $p_a$ acts as reference vector and hence can be taken to be any null vector $r$. This is an indication of the diffeomorphism symmetry of gravity amplitudes. We can thus rewrite the soft factor as \begin{equation}
    \text {Soft}_{\text {subleading }}^{\mathrm{SG}}(a,s,b)=\sum_{i=1}^n\frac{[is]\langle ir\rangle}{\langle si\rangle\langle sr\rangle}\tilde{h}_{s}^{\dot{\alpha}} \frac{\partial}{\partial \tilde{h}_{i}^{\dot{\alpha}}}+\frac{\langle si\rangle [ir]}{[is][sr]}\delta^8(\eta^s)h_{s}^{\alpha} \frac{\partial}{\partial h_{i}^{\alpha}}
    \label{eq:gravitonsoftlim}
\end{equation} 
\subsection{Leading soft gravitino limit}
To calculate the soft limit of gravitinos, we use the results of \cite{Liu:2014vva}. Under the holomorphic soft limit of the superfield, we have 
\begin{equation}
 \mathcal{M}_{n+1}\left(\Psi_s,\Psi_1, \ldots, \Psi_n\right)=\left(\frac{1}{\varepsilon^3} \mathcal{S}^{(0)}+\frac{1}{\varepsilon^2} \mathcal{S}^{(1)}+\frac{1}{\varepsilon} \mathcal{S}^{(2)}\right) \mathcal{M}_n\left(\Psi_1, \ldots, \Psi_n\right)+O\left(\varepsilon^0\right) .
 \label{eq:softsupamp}
\end{equation}
The leading soft factor\footnote{note that we have made explicit the reference vector $r$ which was taken to be $p_n$ in \cite{Liu:2014vva}} is same with the one in pure gravity:
\begin{equation}
    \mathcal{S}^{(0)}=\sum_{i=1}^{n} \frac{[si]\langle ri\rangle^2}{\langle si\rangle\langle rs\rangle^2}=S^{(0)}.
\end{equation}
The sub-leading soft operator is given by
\begin{equation}
\mathcal{S}^{(1)}=\sum_{i=1}^{n} \frac{[si]\langle ri\rangle}{\langle si\rangle\langle rs\rangle}\left(\tilde{\lambda}_{s \dot{\alpha}} \frac{\partial}{\partial \tilde{\lambda}_{i\dot{\alpha}}}+\eta_{sA} \frac{\partial}{\partial \eta_{iA}}\right)=S^{(1)}+\eta_{sA} \mathcal{S}^{A(1)}.    
\end{equation}
where \[
\mathcal{S}^{A(1)}= \sum_{i=1}^{n} \frac{[si]\langle ri\rangle}{\langle si\rangle\langle rs\rangle}\frac{\partial}{\partial \eta_{iA}}
\]
Here the leading soft gravitino operator  involves
the first order derivatives with respect to the Grassmannian variables $\eta_{i}$'s. These term will preserves the total helicity as well as $\mathrm{SU}(8)$ $R$-symmetry.\\
The sub-sub-leading soft factor is given by 
\begin{equation}
  \begin{aligned}
\mathcal{S}^{(2)} &=S^{(2)}+\eta_{sA} \mathcal{S}^{A(2)}+\frac{1}{2} \eta_{sA} \eta_{sB} \mathcal{S}^{AB(2)}
\end{aligned}  
\end{equation}
where
\begin{equation}
\begin{split}
S^{(2)} &=\frac{1}{2} \sum_{i=1}^n \frac{[si]}{\langle si\rangle} \tilde{\lambda}_{s \dot{\alpha}} \tilde{\lambda}_{s \dot{\beta}} \frac{\partial^2}{\partial \tilde{\lambda}_{i \dot{\alpha}} \partial \tilde{\lambda}_{i \dot{\beta}}}, \\
\mathcal{S}^{A(2)} &=\sum_{i=1}^n \frac{[si]}{\langle si\rangle} \tilde{\lambda}_{s \dot{\alpha}} \frac{\partial^2}{\partial \tilde{\lambda}_{i\dot{\alpha}} \partial \eta_{aA}}, \\
\mathcal{S}^{AB(2)} &=\sum_{i=1}^n \frac{[si]}{\langle si\rangle} \frac{\partial^2}{\partial \eta_{aB} \partial \eta_{aA}} .
\end{split}
\end{equation}
We now expand the generic superamplitude on the left hand side of \eqref{eq:softsupamp} in the grassmann odd variable $\eta_s$ of the soft superfield:
\begin{equation}
    \begin{aligned}
\mathcal{M}_{n+1}\left(\Psi_s,\Psi_1, \ldots, \Psi_n\right)=& \mathcal{M}_{n+1}\left(H_{s+},\Psi_1, \ldots, \Psi_n\right)+\eta_{sA} \mathcal{M}_{n+1}\left(S^A_{s+},\Psi_1, \ldots, \Psi_n\right) \\
&+\frac{1}{2} \eta_{sA} \eta_{sB} \mathcal{M}_{n+1}\left(G^{AB}_{s+},\Psi_1, \ldots, \Psi_n\right)+\cdots
\end{aligned}
\end{equation}
and compare with the right hand side of \eqref{eq:softsupamp} to get the following soft limits:
\begin{table}[H]
\renewcommand{\arraystretch}{1.5}
\centering
\begin{tabular}{|c|c|}
\hline
\textbf{Soft Superfields} & \textbf{Superamplitude expansion on $\varepsilon \to 0$}\\ \hline
\hline
Soft graviton & $\mathcal{M}_{n+1}\left(H_{s+},\ldots\right)=\left(\frac{1}{\varepsilon^3} S^{(0)}+\frac{1}{\varepsilon^2} S^{(1)}+\frac{1}{\varepsilon} S^{(2)}\right) \mathcal{M}_n+O\left(\varepsilon^0\right)$\\\hline 
  Soft gravitino& $\mathcal{M}_{n+1}\left(S^A_{s+},\ldots\right)=\left(\frac{1}{\varepsilon^2} \mathcal{S}^{A(1)}+\frac{1}{\varepsilon} \mathcal{S}^{A(2)}\right)\mathcal{M}_n+O\left(\varepsilon^0\right) $\\\hline
Soft graviphoton & $\mathcal{M}_{n+1}\left(G^{AB}_{s+}\ldots\right)=\frac{1}{\varepsilon} \mathcal{S}^{AB(2)}\mathcal{M}_n+\mathcal{O}\left(\varepsilon^0\right)$ \\ \hline
Soft graviphotino & $\mathcal{M}_{n+1}\left( \chi^{ ABC}_s\ldots\right)=\frac{0}{\varepsilon}+\mathcal{O}\left(\varepsilon^0\right)$ \\ \hline
  Soft scalar& $\mathcal{M}_{n+1}\left( \Phi^{A B C D}_s,\ldots \right)=\frac{0}{\varepsilon}+\mathcal{O}\left(\varepsilon^0\right)$ \\ \hline
\end{tabular}
\caption{Various soft limit expansion of the superamplitude} \label{tab:softlimits}
\end{table}
One can easily check that the soft graviton limit obtained here coincides with our calculations in Subsection \ref{subsec:softgraviton}. We also see that there are no soft divergences for graviphotino and scalar. 
\section{Conclusion}
In this work we have computed the soft and collinear limits of the maximally supersymmetric $\mathcal{N}=8$ supergravity theory in four spacetime dimensions using the double copy relations in both soft and collinear sectors of $\mathcal{N}=4$ Super Yang-Mills. The computations are done in the celestial basis appropriate for applications to celestial holography. 
An important point in our application of double copy is a different choice of self-duality condition for scalars. The constraints imposed here differs in signs: 
$\alpha_4=\tilde{\alpha}_4=-1$ and $\alpha_8=1$. This choice is motivated by our desire to combine the collinear limits for different factorisations of $\mathcal{N}=4$ SYM to $\mathcal{N}=8$ supergravity. Based on the factorisation of states in the gravity theory in terms of states in the gauge theory, we are able to constrain and determine the R-symmetry indices in the collinear limit. This is also the novelty of this work. \par The goal of this work is twofold: first we would like to construct the dual celestial CFT corresponding to the bulk $\mathcal{N}=8$ supergravity in four dimensions. This requires the collinear limits of bulk amplitudes as they imply the OPEs of (super)conformal operators in the CCFT. Second, we would like to determine the asymptotic symmetries of $\mathcal{N}=8$ supergravity using celestial holography. As discussed in \cite{tab}, our final goal is to determine the contribution of (super)BMS hairs to black hole entropy \footnote{For similar analysis in the context of three dimensional supergravity can be found in \cite{Banerjee:2018hbl} and references there in.}. The first step to such an analysis would be to understand the extension of the BMS group to  \textit{super} BMS group in $\mathcal{N}=8$ supergravity. The corresponding $\mathcal{N}=1$ supergravity case has already been worked out in \cite{Fotopoulos:2020bqj} and a primary construction of the same for higher supersymmetric cases, purely from algebraic perspective, has been addressed in \cite{Banerjee:2022abf}. However a thorough  gravity analysis with higher supersymmetry is still missing. This issue  has been addressed in a companion paper \cite{tab}, where the asymptotic symmetry algebra of the  $\mathcal{N}=8$ supergravity has been derived. 
\section*{Acknowledgement}
We would like to thank Abay Zhakenov for helping us with some Mathematica computations. The work of RKS is supported by the US Department of Energy under grant DE-SC0010008. The work of TR is supported by University Grant Commission, Govt. of India. Finally we highly appreciate the people of India for their support to fundamental research.
\appendix
\section{A brief review of spinor-helicity formalism}\label{app:spinorhelicity}
Recall that helicity spinors are left and right handed representations of the Lorentz group $\mathrm{SO}(1,3)\sim\mathrm{SL}(2,\mathbb{C})$. We denote the left and right handed helicity spinors by $h_{\alpha}$ and $\tilde{h}^{\dot{\alpha}}$ respectively. Lorentz invariant contractions of spinors is defined using the completely antisymmetric rank $2$ tensor $\epsilon^{\alpha\beta}$ defined as 
\begin{equation}
    \epsilon^{\alpha \beta}=-\epsilon_{\alpha \beta}=\epsilon^{\dot{\alpha} \dot{\beta}}=-\epsilon_{\dot{\alpha} \dot{\beta}}=\left(\begin{array}{cc}0 & 1 \\ -1 & 0\end{array}\right).
\end{equation}
The contractions are then defined as 
\begin{equation}
    \begin{array}{l}
\langle \lambda \chi\rangle \equiv \epsilon^{\alpha \beta} \lambda_{\alpha} \chi_{\beta}=\lambda_{\alpha} \chi^{\alpha}=-\lambda^{\alpha} \chi_{\alpha}=-\langle \chi \lambda\rangle \\
{[\lambda \chi] \equiv \epsilon_{\dot{\alpha} \dot{\beta}} \widetilde{\lambda}^{\dot{\alpha}} \widetilde{\chi}^{\dot{\beta}}=\widetilde{\lambda}^{\dot{\alpha}} \widetilde{\chi}_{\dot{\alpha}}=-\widetilde{\lambda}_{\dot{\alpha}} \widetilde{\chi}^{\dot{\alpha}}=-[\chi \lambda]}.
\end{array}
\end{equation}
Wherever we have angular brackets, we understand that it is the contraction of left handed spinor whereas the square bracket is the contraction of the right handed spinor. We thus suggestively denote left handed spinor by $|\lambda\rangle^{\alpha}$ and right handed spinor by $[\lambda|^{\dot{\alpha}}$.
A given null momentum $p^{\mu}$ can be written as a bispinor 
\begin{equation}
    p^{\alpha \dot{\alpha}}=\sigma_{\mu}^{\alpha \dot{\alpha}} p^{\mu}=\left(\begin{array}{cc}
p^{0}+p^{3} & p^{1}-i p^{2} \\
p^{1}+i p^{2} & p^{0}-p^{3}
\end{array}\right)\equiv |p\rangle [p|
\end{equation}
where $\sigma_{\mu}=(1, \sigma_x,\sigma_y,\sigma_z)$ and $|p\rangle, [p|$ are some spinors. For real physical momentum, the two spinors and their contractions are related by complex
conjugation $([p|)^*=|p\rangle)$ and $\langle pq\rangle^*=[qp]$. Given the bispinor of a 4-vector $p^{\mu}$, we can recover the 4-vector as follows:
\[
p^{\mu}=\frac{1}{2} \sigma^{\mu \alpha \dot{\alpha}} p_{\dot{\alpha} \alpha}=\frac{1}{2} \bar{\sigma}_{\dot{\alpha} \alpha}^{\mu}p^{{\alpha} \dot{\alpha}},
\]
where $\bar{\sigma}_{\mu}=(1, -\sigma_x,-\sigma_y,-\sigma_z)$. The inner product of two null momentas $p^{\mu}=|p\rangle [p|$ and $q^{\mu}=|q\rangle [q|$ is given in terms of spinor contractions as 
\begin{equation}
    p\cdot q=\frac{1}{2}[pq]\langle qp\rangle.
\end{equation}
If we have several momenta, which is usually the case in scattering processes, say $p_1,\dots,p_n$, then we shorten the notations further and denote the corresponding spinors by $|i\rangle,[i|$ for $i=1,\dots,n$. The momentum conservation can then be expressed as 
\begin{equation}
    \sum_{j=1}^n\langle ij\rangle [ji]=0
\end{equation}
for $p_i=|i\rangle[i|$. $i$ be any one out of the $n$ external momenta. One can also express polarisations in terms of spinors but we will not need it explicitly in our discussions.

\section{Split Factors}\label{split}
Here we list all of the split factors corresponding to both like and unlike spins in our supergravity theory.\\
\textbf{Gravi-photons splits:} 
\begin{equation}\label{A1}
\begin{split}
&\text{Split}_{0}^{\text {SG}}\left(z, 1^{1+0}, 2^{1+0}\right)= -\frac{[12]}{\langle 12 \rangle},\qquad
\text{Split}_{0}^{\text {SG}}\left(z, 1^{1+0}, 2^{0+1}\right)=- \frac{[12]}{\langle 12 \rangle}\\
&\text{Split}_{0}^{\text {SG}}\left(z, 1^{\frac{1}{2}+\frac{1}{2}}, 2^{\frac{1}{2}+\frac{1}{2}}\right)=-\frac{[12]}{\langle 12 \rangle},\qquad
\text{Split}_{+2}^{\text {SG}}\left(z, 1^{1+0}, 2^{-1+0}\right)=-(1-z)^2\frac{[12]}{\langle 12 \rangle}\\
&\text{Split}_{-2}^{\text {SG}}\left(z, 1^{1+0}, 2^{-1+0}\right)=-z^2 \frac{\langle 12 \rangle}{ [12]} ,\qquad
\text{Split}_{-2}^{\text {SG}}\left(z, 1^{-\frac{1}{2}-\frac{1}{2}}, 2^{\frac{1}{2}+\frac{1}{2}}\right)= -(1-z)^2\frac{\langle 12 \rangle}{ [12]}\\
&\text{Split}_{+2}^{\text {SG}}\left(z, 1^{-\frac{1}{2}-\frac{1}{2}}, 2^{\frac{1}{2}+\frac{1}{2}}\right)=-z^2\frac{[12]}{ \langle 12 \rangle}
\end{split} 
\end{equation}
\textbf{Gravi-photinos splits:}
\begin{equation}\label{A2}
\begin{split}
&\text{Split}_{1}^{\text {SG}}\left(z, 1^{\frac{1}{2}+0}, 2^{\frac{1}{2}+0}\right)=- \sqrt{z(1-z)} \frac{[12]}{\langle 12 \rangle}, \qquad  \text{Split}_{1}^{\text {SG}}\left(z, 1^{\frac{1}{2}+0}, 2^{0+\frac{1}{2}}\right)= -\sqrt{z(1-z)}\frac{[12]}{\langle 12 \rangle}\\
&\text{Split}_{+2}^{\text {SG}}\left(z, 1^{\frac{1}{2}+0}, 2^{-\frac{1}{2}+0}\right)= -\sqrt{z(1-z)^3}\frac{[12]}{\langle 12 \rangle},\qquad \text{Split}_{-2}^{\text {SG}}\left(z, 1^{\frac{1}{2}+0}, 2^{-\frac{1}{2}+0}\right)= -\sqrt{z^3(1-z)}\frac{\langle 12 \rangle}{[12]}\\
&\text{Split}_{1}^{\text {SG}}\left(z, 1^{1 -\frac{1}{2}}, 2^{-\frac{1}{2}+ 1}\right)= -\sqrt{z(1-z)}\frac{[12]}{\langle 12 \rangle },\qquad \text{Split}_{-2}^{\text {SG}}\left(z, 1^{\frac{1}{2}-1}, 2^{-\frac{1}{2} +1}\right)= -\sqrt{z(1-z)^3} \frac{\langle 12 \rangle}{[12]}\\
&\text{Split}_{+2}^{\text {SG}}\left(z, 1^{\frac{1}{2}-1}, 2^{-\frac{1}{2} +1}\right)= -\sqrt{z^3(1-z)}\frac{[12]}{\langle 12 \rangle}.
\end{split} 
\end{equation}
\textbf{Scalars Splits:}
\begin{equation}\label{A3}
\begin{split}
&\text{Split}_{-2}^{\text {SG}}\left(z, 1^{0+0}, 2^{0+0}\right)-z(1-z)\frac{\langle 12 \rangle }{[12]},\qquad \text{Split}_{+2}^{\text {SG}}\left(z, 1^{0+0}, 2^{0+0}\right)= -z(1-z)\frac{[12]}{\langle 12 \rangle}\\
&\text{Split}_{-2}^{\text {SG}}\left(z, 1^{-1+1}, 2^{+1-1}\right)=-z(1-z)\frac{\langle 12 \rangle }{[12]},\qquad \text{Split}_{+2}^{\text {SG}}\left(z, 1^{-1+1}, 2^{+1-1}\right)=-z(1-z)\frac{[12]}{ \langle 12 \rangle}\\
&\text{Split}_{-2}^{\text {SG}}\left(z, 1^{-\frac{1}{2}+\frac{1}{2}}, 2^{+\frac{1}{2}-\frac{1}{2}}\right)=-z(1-z) \frac{\langle 12 \rangle}{[12]}
,\qquad \text{Split}_{+2}^{\text {SG}}\left(z, 1^{-\frac{1}{2}+\frac{1}{2}}, 2^{+\frac{1}{2}-\frac{1}{2}}\right)= -z(1-z) \frac{[12]}{ \langle 12 \rangle}.
\end{split}
\end{equation}
\textbf{Graviton-Gravitino Splits:}
\begin{equation}\label{A4}
\begin{split}
    &\text{Split}_{-\frac{3}{2}}^{\text {SG}}\left(z, 1^{1+1}, 2^{1+\frac{1}{2}}\right)=-\frac{1}{z\sqrt{1-z}}\frac{[12]}{\langle 12 \rangle},\qquad \text{Split}_{-\frac{3}{2}}^{\text {SG}}\left(z, 1^{1+1}, 2^{\frac{1}{2}+1}\right)=- \frac{1}{z\sqrt{(1-z)}} \frac{[12]}{\langle 12 \rangle}\\
    &\text{Split}_{+\frac{3}{2}}^{\text {SG}}\left(z, 1^{1+1}, 2^{-1-\frac{1}{2}}\right)=-\frac{\sqrt{(1-z)^5}}{z}\frac{[12]}{\langle 12 \rangle},\qquad \text{Split}_{+\frac{3}{2}}^{\text {SG}}\left(z, 1^{1+1}, 2^{-\frac{1}{2}-1}\right)=-\frac{\sqrt{(1-z)^5}}{z}\frac{[12]}{\langle 12 \rangle}
\end{split} 
\end{equation}
All other Splits can be written via the helicity flipping relation in Eq. \eqref{flip}.\\ \\
\textbf{Graviton-Graviphoton splits:}
\begin{equation}\label{A5}
\begin{split}
    &\text{Split}_{-1}^{\text {SG}}\left(z, 1^{1+1}, 2^{1+0}\right)=-\frac{1}{z} \frac{[12]}{\langle 12 \rangle},\qquad \text{Split}_{-1}^{\text {SG}}\left(z, 1^{1+1}, 2^{0+1}\right)=-\frac{1}{z} \frac{[12]}{\langle 12 \rangle}\\
    &\text{Split}_{+1}^{\text {SG}}\left(z, 1^{1+1}, 2^{0-1}\right)=-\frac{(1-z)^2}{z}\frac{[12]}{\langle 12 \rangle},\qquad \text{Split}_{+1}^{\text {SG}}\left(z, 1^{1+1}, 2^{-1+0}\right)=-\frac{(1-z)^2}{z}\frac{[12]}{\langle 12 \rangle}\\
     &\text{Split}_{-1}^{\text {SG}}\left(z, 1^{1+1}, 2^{\frac{1}{2}+\frac{1}{2}}\right)=-\frac{1}{z}\frac{[12]}{\langle 12 \rangle},\qquad \text{Split}_{+1}^{\text {SG}}\left(z, 1^{1+1}, 2^{-\frac{1}{2}-\frac{1}{2}}\right)=-\frac{(1-z)^2}{z}\frac{[12]}{\langle 12 \rangle}
\end{split}
\end{equation}
Rest are summarised in  Eq. \eqref{flip}.\\ \\
\textbf{Graviton-Graviphotino splits:}
\begin{equation}\label{A6}
    \begin{split}
         &\text{Split}_{-\frac{1}{2}}^{\text {SG}}\left(z, 1^{1+1}, 2^{\frac{1}{2}+0}\right)=-\frac{\sqrt{(1-z)}}{z}\frac{[12]}{\langle 12 \rangle},\qquad \text{Split}_{\frac{1}{2}}^{\text {SG}}\left(z, 1^{1+1}, 2^{-\frac{1}{2}+0}\right)= -\frac{\sqrt{(1-z)^3}}{z} \frac{[12]}{\langle 12 \rangle}\\
         &\text{Split}_{-\frac{1}{2}}^{\text {SG}}\left(z, 1^{1+1}, 2^{0+\frac{1}{2}}\right)=-\frac{\sqrt{(1-z)}}{z} \frac{[12]}{\langle 12 \rangle},\qquad \text{Split}_{\frac{1}{2}}^{\text {SG}}\left(z, 1^{1+1}, 2^{0-\frac{1}{2}}\right)= -\frac{\sqrt{(1-z)^3}}{z} \frac{[12]}{\langle 12 \rangle}\\
         &\text{Split}_{-\frac{1}{2}}^{\text {SG}}\left(z, 1^{1+1}, 2^{1-\frac{1}{2}}\right)=-\frac{\sqrt{(1-z)}}{z} \frac{[12]}{\langle 12 \rangle},\qquad \text{Split}_{\frac{1}{2}}^{\text {SG}}\left(z, 1^{1+1}, 2^{-1+\frac{1}{2}}\right)= -\frac{\sqrt{(1-z)^3}}{z} \frac{[12]}{\langle 12 \rangle}\\
         &\text{Split}_{-\frac{1}{2}}^{\text {SG}}\left(z, 1^{1+1}, 2^{-\frac{1}{2}+1}\right)=-\frac{\sqrt{(1-z)}}{z} \frac{[12]}{\langle 12 \rangle},\qquad \text{Split}_{\frac{1}{2}}^{\text {SG}}\left(z, 1^{1+1}, 2^{\frac{1}{2}-1}\right)= -\frac{\sqrt{(1-z)^3}}{z} \frac{[12]}{\langle 12 \rangle}
\end{split}
\end{equation}
\textbf{Graviton-Scalar Splits:}
\begin{equation}\label{A7}
    \begin{split}
         &\text{Split}_{0}^{\text {SG}}\left(z, 1^{1+1}, 2^{0+0}\right)=-\frac{(1-z)}{z} \frac{[12]}{\langle 12 \rangle}\\
         &\text{Split}_{0}^{\text {SG}}\left(z, 1^{1+1}, 2^{1-1}\right)=-\frac{(1-z)}{z} \frac{[12]}{\langle 12 \rangle}\\
         &\text{Split}_{0}^{\text {SG}}\left(z, 1^{1+1}, 2^{\frac{1}{2}-\frac{1}{2}}\right)=-\frac{(1-z)}{z} \frac{[12]}{\langle 12 \rangle}
\end{split}
\end{equation}
\textbf{Gravitino-Graviphoton Splits:}
\begin{equation}\label{A8}
    \begin{split}
        &\text{Split}_{-\frac{1}{2}}^{\text {SG}}\left(z, 1^{\frac{1}{2}+1}, 2^{0+1}\right)=-\frac{1}{\sqrt{z}} \frac{[12]}{\langle 12 \rangle},\qquad \text{Split}_{-\frac{1}{2}}^{\text {SG}}\left(z, 1^{\frac{1}{2}+1}, 2^{1+0}\right)=-\frac{1}{\sqrt{z}} \frac{[12]}{\langle 12 \rangle}\\
        &\text{Split}_{-\frac{1}{2}}^{\text {SG}}\left(z, 1^{\frac{1}{2}+1}, 2^{\frac{1}{2}+\frac{1}{2}}\right)=-\frac{1}{\sqrt{z}} \frac{[12]}{\langle 12 \rangle},\qquad \text{Split}_{-\frac{1}{2}}^{\text {SG}}\left(z, 1^{1+\frac{1}{2}}, 2^{0+1}\right)=-\frac{1}{\sqrt{z}} \frac{[12]}{\langle 12 \rangle}\\
        &\text{Split}_{-\frac{1}{2}}^{\text {SG}}\left(z, 1^{1+\frac{1}{2}}, 2^{1+0}\right)=-\frac{1}{\sqrt{z}} \frac{[12]}{\langle 12 \rangle},\qquad \text{Split}_{-\frac{1}{2}}^{\text {SG}}\left(z, 1^{1+\frac{1}{2}}, 2^{\frac{1}{2}+\frac{1}{2}}\right)=-\frac{1}{\sqrt{z}} \frac{[12]}{\langle 12 \rangle}\\
        &\text{Split}_{+\frac{3}{2}}^{\text {SG}}\left(z, 1^{\frac{1}{2}+1}, 2^{0-1}\right)=- \frac{(1-z)^2}{\sqrt{z}}\frac{[12]}{\langle 12 \rangle},\qquad \text{Split}_{+\frac{3}{2}}^{\text {SG}}\left(z, 1^{1+\frac{1}{2}}, 2^{-1+0}\right)=- \frac{(1-z)^2}{\sqrt{z}}\frac{[12]}{\langle 12 \rangle}
    \end{split}
\end{equation}
\textbf{Gravitino-Graviphotino Splits:}
\begin{equation}\label{A9}
    \begin{split}
         &\text{Split}_{0}^{\text {SG}}\left(z, 1^{\frac{1}{2}+1}, 2^{0+\frac{1}{2}}\right)=-\sqrt{\frac{(1-z)}{z}} \frac{[12]}{\langle 12 \rangle},\qquad \text{Split}_{0}^{\text {SG}}\left(z, 1^{\frac{1}{2}+1}, 2^{\frac{1}{2}+0}\right)=-\sqrt{\frac{(1-z)}{z}} \frac{[12]}{\langle 12 \rangle}\\
         &\text{Split}_{0}^{\text {SG}}\left(z, 1^{\frac{1}{2}+1}, 2^{1-\frac{1}{2}}\right)=-\sqrt{\frac{(1-z)}{z}} \frac{[12]}{\langle 12 \rangle},\qquad \text{Split}_{0}^{\text {SG}}\left(z, 1^{\frac{1}{2}+1}, 2^{-\frac{1}{2}+1}\right)=-\sqrt{\frac{(1-z)}{z}} \frac{[12]}{\langle 12 \rangle}\\
         &\text{Split}_{+1}^{\text {SG}}\left(z, 1^{\frac{1}{2}+1}, 2^{0-\frac{1}{2}}\right)=-\sqrt{\frac{(1-z)^3}{z}} \frac{[12]}{\langle 12 \rangle},\qquad\text{Split}_{+1}^{\text {SG}}\left(z, 1^{\frac{1}{2}+1}, 2^{-\frac{1}{2}+0}\right)=-\sqrt{\frac{(1-z)^3}{z}} \frac{[12]}{\langle 12 \rangle}\\
         &\text{Split}_{+1}^{\text {SG}}\left(z, 1^{\frac{1}{2}+1}, 2^{\frac{1}{2}-1}\right)=-\sqrt{\frac{(1-z)^3}{z}} \frac{[12]}{\langle 12 \rangle},\qquad\text{Split}_{0}^{\text {SG}}\left(z, 1^{1+\frac{1}{2}}, 2^{0+\frac{1}{2}}\right)=-\sqrt{\frac{(1-z)}{z}} \frac{[12]}{\langle 12 \rangle}\\
         &\text{Split}_{0}^{\text {SG}}\left(z, 1^{1+\frac{1}{2}}, 2^{\frac{1}{2}+0}\right)=-\sqrt{\frac{(1-z)}{z}} \frac{[12]}{\langle 12 \rangle},\qquad \text{Split}_{0}^{\text {SG}}\left(z, 1^{1+\frac{1}{2}}, 2^{1-\frac{1}{2}}\right)=-\sqrt{\frac{(1-z)}{z}} \frac{[12]}{\langle 12 \rangle}\\
         &\text{Split}_{0}^{\text {SG}}\left(z, 1^{1+\frac{1}{2}}, 2^{-\frac{1}{2}+1}\right)=-\sqrt{\frac{(1-z)}{z}} \frac{[12]}{\langle 12 \rangle},\qquad \text{Split}_{+1}^{\text {SG}}\left(z, 1^{1+\frac{1}{2}}, 2^{0-\frac{1}{2}}\right)=-\sqrt{\frac{(1-z)^3}{z}} \frac{[12]}{\langle 12 \rangle}\\
         &\text{Split}_{+1}^{\text {SG}}\left(z, 1^{1+\frac{1}{2}}, 2^{-\frac{1}{2}+0}\right)=-\sqrt{\frac{(1-z)^3}{z}} \frac{[12]}{\langle 12 \rangle},\qquad \text{Split}_{+1}^{\text {SG}}\left(z, 1^{1+\frac{1}{2}}, 2^{-1+\frac{1}{2}}\right)=-\sqrt{\frac{(1-z)^3}{z}} \frac{[12]}{\langle 12 \rangle}
 \end{split}
\end{equation}
\textbf{Gravitino-Scalar Splits:}
\begin{equation}\label{A10}
    \begin{split}
        &\text{Split}_{\frac{1}{2}}^{\text {SG}}\left(z, 1^{\frac{1}{2}+1}, 2^{0+0}\right)=-\frac{(1-z)}{\sqrt{z}} \frac{[12]}{\langle 12 \rangle},\qquad \text{Split}_{\frac{1}{2}}^{\text {SG}}\left(z, 1^{\frac{1}{2}+1}, 2^{1-1}\right)=-\frac{(1-z)}{\sqrt{z}} \frac{[12]}{\langle 12 \rangle}\\
        &\text{Split}_{\frac{1}{2}}^{\text {SG}}\left(z, 1^{\frac{1}{2}+1}, 2^{\frac{1}{2}-\frac{1}{2}}\right)=-\frac{(1-z)}{\sqrt{z}} \frac{[12]}{\langle 12 \rangle},\qquad \text{Split}_{\frac{1}{2}}^{\text {SG}}\left(z, 1^{1+\frac{1}{2}}, 2^{0+0}\right)=-\frac{(1-z)}{\sqrt{z}} \frac{[12]}{\langle 12 \rangle}
    \end{split}
\end{equation}
Similarly for other factorisations of Gravitino we have the same split factors.\\
\textbf{Graviphoton-Graviphotino Splits:}
\begin{equation}\label{A11}
    \begin{split}
      &\text{Split}_{\frac{1}{2}}^{\text {SG}}\left(z, 1^{0+1}, 2^{0+\frac{1}{2}}\right)=-\sqrt{(1-z)}\frac{[12]}{\langle 12 \rangle},\qquad\text{Split}_{\frac{1}{2}}^{\text {SG}}\left(z, 1^{0+1}, 2^{\frac{1}{2}+0}\right)=-\sqrt{(1-z)}\frac{[12]}{\langle 12 \rangle}\\
     &\text{Split}_{\frac{1}{2}}^{\text {SG}}\left(z, 1^{0+1}, 2^{1-\frac{1}{2}}\right)=-\sqrt{(1-z)}\frac{[12]}{\langle 12 \rangle},\qquad\text{Split}_{\frac{1}{2}}^{\text {SG}}\left(z, 1^{\frac{1}{2}+\frac{1}{2}}, 2^{0+\frac{1}{2}}\right)=-\sqrt{(1-z)}\frac{[12]}{\langle 12 \rangle}\\
    &\text{Split}_{\frac{1}{2}}^{\text {SG}}\left(z, 1^{\frac{1}{2}+\frac{1}{2}}, 2^{\frac{1}{2}+0}\right)=-\sqrt{(1-z)}\frac{[12]}{\langle 12 \rangle},\qquad\text{Split}_{\frac{1}{2}}^{\text {SG}}\left(z, 1^{\frac{1}{2}+\frac{1}{2}}, 2^{1-\frac{1}{2}}\right)=-\sqrt{(1-z)}\frac{[12]}{\langle 12 \rangle}\\
     &\text{Split}_{\frac{1}{2}}^{\text {SG}}\left(z, 1^{\frac{1}{2}+\frac{1}{2}}, 2^{-\frac{1}{2}+1}\right)=-\sqrt{(1-z)}\frac{[12]}{\langle 12 \rangle},\qquad\text{Split}_{\frac{1}{2}}^{\text {SG}}\left(z, 1^{1+0}, 2^{0+\frac{1}{2}}\right)=-\sqrt{(1-z)}\frac{[12]}{\langle 12 \rangle}\\
    &\text{Split}_{\frac{1}{2}}^{\text {SG}}\left(z, 1^{1+0}, 2^{\frac{1}{2}+0}\right)=-\sqrt{(1-z)}\frac{[12]}{\langle 12 \rangle},\qquad\text{Split}_{\frac{1}{2}}^{\text {SG}}\left(z, 1^{1+0}, 2^{-\frac{1}{2}+1}\right)=-\sqrt{(1-z)}\frac{[12]}{\langle 12 \rangle}\\
       &\text{Split}_{\frac{3}{2}}^{\text {SG}}\left(z, 1^{0+1}, 2^{0-\frac{1}{2}}\right)=-(1-z)^{\frac{3}{2}}\frac{[12]}{\langle 12 \rangle},\qquad\text{Split}_{\frac{3}{2}}^{\text {SG}}\left(z, 1^{0+1}, 2^{\frac{1}{2}-1}\right)=-(1-z)^{\frac{3}{2}}\frac{[12]}{\langle 12 \rangle}\\
     &\text{Split}_{\frac{3}{2}}^{\text {SG}}\left(z, 1^{\frac{1}{2}+\frac{1}{2}}, 2^{0-\frac{1}{2}}\right)=-(1-z)^{\frac{3}{2}}\frac{[12]}{\langle 12 \rangle},\qquad\text{Split}_{\frac{3}{2}}^{\text {SG}}\left(z, 1^{\frac{1}{2}+\frac{1}{2}}, 2^{-\frac{1}{2}+0}\right)=-(1-z)^{\frac{3}{2}}\frac{[12]}{\langle 12 \rangle}\\
    &\text{Split}_{\frac{3}{2}}^{\text {SG}}\left(z, 1^{1+0}, 2^{-\frac{1}{2}+0}\right)=-(1-z)^{\frac{3}{2}}\frac{[12]}{\langle 12 \rangle},\qquad\text{Split}_{\frac{3}{2}}^{\text {SG}}\left(z, 1^{1+0}, 2^{-1+\frac{1}{2}}\right)=-(1-z)^{\frac{3}{2}}\frac{[12]}{\langle 12 \rangle}
    \end{split}
\end{equation}
\textbf{Graviphoton-Scalar Splits:}
\begin{equation}\label{A12}
    \begin{split}
      &\text{Split}_{1}^{\text {SG}}\left(z, 1^{0+1}, 2^{0+0}\right)=-(1-z)\frac{[12]}{\langle 12 \rangle},\qquad\text{Split}_{1}^{\text {SG}}\left(z, 1^{0+1}, 2^{1-1}\right)=-(1-z)\frac{[12]}{\langle 12 \rangle}\\
      &\text{Split}_{1}^{\text {SG}}\left(z, 1^{0+1}, 2^{\frac{1}{2}-\frac{1}{2}}\right)=-(1-z)\frac{[12]}{\langle 12 \rangle},\qquad\text{Split}_{1}^{\text {SG}}\left(z, 1^{1+0}, 2^{0+0}\right)=-(1-z)\frac{[12]}{\langle 12 \rangle}\\
      &\text{Split}_{1}^{\text {SG}}\left(z, 1^{\frac{1}{2}+\frac{1}{2}}, 2^{0+0}\right)=-(1-z)\frac{[12]}{\langle 12 \rangle},\qquad \text{Split}_{+1}^{\text {SG}}\left(z, 1^{\frac{1}{2}+\frac{1}{2}}, 2^{\frac{1}{2}-\frac{1}{2}}\right)=-(1-z)\frac{[12]}{\langle 12 \rangle}\\
      &\text{Split}_{-1}^{\text {SG}}\left(z, 1^{-1+0}, 2^{1-1}\right)=-(1-z)\frac{\langle 12 \rangle}{[12]},\qquad\text{Split}_{-1}^{\text {SG}}\left(z, 1^{-1+0}, 2^{\frac{1}{2}-\frac{1}{2}}\right)=-(1-z)\frac{\langle 12 \rangle}{[12]}\\
      &\text{Split}_{-1}^{\text {SG}}\left(z, 1^{-\frac{1}{2}-\frac{1}{2}}, 2^{\frac{1}{2}-\frac{1}{2}}\right)=-(1-z)\frac{\langle 12 \rangle}{[12]}
 \end{split}
\end{equation}
\textbf{Graviphotino-Scalar Splits:}
\begin{equation}\label{A13}
    \begin{split}
        &\text{Split}_{\frac{3}{2}}^{\text {SG}}\left(z, 1^{0+\frac{1}{2}}, 2^{0+0}\right)=-z^{\frac{1}{2}}(1-z) \frac{[12]}{\langle 12 \rangle},\qquad\text{Split}_{\frac{3}{2}}^{\text {SG}}\left(z, 1^{0+\frac{1}{2}}, 2^{\frac{1}{2}-\frac{1}{2}}\right)=-z^{\frac{1}{2}}(1-z) \frac{[12]}{\langle 12 \rangle}\\
      &\text{Split}_{\frac{3}{2}}^{\text {SG}}\left(z, 1^{\frac{1}{2}+0}, 2^{0+0}\right)=-z^{\frac{1}{2}}(1-z) \frac{[12]}{\langle 12 \rangle},\qquad\text{Split}_{-\frac{3}{2}}^{\text {SG}}\left(z, 1^{-\frac{1}{2}+0}, 2^{\frac{1}{2}-\frac{1}{2}}\right)=- z^{\frac{1}{2}}(1-z) \frac{\langle 12 \rangle }{[12]}\\
      &\text{Split}_{-\frac{3}{2}}^{\text {SG}}\left(z, 1^{-1+\frac{1}{2}}, 2^{1-1}\right)=- z^{\frac{1}{2}}(1-z) \frac{\langle 12 \rangle }{[12]},\qquad \text{Split}_{-\frac{3}{2}}^{\text {SG}}\left(z, 1^{-1+\frac{1}{2}}, 2^{\frac{1}{2}-\frac{1}{2}}\right)=- z^{\frac{1}{2}}(1-z) \frac{\langle 12 \rangle }{[12]}
    \end{split}
\end{equation}
\section{Explicit computations of Amplitudes}\label{details}
In this appendix, we explicitly calculate the collinear limits of states various spin combinations. 
\subsection{Like spins}\label{like}
The collinear limits of gravitons is calculated in Section \ref{likespin} in detail. So we start with collinear limit of gravitinos.\\\\
\noindent\textbf{\textit{Gravitinos}}\\
The factorisation of R-symmetry indices has the form 
\[
\begin{cases}\left(a;\frac{3}{2}\right)=\left(a;\frac{1}{2}\right)\otimes 1 \\
\left(r;\frac{3}{2}\right)=1\otimes\left(r;\frac{1}{2}\right).
\end{cases}
\]
We then have
\begin{equation}
    \begin{split}
M_n\left(1^{a;+\frac{3}{2}}, 2^{b;+\frac{3}{2}}, \cdots, n\right)&=M_n\left(1^{\left(a;\frac{1}{2}\right)\otimes 1}, 2^{\left(b;\frac{1}{2}\right)\otimes 1}, \cdots, n\right)\\
&=\text{Split}_{-1}^{\text {SG}}\left(z, 1^{\frac{1}{2}+1}, 2^{\frac{1}{2}+1}\right) \times M_{n-1}\left(p^{ab;+1},\cdots, n\right)\\
&= \frac{\omega_p}{\sqrt{\omega_1 \omega_2}} \frac{\bar{z}_{12}}{z_{12}} \; M_{n-1}\left(p^{ab;+1},\cdots, n\right)
\end{split}
\label{gravitino1}
\end{equation}
\begin{equation}
    \begin{split}
M_n\left(1^{a;+\frac{3}{2}}, 2^{r;+\frac{3}{2}}, \cdots, n\right)&=M_n\left(1^{\left(a;\frac{1}{2}\right)\otimes 1}, 2^{1\otimes\left(r;\frac{1}{2}\right)}, \cdots, n\right)\\
&=\text{Split}_{-1}^{\text {SG}}\left(z, 1^{\frac{1}{2}+1}, 2^{1+\frac{1}{2}}\right) \times M_{n-1}\left(p^{ar;+1},\cdots, n\right)\\
&= \frac{\omega_p}{\sqrt{\omega_1 \omega_2}} \frac{\bar{z}_{12}}{z_{12}} \; M_{n-1}\left(p^{ar;+1},\cdots, n\right)
\end{split}
\label{gravitino2}
\end{equation}
The collinear limits remains the same under $(a,b)\to (r,s)$. All these can be combined and we can write
\begin{equation}
\begin{split}
M_n\left(1^{A;+\frac{3}{2}}, 2^{B;+\frac{3}{2}}, \cdots, n\right)=\frac{\omega_p}{\sqrt{\omega_1 \omega_2}} \frac{\bar{z}_{12}}{z_{12}} \; M_{n-1}\left(p^{AB;+1},\cdots, n\right)  
\end{split}
\end{equation}
For opposite helicities, we have
\begin{equation}
\begin{split}
M_n\left(1^{a;+\frac{3}{2}}, 2_b^{-\frac{3}{2}}, \cdots, n\right)&=\delta^{a}_b\text{Split}_{-2}^{\text {SG}}\left(z, 1^{\frac{1}{2}+1}, 2^{-\frac{1}{2}-1}\right)M_{n-1}\left(p^{+2},\cdots, n\right)\\&+\delta^{a}_b\text{Split}_{+2}^{\text {SG}}\left(z, 1^{\frac{1}{2}+1}, 2^{-\frac{1}{2}-1}\right)M_{n-1}\left(p^{-2},\cdots, n\right) \\
&= \delta^{a}_b\frac{\omega_2^{\frac{5}{2}}}{\omega_1^{\frac{1}{2}}\omega_p^2}\frac{\bar{z}_{12}}{z_{12}}M_{n-1}\left(p^{-2},\cdots, n\right)+\delta^{a}_b\frac{\omega_1^{\frac{5}{2}}}{\omega_2^{\frac{1}{2}}\omega_p^2}\frac{z_{12}}{\bar{z}_{12}}M_{n-1}\left(p^{+2},\cdots, n\right).
\end{split}
\end{equation}
The collinear limit remains the same under $(a,b)\to (r,s)$. 
Infact since there are no other nontrivial split factors for other factorisations, the above collinear limit is true for any $1\leq A,B\leq 8$:
\begin{equation}
\begin{split}
M_n\left(1^{A;+\frac{3}{2}}, 2_B^{-\frac{3}{2}}, \cdots, n\right)=\delta^{A}_B\frac{\omega_2^{\frac{5}{2}}}{\omega_1^{\frac{1}{2}}\omega_p^2}\frac{\bar{z}_{12}}{z_{12}}M_{n-1}\left(p^{-2},\cdots, n\right)+\delta^{A}_B\frac{\omega_1^{\frac{5}{2}}}{\omega_2^{\frac{1}{2}}\omega_p^2}\frac{z_{12}}{\bar{z}_{12}}M_{n-1}\left(p^{+2},\cdots, n\right).
\end{split}
\label{eq:gravitino}
\end{equation}
\textbf{\textit{Graviphotons}}\\
The factorizations of R-symmetry indices are as follows, 
\[
\begin{cases}\left(ab;1\right)=\left(ab;0\right)\otimes 1 \\
\left(ar;1\right)=\left(a,\frac{1}{2}\right)\otimes\left(r;\frac{1}{2}\right)\\
\left(rs;1\right)=1\otimes\left(rs;0\right)
\end{cases}
\]
Using the split factors from Appendix \ref{split} we have
\begin{equation}
    \begin{split}
M_n\left(1^{ab;+1}, 2^{cd;+1}, \cdots, n\right)&=M_n\left(1^{ab; (1\otimes 0)}, 2^{cd; (1\otimes0)}, \cdots, n\right)\\
&= \text{Split}_{0}^{\text {SG}}\left(z, 1^{1+0}, 2^{1+0}\right) \times M_{n-1}\left(p^{abcd;0},\cdots, n\right)\\
&= \frac{\bar{z}_{12}}{z_{12}} \times M_{n-1}\left(p^{abcd;0},\cdots, n\right)
\end{split}
\end{equation}
Similarly 
\[
M_n\left(1^{rs;+1}, 2^{tu;+1}, \cdots, n\right)=\frac{\bar{z}_{12}}{z_{12}} \times M_{n-1}\left(p^{rstu;0},\cdots, n\right)
\]
Next
\begin{equation}
\begin{split}
M_n\left(1^{rs;+1}, 2^{ab; +1}, \cdots, n\right)&=M_n\left(1^{rs; (1\otimes 0)}, 2^{ab; (0\otimes1)}, \cdots, n\right)\\
&= \frac{\bar{z}_{12}}{z_{12}}\times M_{n-1}\left(p^{rsab;0},\cdots, n\right)
\end{split}
\end{equation}
Similarly
\[
M_n\left(1^{ab;+1}, 2^{rs; +1}, \cdots, n\right)= \frac{\bar{z}_{12}}{z_{12}}\times M_{n-1}\left(p^{abrs;0},\cdots, n\right)
\]
Next
\begin{equation}
    \begin{split}
M_n\left(1^{ar; +1}, 2^{bs; +1}, \cdots, n\right)&=M_n\left(1^{ar; (\frac{1}{2}\otimes\frac{1}{2})}, 2^{bs; (\frac{1}{2}\otimes\frac{1}{2})}, \cdots, n\right)\\
&= \frac{\bar{z}_{12}}{z_{12}} \times M_{n-1}\left(p^{arbs;0},\cdots, n\right)
\end{split}
\end{equation}
These can be combined to write the collinear limit uniformly as 
\begin{equation}
    \begin{split}
        M_n\left(1^{AB; +1}, 2^{CD; +1}, \cdots, n\right)&=\frac{\bar{z}_{12}}{z_{12}} \times M_{n-1}\left(p^{ABCD;0},\cdots, n\right)
    \end{split}
\end{equation}
where $1\leq A,B\leq 8$.
For opposite helicities we have,
\begin{equation}
    \begin{split}
M_n\left(1^{ar;+1}, 2_{bs}^{-1}, \cdots, n\right)&=M_n\left(1^{ar; (\frac{1}{2}\otimes\frac{1}{2})}, 2_{bs}^{(-\frac{1}{2}\otimes-\frac{1}{2})}, \cdots, n\right)\\
&=-\delta^{a}_b\delta^{r}_s \Big[\text{Split}_{-2}^{\text {SG}}\left(z, 1^{\frac{1}{2}+\frac{1}{2}}, 2^{-\frac{1}{2}-\frac{1}{2}}\right) \times M_{n-1}\left(p^{+2},\cdots, n\right)\\
& \qquad + \text{Split}_{+2}^{\text {SG}}\left(z, 1^{\frac{1}{2}+\frac{1}{2}}, 2^{-\frac{1}{2}-\frac{1}{2}}\right) \times M_{n-1}\left(p^{-2},\cdots, n\right) \Big]\\
&=-\delta^{a}_b\delta^{r}_s\Big[\frac{\omega_2^2 }{\omega_p^2}\frac{\bar{z}_{12}}{z_{12}} \times M_{n-1}\left(p^{-2},\cdots, n\right)+ \frac{\omega_1^2 }{\omega_p^2}\frac{z_{12}}{\bar{z}_{12}} \times M_{n-1}\left(p^{+2},\cdots, n\right)\Big]
\end{split}
\label{eq:+1-1arbs}
\end{equation}
 Note that the negative sign in the first comes from the negative sign in the factorisation of negative helicity graviphotons.\\ 
Similarly, 
\begin{equation}
    \begin{split}
M_n\left(1^{ab;+1}, 2_{cd}^{-1}, \cdots, n\right)&=M_n\left(1^{ab; (1\otimes0)}, 2_{cd}^{(-1\otimes0)}, \cdots, n\right)\\
&=\frac{1}{2!}\alpha_4\epsilon_{cdef}\epsilon^{abef} \Big[\text{Split}_{-2}^{\text {SG}}\left(z, 1^{1+0}, 2^{-1+0}\right) \times M_{n-1}\left(p^{+2},\cdots, n\right)\\
& \qquad +  \text{Split}_{+2}^{\text {SG}}\left(z, 1^{1+0}, 2^{-1+0}\right) \times M_{n-1}\left(p^{-2},\cdots, n\right) \Big]\\
&=\alpha_4\delta^{ab}_{cd}\Big[\frac{\omega_2^2 }{\omega_p^2}\frac{\bar{z}_{12}}{z_{12}} \times M_{n-1}\left(p^{-2},\cdots, n\right)+ \frac{\omega_1^2 }{\omega_p^2}\frac{z_{12}}{\bar{z}_{12}} \times M_{n-1}\left(p^{+2},\cdots, n\right)\Big]
\end{split}
\label{eq:collimgg10}
\end{equation}
where the generalised Kronecker delta $\delta^{a_1\dots a_n}_{b_1\dots b_n}$ is defined as 
\begin{equation}
    \delta^{a_1\dots a_n}_{b_1\dots b_n}=\sum_{\sigma\in S_n}\text{sign}(\sigma)\delta^{a_{\sigma(1)}}_{b_1}\dots\delta^{a_{\sigma(n)}}_{b_n},
\end{equation}
and we used the self-duality condition Eq.\eqref{eq:selfdualcond}.
This collinear limit remains the same under $(a,b,c,d)\to (r,s,t,u)$ with $\alpha_4$ replaced by $\tilde{\alpha}_4$. Thus if we pick $\alpha_4=\tilde{\alpha}_4=-1$, then using the fact that $\delta^a_r=0$, we can write the collinear limit of two opposite helicity gauge bosons collectively as 
\begin{equation}
\begin{split}\label{eq:collimggcomb}
    M_n\left(1^{AB;+1}, 2_{CD}^{-1}, \cdots, n\right)&=-\delta^{AB}_{CD}\Big[\frac{\omega_2^2 }{\omega_p^2}\frac{\bar{z}_{12}}{z_{12}} \times M_{n-1}\left(p^{-2},\cdots, n\right)\\
  &\hspace{3cm}+ \frac{\omega_1^2 }{\omega_p^2}\frac{z_{12}}{\bar{z}_{12}} \times M_{n-1}\left(p^{+2},\cdots, n\right)\Big].
\end{split}
\end{equation}
Our choice of the parameters $\alpha_4$ and $\tilde{\alpha}_4$ is purely motivated by our desire to combine the collinear limits for different factorisations of the gauge bosons in supergravity. Other choices of the parameters will introduce some extra negative signs in some of the collinear limits. \\ \\
\textbf{\textit{Graviphotinos}}\\
The factorisation R-symmetry indices is given by
\[
\begin{split}
&\begin{cases} 
\left(abr;\frac{1}{2}\right)= \left( ab; 0 \right)\otimes\left(r;\frac{1}{2}\right)&\\
\left(ars;\frac{1}{2}\right)=\left(a;\frac{1}{2}\right)\otimes \left( rs; 0 \right)
\end{cases}
\\&
\begin{cases}
\left(rst;\frac{1}{2}\right)=-\epsilon^{rstu}(1\otimes (u;-\frac{1}{2})) &\\
\left(abc;\frac{1}{2}\right)=-\epsilon^{abcd}((d;-\frac{1}{2})\otimes 1)
\end{cases} \quad (\text{sum over}~ u,d)
\end{split}
\]
Using this factoriation and the split factors in Appendix \ref{split} the collinear limits of various combination of R-symmetry indices is calculated below. We have  
\begin{equation}
\begin{split}
M_n\left(1^{ars;+\frac{1}{2}}, 2^{btu; \frac{1}{2}}, \cdots, n\right)&=M_n\left(1^{ars; (\frac{1}{2}\otimes0)}, 2^{btu; (\frac{1}{2}\otimes0)}, \cdots, n\right)\\
&=\epsilon^{rstu}\epsilon^{abcd}\;\text{Split}_{1}^{\text {SG}}\left(z, 1^{\frac{1}{2}+0}, 2^{\frac{1}{2}+0}\right)\times M_{n-1}\left(p_{cd}^{-1},\cdots, n\right)\\
&=\epsilon^{rstu} \epsilon^{abcd}\; \frac{\sqrt{\omega_1 \omega_2}}{\omega_p}\frac{\bar{z}_{12}}{z_{12}}\times M_{n-1}\left(p_{cd}^{-1},\cdots, n\right).
\end{split}
\end{equation}
Here the $\epsilon^{rstu}$ factor appears because of the collinear split factor between two scalars in $\mathcal{N}=4$ SYM.\\
 Similarly for other non-trivial factorisation we have,
\begin{equation}
\begin{split}
M_n\left(1^{ars;+\frac{1}{2}}, 2^{bct; +\frac{1}{2}}, \cdots, n\right)&=M_n\left(1^{ars; (\frac{1}{2}\otimes 0)}, 2^{bct; (0\otimes\frac{1}{2})}, \cdots, n\right)\\
&=\epsilon^{abcd}\epsilon^{rstu} \; \frac{\sqrt{\omega_1 \omega_2}}{\omega_p}\frac{\bar{z}_{12}}{z_{12}}\times M_{n-1}\left(p^{-1}_{du},\cdots, n\right)
\end{split}
\end{equation}
 \begin{equation}
\begin{split}
M_n\left(1^{rst;+\frac{1}{2}}, 2^{abc; +\frac{1}{2}}, \cdots, n\right)&=M_n\left(1^{rst; (1\otimes- \frac{1}{2})}, 2^{abc; (- \frac{1}{2}\otimes1)}, \cdots, n\right)\\
&=\epsilon^{rstu}\epsilon^{abcd} \; \frac{\sqrt{\omega_1 \omega_2}}{\omega_p}\frac{\bar{z}_{12}}{z_{12}}\times M_{n-1}\left(p_{ud}^{-1},\cdots, n\right)
\end{split}
\end{equation}
\begin{equation}
\begin{split}
    M_n\left(1^{ars;+\frac{1}{2}}, 2_{btu}^{-\frac{1}{2}}, \cdots, n\right)&=M_n\left(1^{ars; (\frac{1}{2}\otimes 0)}, 2_{btu}^{(-\frac{1}{2}\otimes0)}, \cdots, n\right)\\
&=\epsilon_{tuvw}\epsilon^{rsvw} \delta^{a}_{b} \Big[ \text{Split}_{-2}^{\text {SG}}\left(z, 1^{\frac{1}{2}+0}, 2^{-\frac{1}{2}+0}\right)\times M_{n-1}\left(p^{+2},\cdots, n\right)\\
&\qquad + \text{Split}_{+2}^{\text {SG}}\left(z, 1^{\frac{1}{2}+0}, 2^{-\frac{1}{2}+0}\right)\times M_{n-1}\left(p^{-2},\cdots, n\right)\Big]\\
&=\epsilon_{tuvw}\epsilon^{rsvw} \delta^{a}_{b}\Big[ \frac{\omega_1^{\frac{3}{2}} \omega_2^{\frac{1}{2}}}{\omega_p^2}\frac{z_{12}}{\bar{z}_{12}}\times M_{n-1}\left(p^{+2},\cdots, n\right)\\
&\qquad + \frac{\omega_2^{\frac{3}{2}} \omega_1^{\frac{1}{2}}}{\omega_p^2}\frac{\bar{z}_{12}}{z_{12}}\times M_{n-1}\left(p^{-2},\cdots, n\right) \Big]
\end{split}
\end{equation}\\\\
\textbf{\textit{Scalars}}\\
The three possible channels are $0= 0 \otimes 0$, $0 = \pm 1 \otimes \mp 1$ and $0 = \pm\frac{1}{2} \otimes \mp\frac{1}{2}$. We have the non-trivial splits are given in Appendix \ref{A3}.
The factorization of R-symmetry indices are,
\[
\left(abrs;0\right)=\left(ab;0\right)\otimes (rs; 0)
\]
\[
\begin{cases}\left(abcd;0\right)=-\epsilon^{abcd} (-1\otimes 1) \\ \left(rstu;0\right)=-\epsilon^{rstu} (1\otimes -1 )
\end{cases}
\]
\[
\begin{cases}\left(abcr;0\right)=-\epsilon^{abcd}\left(d;-\frac{1}{2}\right)\otimes (r;\frac{1}{2})\\ \left(arst;0\right)=-\epsilon^{rstu}(a;\frac{1}{2})\otimes\left(u;-\frac{1}{2}\right) 
\end{cases}
\]
The collinear amplitudes are then given by
\begin{equation}
\begin{split}
M_n\left(1^{abrs;0}, 2^{cdtu; 0}, \cdots, n\right)&=M_n\left(1^{abrs; (0\otimes0)}, 2^{cdtu; (0\otimes0)}, \cdots, n\right)\\
&=\epsilon^{abcd} \epsilon^{rstu}\Big[\text{Split}_{-2}^{\text {SG}}\left(z, 1^{0+0}, 2^{0+0}\right)\times M_{n-1}\left(p^{+2},\cdots, n\right)\\
&\qquad + \text{Split}_{+2}^{\text {SG}}\left(z, 1^{0+0}, 2^{0+0}\right)\times M_{n-1}\left(p^{-2},\cdots, n\right)\Big]\\
&= \epsilon^{abcd} \epsilon^{rstu}\Big[\frac{\omega_1 \omega_2}{\omega_p^2}\frac{z_{12}}{\bar{z}_{12}}\times M_{n-1}\left(p^{+2},\cdots, n\right)\\
& \qquad + \frac{\omega_1 \omega_2}{\omega_p^2}\frac{\bar{z}_{12}}{z_{12}}\times M_{n-1}\left(p^{-2},\cdots, n\right)\Big]
\end{split}
\end{equation}
 \begin{equation}
\begin{split}
M_n\left(1^{abcd;0}, 2^{rstu; 0}, \cdots, n\right)&=M_n\left(1^{abcd; (-1\otimes1)}, 2^{rstu; (+1\otimes-1)}, \cdots, n\right)\\
&=\epsilon^{abcd}\epsilon^{rstu}\Big[\text{Split}_{-2}^{\text {SG}}\left(z, 1^{-1+1}, 2^{+1-1}\right)\times M_{n-1}\left(p^{+2},\cdots, n\right)\\
&\qquad + \text{Split}_{+2}^{\text {SG}}\left(z, 1^{-1+1}, 2^{+1-1}\right)\times M_{n-1}\left(p^{-2},\cdots, n\right)\Big]\\
&=\epsilon^{abcd}\epsilon^{rstu}\Big[ \frac{\omega_2\omega_1}{\omega_p^2 }\frac{z_{12}}{\bar{z}_{12}}\times M_{n-1}\left(p^{+2},\cdots, n\right)\\
& \qquad + \frac{\omega_1\omega_2}{\omega_p^2}\frac{\bar{z}_{12}}{z_{12}}\times M_{n-1}\left(p^{-2},\cdots, n\right)\Big]
\end{split}
\end{equation}
\begin{equation}
\begin{split}
M_n\left(1^{abcu;0}, 2^{drst; 0}, \cdots, n\right)&=\epsilon^{abce}\epsilon^{rstv}M_n\left(1^{(e,-\frac{1}{2})\otimes(u,\frac{1}{2})}, 2^{(d,+\frac{1}{2})\otimes(v,-\frac{1}{2})}, \cdots, n\right)\\
&=\epsilon^{abce}\epsilon^{rstv}\delta^d_e\delta^u_v\Big[\text{Split}_{-2}^{\text {SG}}\left(z, 1^{-\frac{1}{2}+\frac{1}{2}}, 2^{\frac{1}{2}-\frac{1}{2}}\right)\times M_{n-1}\left(p^{+2},\cdots, n\right)\\
&\qquad + \text{Split}_{+2}^{\text {SG}}\left(z, 1^{-\frac{1}{2}+\frac{1}{2}}, 2^{+\frac{1}{2}-\frac{1}{2}}\right)\times M_{n-1}\left(p^{-2},\cdots, n\right)\Big]\\
&=\epsilon^{abcd}\epsilon^{rstu}\Big[ \frac{\omega_2\omega_1}{\omega_p^2}\frac{z_{12}}{\bar{z}_{12}}\times M_{n-1}\left(p^{+2},\cdots, n\right)\\
& \qquad + \frac{\omega_1\omega_2}{\omega_p^2}\frac{\bar{z}_{12}}{z_{12}}\times M_{n-1}\left(p^{-2},\cdots, n\right)\Big]
\end{split}
\end{equation}
Similarly,
\[ \begin{split}
   M_n\left(1^{arst;0}, 2^{bcdu; 0}, \cdots, n\right)&=\epsilon^{rstu}\epsilon^{abcd}\Big[ \frac{\omega_2\omega_1}{\omega_p^2}\frac{z_{12}}{\bar{z}_{12}}\times M_{n-1}\left(p^{+2},\cdots, n\right)\\
& \qquad + \frac{\omega_1\omega_2}{\omega_p^2}\frac{\bar{z}_{12}}{z_{12}}\times M_{n-1}\left(p^{-2},\cdots, n\right)\Big]
\end{split}\]
\subsection{For Unlike Spins}\label{unlike}
We now use the splits for mixed helicities listed in Appendix \ref{split} and the factorisation of R-symmetry indices mentioned in the calculation of collinear limit for like spins.\\\\ 
\textit{\textbf{Graviton-Gravitino}}\\
We have
\begin{equation}
\begin{split}
M_n\left(1^{+2}, 2^{r; +\frac{3}{2}}, \cdots, n\right)&=M_n\left(1^{ (1\otimes 1)}, 2^{1 \otimes (r; \frac{1}{2})}, \cdots, n\right)\\
&=\text{Split}_{-\frac{3}{2}}^{\text {SG}}\left(z, 1^{1+1}, 2^{1+\frac{1}{2}}\right)\times M_{n-1}\left(p^{r; +\frac{3}{2}},\cdots, n\right)\\
&= \frac{\omega_p^{\frac{3}{2}}}{\omega_2^{\frac{1}{2}} \omega_1} \frac{\bar{z}_{12}}{z_{12}}\times M_{n-1}\left(p^{r; +\frac{3}{2}},\cdots, n\right)
\end{split}
\end{equation}
\begin{equation}
\begin{split}
M_n\left(1^{+2}, 2_{r}^{ -\frac{3}{2}}, \cdots, n\right)&=M_n\left(1^{ (1\otimes 1)}, 2^{-1 \otimes (r; -\frac{1}{2})}, \cdots, n\right)\\
&=\text{Split}_{+\frac{3}{2}}^{\text {SG}}\left(z, 1^{1+1}, 2^{-1-\frac{1}{2}}\right)\times M_{n-1}\left(p_{r}^{ -\frac{3}{2}},\cdots, n\right)\\
&= \frac{\omega_2^{\frac{5}{2}}}{\omega_p^{\frac{3}{2}}\omega_1} \frac{\bar{z}_{12}}{z_{12}}\times M_{n-1}\left(p_{r}^{ -\frac{3}{2}},\cdots, n\right)
\end{split}
\end{equation}
 Similarly we have
\[
    \begin{split}
        M_n\left(1^{+2}, 2^{a; +\frac{3}{2}}, \cdots, n\right)&=  \frac{\omega_p^{\frac{3}{2}}}{\omega_2^{\frac{1}{2}} \omega_1} \frac{\bar{z}_{12}}{z_{12}}\times M_{n-1}\left(p^{a; +\frac{3}{2}},\cdots, n\right)
    \end{split}
\]
\[ \begin{split}
    M_n\left(1^{+2}, 2_{a}^{ -\frac{3}{2}}, \cdots, n\right)&= \frac{\omega_2^{\frac{5}{2}}}{\omega_p^{\frac{3}{2}}\omega_1} \frac{\bar{z}_{12}}{z_{12}} \times M_{n-1}\left(p_{a}^{ -\frac{3}{2}},\cdots, n\right)
\end{split} \] 
Other helicity combination of graviton and gravitino can be obtained by flipping the indices along with $z_{12}\leftrightarrow \Bar{z}_{12}$. 
\\\\
\textbf{\textit{Graviton-Graviphoton}}\\
\begin{equation}
\begin{split}
M_n\left(1^{+2}, 2^{ab; +1}, \cdots, n\right)&=M_n\left(1^{ (1\otimes 1)}, 2^{(ab;0) \otimes 1}, \cdots, n\right)\\
&=\text{Split}_{-1}^{\text {SG}}\left(z, 1^{1+1}, 2^{0+1}\right)\times M_{n-1}\left(p^{ab; +1},\cdots, n\right)\\
&= \frac{\omega_p}{\omega_1} \frac{\bar{z}_{12}}{z_{12}}\times M_{n-1}\left(p^{ab; +1},\cdots, n\right)
\end{split}
\end{equation}
\begin{equation}
\begin{split}
M_n\left(1^{+2}, 2^{rs; +1}, \cdots, n\right)&=M_n\left(1^{ (1\otimes 1)}, 2^{ 1 \otimes (rs;0)}, \cdots, n\right)\\
&=\text{Split}_{-1}^{\text {SG}}\left(z, 1^{1+1}, 2^{1+0}\right)\times M_{n-1}\left(p^{rs; +1},\cdots, n\right)\\
&= \frac{\omega_p}{\omega_1}\frac{\bar{z}_{12}}{z_{12}}\times M_{n-1}\left(p^{rs; +1},\cdots, n\right)
\end{split}
\end{equation}
\begin{equation}
\begin{split}
M_n\left(1^{+2}, 2^{ar; +1}, \cdots, n\right)&=M_n\left(1^{ (1\otimes 1)}, 2^{(a;\frac{1}{2}) \otimes (r;\frac{1}{2})}, \cdots, n\right)\\
&=\text{Split}_{-1}^{\text {SG}}\left(z, 1^{1+1}, 2^{\frac{1}{2}+\frac{1}{2}}\right)\times M_{n-1}\left(p^{ar; +1},\cdots, n\right)\\
&= \frac{\omega_p}{\omega_1}\frac{\bar{z}_{12}}{z_{12}}\times M_{n-1}\left(p^{ar; +1},\cdots, n\right)
\end{split}
\end{equation}
\begin{equation}
\begin{split}
M_n\left(1^{+2}, 2_{ab}^{ -1}, \cdots, n\right)&=M_n\left(1^{ (1\otimes 1)}, 2^{(ab;0) \otimes -1}, \cdots, n\right)\\
&=\text{Split}_{+1}^{\text {SG}}\left(z, 1^{1+1}, 2^{0-1}\right)\times M_{n-1}\left(p_{ab}^{ -1},\cdots, n\right)\\
&= \frac{\omega_2^2}{\omega_p \omega_1} \frac{\bar{z}_{12}}{z_{12}}\times M_{n-1}\left(p_{ab}^{ -1},\cdots, n\right)
\end{split}
\end{equation}
\begin{equation}
\begin{split}
M_n\left(1^{+2}, 2_{rs}^{ -1}, \cdots, n\right)&=M_n\left(1^{ (1\otimes 1)}, 2^{ -1\otimes (rs;0)}, \cdots, n\right)\\
&=\text{Split}_{+1}^{\text {SG}}\left(z, 1^{1+1}, 2^{-1+0}\right)\times M_{n-1}\left(p_{rs}^{-1},\cdots, n\right)\\
&=  \frac{\omega_2^2}{\omega_p \omega_1} \frac{\bar{z}_{12}}{z_{12}}\times M_{n-1}\left(p_{rs}^{-1},\cdots, n\right)
\end{split}
\end{equation}
\begin{equation}
\begin{split}
M_n\left(1^{+2}, 2_{ar}^{-1}, \cdots, n\right)&=M_n\left(1^{ (1\otimes 1)}, 2^{(a;-\frac{1}{2}) \otimes (r;-\frac{1}{2})}, \cdots, n\right)\\
&=\text{Split}_{+1}^{\text {SG}}\left(z, 1^{1+1}, 2^{-\frac{1}{2}-\frac{1}{2}}\right)\times M_{n-1}\left(p_{ar}^{ -1},\cdots, n\right)\\
&= \frac{\omega_2^2}{\omega_1 \omega_p} \frac{\bar{z}_{12}}{z_{12}}\times M_{n-1}\left(p_{ar}^{-1},\cdots, n\right)
\end{split}
\end{equation}
Similarly we can calculate the collinear helicity combinations of Graviphotons with negative helicity Gravitons.\\
Hence
\begin{equation}
    \begin{split}
        M_n\left(1^{+2}, 2_{AB}^{-1}, \cdots, n\right)= \frac{\omega_2^2}{\omega_1 \omega_p} \frac{\bar{z}_{12}}{z_{12}}\times M_{n-1}\left(p_{AB}^{-1},\cdots, n\right)
    \end{split}
\end{equation}
\textit{\textbf{Graviton-Graviphotino}}\\
 \begin{equation}
\begin{split}
M_n\left(1^{+2}, 2^{abr; +\frac{1}{2}}, \cdots, n\right)&=M_n\left(1^{ (1\otimes 1)}, 2^{(ab;0)\otimes(r; \frac{1}{2})}, \cdots, n\right)\\
&=\text{Split}_{-\frac{1}{2}}^{\text {SG}}\left(z, 1^{1+1}, 2^{0+\frac{1}{2}}\right)\times M_{n-1}\left(p^{abr; +\frac{1}{2}},\cdots, n\right)\\
&= \frac{\sqrt{\omega_2 \omega_p}}{\omega_1} \frac{\bar{z}_{12}}{z_{12}}\times M_{n-1}\left(p^{abr; +\frac{1}{2}},\cdots, n\right)
\end{split}
\end{equation}
Similarly,
\[ \begin{split}
    M_n\left(1^{+2}, 2^{ars; +\frac{1}{2}}, \cdots, n\right)&= \frac{\sqrt{\omega_2 \omega_p}}{\omega_1} \frac{\bar{z}_{12}}{z_{12}}\times M_{n-1}\left(p^{ars; +\frac{1}{2}},\cdots, n\right)
\end{split}\]
\begin{equation}
\begin{split}
M_n\left(1^{+2}, 2^{abc; +\frac{1}{2}}, \cdots, n\right)&=-\epsilon^{abcd}M_n\left(1^{ (1\otimes 1)}, 2^{(d; -\frac{1}{2})\otimes 1}, \cdots, n\right)\\
&=-\frac{1}{3!}\epsilon^{abcd}\epsilon_{defg}\text{Split}_{-\frac{1}{2}}^{\text {SG}}\left(z, 1^{1+1}, 2^{-\frac{1}{2}+1}\right)\times M_{n-1}\left(p^{efg; +\frac{1}{2}},\cdots, n\right)\\
&=\frac{1}{3!}\delta^{abc}_{efg}\frac{\sqrt{\omega_2 \omega_p}}{\omega_1} \frac{\bar{z}_{12}}{z_{12}}\times M_{n-1}\left(p^{efg; +\frac{1}{2}},\cdots, n\right)\\&=\frac{\sqrt{\omega_2 \omega_p}}{\omega_1} \frac{\bar{z}_{12}}{z_{12}}\times M_{n-1}\left(p^{abc; +\frac{1}{2}},\cdots, n\right)
\end{split}
\end{equation}
Here we are using change of basis as a redefinition for the fields in SYM:
\[
\Gamma_a^- \equiv \frac{1}{3!}\epsilon_{abcd}\Gamma^{-bcd}
\]
Similarly,
\[ \begin{split}
    M_n\left(1^{+2}, 2^{rst; +\frac{1}{2}}, \cdots, n\right)&= \frac{\sqrt{\omega_2 \omega_p}}{\omega_1} \frac{\bar{z}_{12}}{z_{12}}\times M_{n-1}\left(p^{rst; +\frac{1}{2}},\cdots, n\right)
\end{split}\]
 \begin{equation}
\begin{split}
M_n\left(1^{+2}, 2_{abr}^{ -\frac{1}{2}}, \cdots, n\right)&=M_n\left(1^{ (1\otimes 1)}, 2^{(ab;0) \otimes (r; -\frac{1}{2})}, \cdots, n\right)\\
&=\text{Split}_{+\frac{1}{2}}^{\text {SG}}\left(z, 1^{1+1}, 2^{0-\frac{1}{2}}\right)\times M_{n-1}\left(p_{abr}^{ -\frac{1}{2}},\cdots, n\right)\\
&= \frac{\omega_2^{\frac{3}{2}}}{\omega_p^{\frac{1}{2}}\omega_1} \frac{\bar{z}_{12}}{z_{12}}\times M_{n-1}\left(p_{abr}^{ -\frac{1}{2}},\cdots, n\right)
\end{split}
\end{equation}
Similarly,
\[ \begin{split}
  M_n\left(1^{+2}, 2_{ars}^{ -\frac{1}{2}}, \cdots, n\right)  &= \frac{\omega_2^{\frac{3}{2}}}{\omega_p^{\frac{1}{2}}\omega_1} \frac{\bar{z}_{12}}{z_{12}}\times M_{n-1}\left(p_{ars}^{ -\frac{1}{2}},\cdots, n\right)
\end{split}\]
 \begin{equation}
\begin{split}
M_n\left(1^{+2}, 2_{abc}^{ -\frac{1}{2}}, \cdots, n\right)  &= \epsilon_{abcd}M_n\left(1^{ (1\otimes 1)}, 2^{(d; \frac{1}{2})\otimes -1}, \cdots, n\right)\\
&=\frac{1}{3!}\epsilon_{abcd}\epsilon^{defg}\text{Split}_{\frac{1}{2}}^{\text {SG}}\left(z, 1^{1+1}, 2^{\frac{1}{2}-1}\right)\times M_{n-1}\left(p^{-\frac{1}{2}}_{efg; },\cdots, n\right)\\
&=-\frac{1}{3!}\delta^{efg}_{abc}\frac{\omega_2^{\frac{3}{2}}}{\omega_p^{\frac{1}{2}}\omega_1} \frac{\bar{z}_{12}}{z_{12}}\times M_{n-1}\left(p^{-\frac{1}{2}}_{efg; },\cdots, n\right)\\&=-\frac{\omega_2^{\frac{3}{2}}}{\omega_p^{\frac{1}{2}}\omega_1} \frac{\bar{z}_{12}}{z_{12}}\times M_{n-1}\left(p_{abc}^{ -\frac{1}{2}},\cdots, n\right)
\end{split}
\end{equation}
Similarly
\[ \begin{split}
  M_n\left(1^{+2}, 2_{rst}^{ -\frac{1}{2}}, \cdots, n\right)  &=- \frac{\omega_2^{\frac{3}{2}}}{\omega_p^{\frac{1}{2}}\omega_1} \frac{\bar{z}_{12}}{z_{12}}\times M_{n-1}\left(p_{rst}^{ -\frac{1}{2}},\cdots, n\right).
\end{split}\]
\textit{\textbf{Graviton-Scalar}}\\
Since the split factors corresponding to all factorisations of the R-symmetry indices is the same, the collinear limit can be uniformly written as 
\begin{equation}
M_n\left(1^{+2}, 2^{ABCD; 0}, \cdots, n\right)= \frac{\omega_2}{\omega_1} \frac{\bar{z}_{12}}{z_{12}}\times M_{n-1}\left(p^{ABCD; 0},\cdots, n\right)
\end{equation}
\textit{\textbf{Gravitino-Graviphoton}}\\
\begin{equation}
\begin{split}
M_n\left(1^{a;+\frac{3}{2}}, 2^{bc; +1}, \cdots, n\right)&=M_n\left(1^{ (a;\frac{1}{2})\otimes 1}, 2^{ (bc; 0)\otimes 1}, \cdots, n\right)\\
&=\text{Split}_{-\frac{1}{2}}^{\text {SG}}\left(z, 1^{\frac{1}{2}+1}, 2^{0+1}\right)\times M_{n-1}\left(p^{abc; \frac{1}{2}},\cdots, n\right)\\
&= \sqrt{\frac{\omega_p}{\omega_1}} \frac{\bar{z}_{12}}{z_{12}}\times M_{n-1}\left(p^{abc; +\frac{1}{2}},\cdots, n\right)
\end{split}
\end{equation}
Similarly,
\[\begin{split}
M_n\left(1^{r;+\frac{3}{2}}, 2^{st; +1}, \cdots, n\right)&=\sqrt{\frac{\omega_p}{\omega_1}} \frac{\bar{z}_{12}}{z_{12}}\times M_{n-1}\left(p^{rst; +\frac{1}{2}},\cdots, n\right) \end{split} \]
\[\begin{split}
M_n\left(1^{a;+\frac{3}{2}}, 2^{rs; +1}, \cdots, n\right)
&= \sqrt{\frac{\omega_p}{\omega_1}} \frac{\bar{z}_{12}}{z_{12}}\times M_{n-1}\left(p^{ars; +\frac{1}{2}},\cdots, n\right)
\end{split} \]
\[\begin{split}
M_n\left(1^{r;+\frac{3}{2}}, 2^{ab; +1}, \cdots, n\right)
&= \sqrt{\frac{\omega_p}{\omega_1}} \frac{\bar{z}_{12}}{z_{12}}\times M_{n-1}\left(p^{rab; +\frac{1}{2}},\cdots, n\right)
\end{split} \]
In conclusion, we can write 
\begin{equation}
M_n\left(1^{A;+\frac{3}{2}}, 2^{BC; +1}, \cdots, n\right)
= \sqrt{\frac{\omega_p}{\omega_1}} \frac{\bar{z}_{12}}{z_{12}}\times M_{n-1}\left(p^{ABC; +\frac{1}{2}},\cdots, n\right)
\end{equation}
\begin{equation}
\begin{split}
M_n\left(1^{a;+\frac{3}{2}}, 2_{bc}^{-1}, \cdots, n\right)&=M_n\left(1^{ (a;\frac{1}{2})\otimes 1}, 2^{(bc; 0)\otimes 1}, \cdots, n\right)\\
&=\frac{1}{2!}\epsilon_{bcde}\epsilon^{adef} \; \text{Split}_{\frac{3}{2}}^{\text {SG}}\left(z, 1^{\frac{1}{2}+1}, 2^{0-1}\right)\times M_{n-1}\left(p_{f}^{ -\frac{3}{2}},\cdots, n\right)\\
&=\delta^{af}_{bc}\frac{\omega_2^2}{\omega_p^{\frac{3}{2}} \omega_1^{\frac{1}{2}}} \frac{\bar{z}_{12}}{z_{12}}\times M_{n-1}\left(p_{f}^{ -\frac{3}{2}},\cdots, n\right)\\&=
\frac{\omega_2^2}{\omega_p^{\frac{3}{2}} \omega_1^{\frac{1}{2}}} \frac{\bar{z}_{12}}{z_{12}}2!\delta^a_{[b}\times M_{n-1}\left(p_{c]}^{ -\frac{3}{2}},\cdots, n\right)
\end{split}
\end{equation}
\[ \begin{split}
 M_n\left(1^{r;+\frac{3}{2}}, 2_{st}^{-1}, \cdots, n\right)  &= \frac{\omega_2^2}{\omega_p^{\frac{3}{2}} \omega_1^{\frac{1}{2}}} \frac{\bar{z}_{12}}{z_{12}} 2!\delta^r_{[s}\times M_{n-1}\left(p_{t]}^{ -\frac{3}{2}},\cdots, n\right). 
\end{split} \]
Hence for any $1\leq A,B\leq 8$ we have,
\begin{equation}
    \begin{split}
 M_n\left(1^{A;+\frac{3}{2}}, 2_{BC}^{-1}, \cdots, n\right)  &= \frac{\omega_2^2}{\omega_p^{\frac{3}{2}} \omega_1^{\frac{1}{2}}} \frac{\bar{z}_{12}}{z_{12}} 2!\delta^A_{[B}\times M_{n-1}\left(p_{C]}^{ -\frac{3}{2}},\cdots, n\right). 
\end{split}
\end{equation}
\textit{\textbf{Gravitino-Graviphotino}}\\
\begin{equation}
\begin{split}
M_n\left(1^{a;+\frac{3}{2}}, 2^{brs; +\frac{1}{2}}, \cdots, n\right)&=M_n\left(1^{ (a;\frac{1}{2})\otimes 1}, 2^{ (b; \frac{1}{2})\otimes(rs;0)}, \cdots, n\right)\\
&=\text{Split}_{0}^{\text {SG}}\left(z, 1^{\frac{1}{2}+1}, 2^{\frac{1}{2}+0}\right)\times M_{n-1}\left(p^{abrs; 0},\cdots, n\right)\\
&= \sqrt{\frac{\omega_2}{\omega_1}} \frac{\bar{z}_{12}}{z_{12}}\times M_{n-1}\left(p^{abrs; 0},\cdots, n\right)
\end{split}
\end{equation}
Similarly for all other factorisations the split factors will remain the same for two same helicity Gravitino and Graviphotino pair,
\[ \begin{split}
    M_n\left(1^{a;+\frac{3}{2}}, 2^{bcr; +\frac{1}{2}}, \cdots, n\right)&=\sqrt{\frac{\omega_2}{\omega_1}} \frac{\bar{z}_{12}}{z_{12}}\times M_{n-1}\left(p^{abcr; 0},\cdots, n\right)
\end{split}\] 
\[ \begin{split}
    M_n\left(1^{r;+\frac{3}{2}}, 2^{sta; +\frac{1}{2}}, \cdots, n\right)&=\sqrt{\frac{\omega_2}{\omega_1}} \frac{\bar{z}_{12}}{z_{12}}\times M_{n-1}\left(p^{rsta; 0},\cdots, n\right)
\end{split}\] 
\[ \begin{split}
    M_n\left(1^{r;+\frac{3}{2}}, 2^{abs; +\frac{1}{2}}, \cdots, n\right)&=\sqrt{\frac{\omega_2}{\omega_1}} \frac{\bar{z}_{12}}{z_{12}}\times M_{n-1}\left(p^{abrs; 0},\cdots, n\right)
\end{split}\] 
\[ \begin{split}
    M_n\left(1^{a;+\frac{3}{2}}, 2^{bcd; +\frac{1}{2}}, \cdots, n\right)&=\sqrt{\frac{\omega_2}{\omega_1}} \frac{\bar{z}_{12}}{z_{12}}\times M_{n-1}\left(p^{abcd; 0},\cdots, n\right)
\end{split}\] 
\[ \begin{split}
    M_n\left(1^{a;+\frac{3}{2}}, 2^{rst; +\frac{1}{2}}, \cdots, n\right)&=\sqrt{\frac{\omega_2}{\omega_1}} \frac{\bar{z}_{12}}{z_{12}}\times M_{n-1}\left(p^{arst; 0},\cdots, n\right)
\end{split}\] 
Collecting all of them, we can write 
\begin{equation}\label{drs}
  M_n\left(1^{A;+\frac{3}{2}}, 2^{BCD; +\frac{1}{2}}, \cdots, n\right)=\sqrt{\frac{\omega_2}{\omega_1}} \frac{\bar{z}_{12}}{z_{12}}\times M_{n-1}\left(p^{ABCD; 0},\cdots, n\right)  
\end{equation}
 \begin{equation}
\begin{split}
M_n\left(1^{a;+\frac{3}{2}}, 2_{bcr}^{ -\frac{1}{2}}, \cdots, n\right)&=M_n\left(1^{ (a;\frac{1}{2})\otimes 1}, 2^{ (bc; 0)\otimes(r;-\frac{1}{2})}, \cdots, n\right)\\
&=-\frac{1}{2!}\epsilon_{bcde}\epsilon^{deaf} \; \text{Split}_{+1}^{\text {SG}}\left(z, 1^{\frac{1}{2}+1}, 2^{0-\frac{1}{2}}\right)\times M_{n-1}\left(p^{-1}_{fr},\cdots, n\right)\\
&=-\delta^{af}_{bc}\; \frac{\omega_2^{\frac{3}{2}}}{\omega_p\omega_1^{\frac{1}{2}}} \frac{\bar{z}_{12}}{z_{12}}\times M_{n-1}\left(p^{-1}_{fr},\cdots, n\right)\\&=-2!\frac{\omega_2^{\frac{3}{2}}}{\omega_p\omega_1^{\frac{1}{2}}} \frac{\bar{z}_{12}}{z_{12}} \delta^a_{[b}\times M_{n-1}\left(p^{-1}_{c]r},\cdots, n\right)\\&=-3!\frac{\omega_2^{\frac{3}{2}}}{\omega_p\omega_1^{\frac{1}{2}}} \frac{\bar{z}_{12}}{z_{12}} \delta^a_{[b}\times M_{n-1}\left(p^{-1}_{cr]},\cdots, n\right)
\end{split}
\end{equation}
where we used the fact that $\delta^a_r=0$ to write 
\begin{equation}
\begin{split}
      2!\delta^a_{[b}p^{-1}_{c]r}&=\delta^a_bp^{-1}_{cr}-\delta^a_bp^{-1}_{rc}+\delta^a_cp^{-1}_{rb}-\delta^a_cp^{-1}_{br}+\delta^a_rp^{-1}_{bc}-\delta^a_rp^{-1}_{cb}\\&=3!\delta_{[b}^ap^{-1}_{cr]}.
\end{split}
\end{equation}
\begin{equation}
\begin{split}
M_n\left(1^{a;+\frac{3}{2}}, 2_{brs}^{ -\frac{1}{2}}, \cdots, n\right)&=M_n\left(1^{ (a;\frac{1}{2})\otimes 1}, 2^{ (b; -\frac{1}{2})\otimes(rs;0)}, \cdots, n\right)\\
&=\delta_{b}^a \; \text{Split}_{+1}^{\text {SG}}\left(z, 1^{\frac{1}{2}+1}, 2^{-\frac{1}{2}+0}\right)\times M_{n-1}\left(p^{-1}_{rs},\cdots, n\right)\\
&= \frac{\omega_2^{\frac{3}{2}}}{\omega_p\omega_1^{\frac{1}{2}}}  \frac{\bar{z}_{12}}{z_{12}} \delta_{b}^{a}\times M_{n-1}\left(p^{-1}_{rs},\cdots, n\right).
\end{split}
\end{equation}
Similarly we have
\[
\begin{split}
M_n\left(1^{r;+\frac{3}{2}}, 2_{ast}^{ -\frac{1}{2}}, \cdots, n\right)&=-\frac{1}{2!}\epsilon_{stuv}\epsilon^{uvrw} \; \frac{\omega_2^{\frac{3}{2}}}{\omega_p\omega_1^{\frac{1}{2}}} \frac{\bar{z}_{12}}{z_{12}}\times M_{n-1}\left(p^{-1}_{wa},\cdots, n\right)\\&=-3!\frac{\omega_2^{\frac{3}{2}}}{\omega_p\omega_1^{\frac{1}{2}}} \frac{\bar{z}_{12}}{z_{12}}\delta^{r}_{[s}\times M_{n-1}\left(p^{-1}_{ta]},\cdots, n\right)
\end{split}
\]
\[
M_n\left(1^{t;+\frac{3}{2}}, 2_{rab}^{ -\frac{1}{2}}, \cdots, n\right)=\; \frac{\omega_2^{\frac{3}{2}}}{\omega_p\omega_1^{\frac{1}{2}}}  \frac{\bar{z}_{12}}{z_{12}}\delta_{r}^t \times M_{n-1}\left(p^{-1}_{ab},\cdots, n\right)
\]
\begin{equation}
\begin{split}
M_n\left(1^{a;+\frac{3}{2}}, 2_{bcd}^{ -\frac{1}{2}}, \cdots, n\right)&=\epsilon_{bcde}M_n\left(1^{ (a;\frac{1}{2})\otimes 1}, 2^{ (e; +\frac{1}{2})\otimes -1}, \cdots, n\right)\\
&=-\frac{1}{2!}\epsilon_{bcde}\epsilon^{aefg} \; \text{Split}_{+1}^{\text {SG}}\left(z, 1^{\frac{1}{2}+1}, 2^{\frac{1}{2}-1}\right)\times M_{n-1}\left(p^{-1}_{fg},\cdots, n\right)\\
&= -\frac{1}{2!}\epsilon_{bcde}\epsilon^{afge} \; \frac{\omega_2^{\frac{3}{2}}}{\omega_p\omega_1^{\frac{1}{2}}} \frac{\bar{z}_{12}}{z_{12}}\times M_{n-1}\left(p^{-1}_{fg},\cdots, n\right)\\
&= -\frac{1}{2!}\delta^{afg}_{bcd} \; \frac{\omega_2^{\frac{3}{2}}}{\omega_p\omega_1^{\frac{1}{2}}} \frac{\bar{z}_{12}}{z_{12}}\times M_{n-1}\left(p^{-1}_{fg},\cdots, n\right)\\&=3\frac{\omega_2^{\frac{3}{2}}}{\omega_p\omega_1^{\frac{1}{2}}} \frac{\bar{z}_{12}}{z_{12}}\delta^a_{[b}\times M_{n-1}\left(p^{-1}_{cd]},\cdots, n\right)
\end{split}
\end{equation}
Note that the second $\epsilon^{aefg}$ comes from the fact that we are lowering the index of he scalar in $\mathcal{N}=4$ SYM in the factorisation of the negative helicity gluon. The factorisation looks as 
\begin{equation}
    G^{-1}_{fg}=\Phi_{fg}\otimes G^{-1}=-\frac{1}{2!}\epsilon_{aefg}\Phi^{ae}\otimes G^{-1}=:-\frac{1}{2!}\epsilon_{aefg} G^{ae-1},
\end{equation}
where $G_{fg}$ is the gluon in $\mathcal{N}=8$ supergravity and $G$ is the gluon in $\mathcal{N}=4$ SYM.
\[
M_n\left(1^{d;+\frac{3}{2}}, 2_{abc}^{ -\frac{1}{2}}, \cdots, n\right)=-3\frac{\omega_2^{\frac{3}{2}}}{\omega_p\omega_1^{\frac{1}{2}}}  \frac{\bar{z}_{12}}{z_{12}}\delta^d_{[a}\times M_{n-1}\left(p^{-1}_{bc]},\cdots, n\right)
\]
Thus we have the collinear limit
\[
M_n\left(1^{u;+\frac{3}{2}}, 2_{rst}^{ -\frac{1}{2}}, \cdots, n\right)=-3\frac{\omega_2^{\frac{3}{2}}}{\omega_p\omega_1^{\frac{1}{2}}}  \frac{\bar{z}_{12}}{z_{12}}\delta^u_{[r}\times M_{n-1}\left(p^{-1}_{st]},\cdots, n\right)
\]

Hence
\begin{equation}
\begin{split}
    M_n\left(1^{A;+\frac{3}{2}}, 2_{BCD}^{ -\frac{1}{2}}, \cdots, n\right)&=-\frac{\omega_2^{\frac{3}{2}}}{\omega_p\omega_1^{\frac{1}{2}}}  \frac{\bar{z}_{12}}{z_{12}}\Big[\delta^A_{B} M_{n-1}\left(p^{-1}_{CD},\cdots, n\right)\\&+\delta^A_{C}
    M_{n-1}\left(p^{-1}_{DB},\cdots, n\right)+\delta^A_{D} M_{n-1}\left(p^{-1}_{BC},\cdots, n\right)\Big]
\end{split}
\end{equation}
Similarly we get the same splitting factors for all other factorisation channels.\\ \\
\textit{\textbf{Gravitino-Scalar}}\\
\begin{equation}
        \begin{split}
M_n\left(1^{a;+\frac{3}{2}}, 2^{bcrs; 0}, \cdots, n\right)&=M_n\left(1^{ (a;\frac{1}{2})\otimes 1}, 2^{ (bc;0)\otimes (rs;0)}, \cdots, n\right)\\
&=\epsilon^{rstu}\epsilon^{abcd}\text{Split}_{\frac{1}{2}}^{\text {SG}}\left(z, 1^{\frac{1}{2}+1}, 2^{0+0}\right)\times M_{n-1}\left(p_{dtu}^{ -\frac{1}{2}},\cdots, n\right)\\
&=\epsilon^{rstu}\epsilon^{abcd} \frac{\omega_2}{\sqrt{\omega_1 \omega_p}} \frac{\bar{z}_{12}}{z_{12}}\times M_{n-1}\left(p_{drs}^{ -\frac{1}{2}},\cdots, n\right)
\end{split}
    \end{equation}
\begin{equation}
        \begin{split}
M_n\left(1^{a;+\frac{3}{2}}, 2^{rstu; 0}, \cdots, n\right)&=M_n\left(1^{ (a;\frac{1}{2})\otimes 1}, 2^{rstu; (1\otimes 1)}, \cdots, n\right)\\
&=-\epsilon^{rstu}\epsilon^{abcd}\text{Split}_{\frac{1}{2}}^{\text {SG}}\left(z, 1^{\frac{1}{2}+1}, 2^{1-1}\right)\times M_{n-1}\left(p_{bcd}^{-\frac{1}{2}},\cdots, n\right)\\
&=-\epsilon^{abcd}\epsilon^{rstu}\frac{\omega_2}{\sqrt{\omega_1 \omega_p}} \frac{\bar{z}_{12}}{z_{12}}\times M_{n-1}\left(p_{bcd}^{-\frac{1}{2}},\cdots, n\right)
\end{split}
    \end{equation}
    where we lowered the index on gluino in the SYM theory.
\begin{equation}
        \begin{split}
M_n\left(1^{r;+\frac{3}{2}}, 2^{abcd; 0}, \cdots, n\right)&=M_n\left(1^{ (r;\frac{1}{2})\otimes 1}, 2^{abcd; (1\otimes 1)}, \cdots, n\right)\\
&=-\epsilon^{abcd}\epsilon^{rstu}\text{Split}_{\frac{1}{2}}^{\text {SG}}\left(z, 1^{\frac{1}{2}+1}, 2^{-1+1}\right)\times M_{n-1}\left(p_{stu}^{-\frac{1}{2}},\cdots, n\right)\\
&=-\epsilon^{abcd}\epsilon^{rstu}\frac{\omega_2}{\sqrt{\omega_1 \omega_p}} \frac{\bar{z}_{12}}{z_{12}}\times M_{n-1}\left(p_{stu}^{-\frac{1}{2}},\cdots, n\right)
\end{split}
    \end{equation}
     Hence we can write the above in simplified form as
    \begin{equation}
        \begin{split}
         M_n\left(1^{A;+\frac{3}{2}}, 2^{BCDE; 0}, \cdots, n\right)&=-\frac{1}{3!}\epsilon^{ABCDEFGH}\frac{\omega_2}{\sqrt{\omega_1 \omega_p}} \frac{\bar{z}_{12}}{z_{12}}\times M_{n-1}\left(p_{FGH}^{-\frac{1}{2}},\cdots, n\right) 
        \end{split}
    \end{equation}
    Similarly we can have the relations for opposite helicity collinear pair.\\ \\
\textit{\textbf{Graviphoton-Graviphotino}}\\
         \begin{equation}
        \begin{split}
M_n\left(1^{ab;+1}, 2^{cdr; +\frac{1}{2}}, \cdots, n\right)&=M_n\left(1^{ (ab;0)\otimes 1}, 2^{ (cd;0)\otimes (r;\frac{1}{2})}, \cdots, n\right)\\
&= \frac{1}{3!}\epsilon^{abcd}\epsilon^{rstu} \; \text{Split}_{\frac{1}{2}}^{\text {SG}}\left(z, 1^{0+1}, 2^{0+\frac{1}{2}}\right)\times M_{n-1}\left(p_{stu}^{ -\frac{1}{2}},\cdots, n\right)\\
&= \frac{1}{3!}\epsilon^{abcd}\epsilon^{rstu} \;\sqrt{\frac{\omega_2}{\omega_p}} \frac{\bar{z}_{12}}{z_{12}}\times M_{n-1}\left(p_{stu}^{ -\frac{1}{2}},\cdots, n\right)
\end{split}
    \end{equation}
    This is true for all other factorisation channels of both positive helicity Graviphoton and Graviphotino collinear pair. Similarly we can have the amplitude for negative helicity collinear pairs.
 \begin{equation}
        \begin{split}
M_n\left(1^{ab;+1}, 2_{cdr; \; -\frac{1}{2}}, \cdots, n\right)&=M_n\left(1^{ (ab;0)\otimes 1}, 2^{ (cd;0)\otimes (r;\frac{1}{2})}, \cdots, n\right)\\
&=-\frac{1}{2!}\epsilon^{abef}\epsilon_{cdef} \; \text{Split}_{\frac{3}{2}}^{\text {SG}}\left(z, 1^{0+1}, 2^{0-\frac{1}{2}}\right)\times M_{n-1}\left(p_{r}^{ -\frac{3}{2}},\cdots, n\right)\\
&=-\delta^{ab}_{cd} \; \frac{\omega_2^{\frac{3}{2}}}{\omega_p^{\frac{3}{2}}} \frac{\bar{z}_{12}}{z_{12}}\times M_{n-1}\left(p_{r}^{ -\frac{3}{2}},\cdots, n\right)
\end{split}
\end{equation}
All other factorisation channels also correspond to the same collinear divergence factor and we get the other amplitudes in the usual way by flipping the helicity and $z_{12}\leftrightarrow\Bar{z}_{12}$.\\ \\
\textit{\textbf{Graviphoton-Scalar}}\\
\begin{equation}
        \begin{split}
M_n\left(1^{ab;+1}, 2^{cdrs; 0}, \cdots, n\right)&=M_n\left(1^{ (ab;0)\otimes 1}, 2^{ (cd;0)\otimes (rs;0)}, \cdots, n\right)\\
&= \epsilon^{abcd} \epsilon^{rstu}\; \text{Split}_{1}^{\text {SG}}\left(z, 1^{0+1}, 2^{0+0}\right)\times M_{n-1}\left(p_{tu}^{-1},\cdots, n\right)\\
&=\epsilon^{abcd} \epsilon^{rstu}\;\frac{\omega_2}{\omega_p}\frac{\bar{z}_{12}}{z_{12}}\times M_{n-1}\left(p_{tu}^{-1},\cdots, n\right)
\end{split}
    \end{equation}
    \begin{equation}
        \begin{split}
M_n\left(1^{rs;+1}, 2^{abtu; 0}, \cdots, n\right)&=M_n\left(1^{ 1 \otimes (rs;0)}, 2^{ (ab;0)\otimes (tu;0)}, \cdots, n\right)\\
&= \epsilon^{rstu} \epsilon^{abcd}\; \text{Split}_{1}^{\text {SG}}\left(z, 1^{1+0}, 2^{0+0}\right)\times M_{n-1}\left(p_{cd}^{-1},\cdots, n\right)\\
&=\epsilon^{rstu} \epsilon^{abcd} \;\frac{\omega_2}{\omega_p}\frac{\bar{z}_{12}}{z_{12}}\times M_{n-1}\left(p_{cd}^{-1},\cdots, n\right)
\end{split}
    \end{equation}
    \begin{equation}
        \begin{split}
M_n\left(1^{ar;+1}, 2^{bcst; 0}, \cdots, n\right)&=M_n\left(1^{ (a;\frac{1}{2}) \otimes (r;\frac{1}{2})}, 2^{ (bc;0)\otimes (st;0)}, \cdots, n\right)\\
&= \epsilon^{abcd} \epsilon^{rstu}\; \text{Split}_{1}^{\text {SG}}\left(z, 1^{\frac{1}{2}+\frac{1}{2}}, 2^{0+0}\right)\times M_{n-1}\left(p_{du}^{-1},\cdots, n\right)\\
&=\epsilon^{abcd} \epsilon^{rstu} \;\frac{\omega_2}{\omega_p}\frac{\bar{z}_{12}}{z_{12}}\times M_{n-1}\left(p_{du}^{-1},\cdots, n\right)
\end{split}
    \end{equation}
    \begin{equation}
        \begin{split}
M_n\left(1^{ab;+1}, 2^{cdef; 0}, \cdots, n\right)&=-M_n\left(1^{ (ab;0)\otimes 1}, 2^{ cdef; (-1\otimes1)}, \cdots, n\right)\\
&=-\epsilon^{cdef}  \epsilon^{abgh} \; \text{Split}_{1}^{\text {SG}}\left(z, 1^{0+1}, 2^{-1+1}\right)\times M_{n-1}\left(p_{gh}^{-1},\cdots, n\right)\\
&=-\epsilon^{cdef}  \epsilon^{abgh} \;\frac{\omega_2}{\omega_p}\frac{\bar{z}_{12}}{z_{12}}\times M_{n-1}\left(p_{gh}^{-1},\cdots, n\right)
\end{split}
    \end{equation}
    \begin{equation}
        \begin{split}
M_n\left(1^{rs;+1}, 2^{cdef; 0}, \cdots, n\right)&=-M_n\left(1^{ 1 \otimes (rs;0) }, 2^{ cdef; (-1 \otimes1)}, \cdots, n\right)\\
&= -\epsilon^{cdef} \epsilon^{rstu}\; \text{Split}_{1}^{\text {SG}}\left(z, 1^{1+0}, 2^{-1+1}\right)\times M_{n-1}\left(p_{tu}^{-1},\cdots, n\right)\\
&=-\epsilon^{cdef} \epsilon^{rstu}\;\frac{\omega_2}{\omega_p}\frac{\bar{z}_{12}}{z_{12}}\times M_{n-1}\left(p_{tu}^{-1},\cdots, n\right)
\end{split}
    \end{equation}
    \begin{equation}
        \begin{split}
M_n\left(1^{ar;+1}, 2^{bcds; 0}, \cdots, n\right)&=-M_n\left(1^{ (a;\frac{1}{2}) \otimes (r;\frac{1}{2})}, 2^{ bcds; (-\frac{1}{2} +\frac{1}{2})}, \cdots, n\right)\\
&= -\epsilon^{abcd} \epsilon^{rstu}\; \text{Split}_{1}^{\text {SG}}\left(z, 1^{\frac{1}{2}+\frac{1}{2}}, 2^{-\frac{1}{2} +\frac{1}{2}}\right)\times M_{n-1}\left(p_{tu}^{-1},\cdots, n\right)\\
&=-\epsilon^{abcd} \epsilon^{rstu} \;\frac{\omega_2}{\omega_p}\frac{\bar{z}_{12}}{z_{12}}\times M_{n-1}\left(p_{tu}^{-1},\cdots, n\right)
\end{split}
    \end{equation}
    Similarly we can write for other remaining factorisation channels.\\
     \begin{equation}
        \begin{split}
M_n\left(1_{ab}^{-1}, 2^{cdef; 0}, \cdots, n\right)&=-M_n\left(1^{ (ab;0)\otimes- 1}, 2^{ cdef; (-1\otimes1)}, \cdots, n\right)\\
&=-\epsilon^{cdef}  \epsilon_{abgh} \; \text{Split}_{-1}^{\text {SG}}\left(z, 1^{0-1}, 2^{-1+1}\right)\times M_{n-1}\left(p^{gh;+1},\cdots, n\right)\\
&=-\epsilon^{cdef}  \epsilon_{abgh} \;\frac{\omega_2}{\omega_p}\frac{z_{12}}{\bar{z}_{12}}\times M_{n-1}\left(p^{gh;+1},\cdots, n\right)
\end{split}
    \end{equation}
    \begin{equation}
        \begin{split}
M_n\left(1_{ar}^{-1}, 2^{bcds; 0}, \cdots, n\right)&=-M_n\left(1^{ (a;-\frac{1}{2}) \otimes (r;-\frac{1}{2})}, 2^{ bcds; (-\frac{1}{2} \otimes \frac{1}{2})}, \cdots, n\right)\\
&=- \epsilon^{bcde} \delta^{s}_{r}\;\epsilon_{aefg}\; \text{Split}_{-1}^{\text {SG}}\left(z, 1^{-\frac{1}{2}-\frac{1}{2}}, 2^{-\frac{1}{2} +\frac{1}{2}}\right)\times M_{n-1}\left(p^{fg;+1},\cdots, n\right)\\
&=-\epsilon^{bcde} \delta^{s}_{r}\epsilon_{aefg}\;\frac{\omega_1}{\omega_p}\frac{z_{12}}{\bar{z}_{12}}\times M_{n-1}\left(p^{fg;+1},\cdots, n\right)
\end{split}
    \end{equation}

    \begin{equation}
        \begin{split}
M_n\left(1_{ar}^{-1}, 2^{bcst; 0}, \cdots, n\right)&=-M_n\left(1^{ (a;\frac{1}{2}) \otimes (r;\frac{1}{2})}, 2^{ (bc;0)\otimes (st;0)}, \cdots, n\right)\\
&=\Big[-\delta^{b}_{a} \delta^{s}_{r}\; \text{Split}_{-1}^{\text {SG}}\left(z, 1^{-\frac{1}{2}-\frac{1}{2}}, 2^{0+0}\right)\times M_{n-1}\left(p^{ct;+1},\cdots, n\right)\\&+\delta^{c}_{a} \delta^{t}_{r}\; \text{Split}_{-1}^{\text {SG}}\left(z, 1^{-\frac{1}{2}-\frac{1}{2}}, 2^{0+0}\right)\times M_{n-1}\left(p^{bs;+1},\cdots, n\right)\\&+\delta^{b}_{a} \delta^{t}_{r}\; \text{Split}_{-1}^{\text {SG}}\left(z, 1^{-\frac{1}{2}-\frac{1}{2}}, 2^{0+0}\right)\times M_{n-1}\left(p^{cs;+1},\cdots, n\right)\\&-\delta^{c}_{a} \delta^{s}_{r}\; \text{Split}_{-1}^{\text {SG}}\left(z, 1^{-\frac{1}{2}-\frac{1}{2}}, 2^{0+0}\right)\times M_{n-1}\left(p^{bt;+1},\cdots, n\right)\Big]\\
&=-\frac{\omega_2}{\omega_p}\frac{z_{12}}{\bar{z}_{12}}4!\delta^{[b}_{a} \delta^{s}_{r}\;  M_{n-1}\left(p^{tc];+1},\cdots, n\right)
\end{split}
    \end{equation}
Note that the above expression contains 16 terms but only four terms are nonzero since $\delta^{a}_{r}=0$. \\ \\
\textit{\textbf{Graviphotino-Scalar}}\\
\begin{equation}
        \begin{split}
M_n\left(1^{abr;+\frac{1}{2}}, 2^{cdst; 0}, \cdots, n\right)&=M_n\left(1^{ (ab;0)\otimes (r;\frac{1}{2})}, 2^{ (cd;0)\otimes (st;0)}, \cdots, n\right)\\
&=  \epsilon^{abcd}\epsilon^{rstu}\; \text{Split}_{\frac{3}{2}}^{\text {SG}}\left(z, 1^{0+\frac{1}{2}}, 2^{0+0}\right)\times M_{n-1}\left(p_{u}^{-\frac{3}{2}},\cdots, n\right)\\
&= \epsilon^{abcd}\epsilon^{rstu} \;\frac{\omega_1^{\frac{1}{2}}  \omega_2}{\omega_p^{\frac{3}{2}}}\frac{\bar{z}_{12}}{z_{12}}\times M_{n-1}\left(p_{u}^{-\frac{3}{2}},\cdots, n\right)
\end{split}
    \end{equation}
\begin{equation}
        \begin{split}
M_n\left(1^{abr;+\frac{1}{2}}, 2^{cstu; 0}, \cdots, n\right)&=-M_n\left(1^{ (ab;0)\otimes (r;\frac{1}{2})}, 2^{ cstu;(\frac{1}{2}\otimes-\frac{1}{2})}, \cdots, n\right)\\
&=  -\epsilon^{abcd}\epsilon^{rstu}\; \text{Split}_{\frac{3}{2}}^{\text {SG}}\left(z, 1^{0+\frac{1}{2}}, 2^{\frac{1}{2}-\frac{1}{2}}\right)\times M_{n-1}\left(p_{d}^{-\frac{3}{2}},\cdots, n\right)\\
&= -\epsilon^{abcd}\epsilon^{rstu} \;\frac{\omega_1^{\frac{1}{2}}  \omega_2}{\omega_p^{\frac{3}{2}}}\frac{\bar{z}_{12}}{z_{12}}\times M_{n-1}\left(p_{d}^{-\frac{3}{2}},\cdots, n\right)
\end{split}
    \end{equation}
    \begin{equation}
        \begin{split}
M_n\left(1^{ars;+\frac{1}{2}}, 2^{bctu; 0}, \cdots, n\right)&=M_n\left(1^{ (a;\frac{1}{2})\otimes (rs;0)}, 2^{ (bc;0)\otimes (tu;0)}, \cdots, n\right)\\
&=  \epsilon^{abcd}\epsilon^{rstu}\; \text{Split}_{\frac{3}{2}}^{\text {SG}}\left(z, 1^{\frac{1}{2}+0}, 2^{0+0}\right)\times M_{n-1}\left(p_{d}^{-\frac{3}{2}},\cdots, n\right)\\
&= \epsilon^{abcd}\epsilon^{rstu} \;\frac{\omega_1^{\frac{1}{2}}  \omega_2}{\omega_p^{\frac{3}{2}}}\frac{\bar{z}_{12}}{z_{12}}\times M_{n-1}\left(p_{d}^{-\frac{3}{2}},\cdots, n\right)
\end{split}
    \end{equation}
 \begin{equation}
        \begin{split}
M_n\left(1_{ars}^{-\frac{1}{2}}, 2^{bctu;0}, \cdots, n\right)&=M_n\left(1^{ (a;\frac{1}{2})\otimes (rs;0)}, 2^{ (bc;0)\otimes (tu;0)}, \cdots, n\right)\\
&=  -\frac{1}{2!}\epsilon^{tuvw}\epsilon_{vwrs}2!\delta^{[b}_{a}\; \text{Split}_{-\frac{3}{2}}^{\text {SG}}\left(z, 1^{-\frac{1}{2}+0}, 2^{0+0}\right)\times M_{n-1}\left(p^{c];+\frac{3}{2}},\cdots, n\right)\\
&= -2!\delta^{tu}_{rs}\frac{\omega_1^{\frac{1}{2}}  \omega_2}{\omega_p^{\frac{3}{2}}}\frac{z_{12}}{\bar{z}_{12}}\delta^{[b}_{a} \times M_{n-1}\left(p^{c];+\frac{3}{2}},\cdots, n\right)
\end{split}
    \end{equation}
\begin{equation}
\begin{split}
M_n\left(1_{ars}^{-\frac{1}{2}}, 2^{btuv~0}, \cdots, n\right)&=-M_n\left(1^{ (a;\frac{1}{2})\otimes (rs;0)}, 2^{ btuv;(\frac{1}{2}\otimes-\frac{1}{2})}, \cdots, n\right)\\
&=-  \delta_{a}^{b}\epsilon^{tuvw} \epsilon_{wrsx}\; \text{Split}_{-\frac{3}{2}}^{\text {SG}}\left(z, 1^{-\frac{1}{2}+0}, 2^{-\frac{1}{2}+\frac{1}{2}}\right)\times M_{n-1}\left(p^{x;+\frac{3}{2}},\cdots, n\right)\\
&=-\delta_{a}^{b}\epsilon^{tuvw} \epsilon_{wrsx}\;\frac{\omega_1^{\frac{1}{2}}  \omega_2}{\omega_p^{\frac{3}{2}}}\frac{z_{12}}{\bar{z}_{12}}\times M_{n-1}\left(p^{x;+\frac{3}{2}},\cdots, n\right)
\end{split}
    \end{equation}
    \begin{equation}
        \begin{split}
M_n\left(1_{rst}^{-\frac{1}{2}}, 2^{avwx; 0}, \cdots, n\right)&=-M_n\left(1^{ rst; (-1\otimes\frac{1}{2})}, 2^{ avwx;(\frac{1}{2}\otimes-\frac{1}{2})}, \cdots, n\right)\\
&=- \epsilon_{rstu} \epsilon^{vwxy}\delta^u_y\; \text{Split}_{-\frac{3}{2}}^{\text {SG}}\left(z, 1^{-1+\frac{1}{2}}, 2^{\frac{1}{2}-\frac{1}{2}}\right)\times M_{n-1}\left(p^{a;+\frac{3}{2}},\cdots, n\right)\\
&=-\epsilon_{rstu} \epsilon^{vwxu} \;\frac{\omega_1^{\frac{1}{2}}  \omega_2}{\omega_p^{\frac{3}{2}}}\frac{z_{12}}{\bar{z}_{12}}\times M_{n-1}\left(p^{a;+\frac{3}{2}},\cdots, n\right)
\end{split}
    \end{equation}
    \begin{equation}
        \begin{split}
M_n\left(1_{rst}^{-\frac{1}{2}}, 2^{uvwx; 0}, \cdots, n\right)&=-M_n\left(1^{rst; (-1\otimes\frac{1}{2})}, 2^{ uvwx;(1\otimes-1)}, \cdots, n\right)\\
&=- \epsilon_{rsty} \epsilon^{uvwx}\; \text{Split}_{-\frac{3}{2}}^{\text {SG}}\left(z, 1^{-1+\frac{1}{2}}, 2^{1-1}\right)\times M_{n-1}\left(p^{y;+\frac{3}{2}},\cdots, n\right)\\
&=-\epsilon_{rsty} \epsilon^{uvwx} \;\frac{\omega_1^{\frac{1}{2}}  \omega_2}{\omega_p^{\frac{3}{2}}}\frac{z_{12}}{\bar{z}_{12}}\times M_{n-1}\left(p^{y;+\frac{3}{2}},\cdots, n\right)
\end{split}
\end{equation}
\bibliography{bms.bib}

\providecommand{\href}[2]{#2}\begingroup\raggedright\begin{thebibliography}{10}

\bibitem{RevModPhys.91.015002}
D.~Poland, S.~Rychkov and A.~Vichi, \emph{The conformal bootstrap: Theory,
  numerical techniques, and applications},
  \href{https://doi.org/10.1103/RevModPhys.91.015002}{\emph{Rev. Mod. Phys.}
  {\bfseries 91} (2019) 015002}.

\bibitem{Poland:2016chs}
D.~Poland and D.~Simmons-Duffin, \emph{{The conformal bootstrap}},
  \href{https://doi.org/10.1038/nphys3761}{\emph{Nature Phys.} {\bfseries 12}
  (2016) 535}.

\bibitem{BELAVIN1984333}
A.~Belavin, A.~Polyakov and A.~Zamolodchikov, \emph{Infinite conformal symmetry
  in two-dimensional quantum field theory},
  \href{https://doi.org/https://doi.org/10.1016/0550-3213(84)90052-X}{\emph{Nuclear
  Physics B} {\bfseries 241} (1984) 333}.

\bibitem{Sahoo:2020ryf}
B.~Sahoo, \emph{{Classical Sub-subleading Soft Photon and Soft Graviton
  Theorems in Four Spacetime Dimensions}},
  \href{https://doi.org/10.1007/JHEP12(2020)070}{\emph{JHEP} {\bfseries 12}
  (2020) 070} [\href{https://arxiv.org/abs/2008.04376}{{\ttfamily
  2008.04376}}].

\bibitem{Saha:2019tub}
A.~P. Saha, B.~Sahoo and A.~Sen, \emph{{Proof of the classical soft graviton
  theorem in $D$ = 4}},
  \href{https://doi.org/10.1007/JHEP06(2020)153}{\emph{JHEP} {\bfseries 06}
  (2020) 153} [\href{https://arxiv.org/abs/1912.06413}{{\ttfamily
  1912.06413}}].

\bibitem{PhysRev.135.B1049}
S.~Weinberg, \emph{Photons and gravitons in $s$-matrix theory: Derivation of
  charge conservation and equality of gravitational and inertial mass},
  \href{https://doi.org/10.1103/PhysRev.135.B1049}{\emph{Phys. Rev.} {\bfseries
  135} (1964) B1049}.

\bibitem{PhysRev.140.B516}
S.~Weinberg, \emph{Infrared photons and gravitons},
  \href{https://doi.org/10.1103/PhysRev.140.B516}{\emph{Phys. Rev.} {\bfseries
  140} (1965) B516}.

\bibitem{PhysRev.166.1287}
D.~J. Gross and R.~Jackiw, \emph{Low-energy theorem for graviton scattering},
  \href{https://doi.org/10.1103/PhysRev.166.1287}{\emph{Phys. Rev.} {\bfseries
  166} (1968) 1287}.

\bibitem{PhysRev.168.1623}
R.~Jackiw, \emph{Low-energy theorems for massless bosons: Photons and
  gravitons}, \href{https://doi.org/10.1103/PhysRev.168.1623}{\emph{Phys. Rev.}
  {\bfseries 168} (1968) 1623}.

\bibitem{He:2014laa}
T.~He, V.~Lysov, P.~Mitra and A.~Strominger, \emph{{BMS supertranslations and
  Weinberg\textquoteright{}s soft graviton theorem}},
  \href{https://doi.org/10.1007/JHEP05(2015)151}{\emph{JHEP} {\bfseries 05}
  (2015) 151} [\href{https://arxiv.org/abs/1401.7026}{{\ttfamily 1401.7026}}].

\bibitem{Bern:2014vva}
Z.~Bern, S.~Davies, P.~Di~Vecchia and J.~Nohle, \emph{{Low-Energy Behavior of
  Gluons and Gravitons from Gauge Invariance}},
  \href{https://doi.org/10.1103/PhysRevD.90.084035}{\emph{Phys. Rev. D}
  {\bfseries 90} (2014) 084035}
  [\href{https://arxiv.org/abs/1406.6987}{{\ttfamily 1406.6987}}].

\bibitem{Campiglia:2014yka}
M.~Campiglia and A.~Laddha, \emph{{Asymptotic symmetries and subleading soft
  graviton theorem}},
  \href{https://doi.org/10.1103/PhysRevD.90.124028}{\emph{Phys. Rev. D}
  {\bfseries 90} (2014) 124028}
  [\href{https://arxiv.org/abs/1408.2228}{{\ttfamily 1408.2228}}].

\bibitem{Laddha:2017ygw}
A.~Laddha and A.~Sen, \emph{{Sub-subleading Soft Graviton Theorem in Generic
  Theories of Quantum Gravity}},
  \href{https://doi.org/10.1007/JHEP10(2017)065}{\emph{JHEP} {\bfseries 10}
  (2017) 065} [\href{https://arxiv.org/abs/1706.00759}{{\ttfamily
  1706.00759}}].

\bibitem{Klose:2015xoa}
T.~Klose, T.~McLoughlin, D.~Nandan, J.~Plefka and G.~Travaglini,
  \emph{{Double-Soft Limits of Gluons and Gravitons}},
  \href{https://doi.org/10.1007/JHEP07(2015)135}{\emph{JHEP} {\bfseries 07}
  (2015) 135} [\href{https://arxiv.org/abs/1504.05558}{{\ttfamily
  1504.05558}}].

\bibitem{Cachazo:2014fwa}
F.~Cachazo and A.~Strominger, \emph{{Evidence for a New Soft Graviton
  Theorem}},  \href{https://arxiv.org/abs/1404.4091}{{\ttfamily 1404.4091}}.

\bibitem{Casali:2014xpa}
E.~Casali, \emph{{Soft sub-leading divergences in Yang-Mills amplitudes}},
  \href{https://doi.org/10.1007/JHEP08(2014)077}{\emph{JHEP} {\bfseries 08}
  (2014) 077} [\href{https://arxiv.org/abs/1404.5551}{{\ttfamily 1404.5551}}].

\bibitem{PhysRevD.97.106019}
A.~A.~H, A.~Kundu and K.~Ray, \emph{Double soft graviton theorems and
  bondi-metzner-sachs symmetries},
  \href{https://doi.org/10.1103/PhysRevD.97.106019}{\emph{Phys. Rev. D}
  {\bfseries 97} (2018) 106019}.

\bibitem{PhysRevLett.120.201601}
Y.~Hamada and G.~Shiu, \emph{Infinite set of soft theorems in gauge-gravity
  theories as ward-takahashi identities},
  \href{https://doi.org/10.1103/PhysRevLett.120.201601}{\emph{Phys. Rev. Lett.}
  {\bfseries 120} (2018) 201601}.

\bibitem{AtulBhatkar:2019vcb}
S.~Atul~Bhatkar, \emph{{Ward identity for loop level soft photon theorem for
  massless QED coupled to gravity}},
  \href{https://doi.org/10.1007/JHEP10(2020)110}{\emph{JHEP} {\bfseries 10}
  (2020) 110} [\href{https://arxiv.org/abs/1912.10229}{{\ttfamily
  1912.10229}}].

\bibitem{Campiglia:2019wxe}
M.~Campiglia and A.~Laddha, \emph{{Loop Corrected Soft Photon Theorem as a Ward
  Identity}}, \href{https://doi.org/10.1007/JHEP10(2019)287}{\emph{JHEP}
  {\bfseries 10} (2019) 287}
  [\href{https://arxiv.org/abs/1903.09133}{{\ttfamily 1903.09133}}].

\bibitem{Hijano:2020szl}
E.~Hijano and D.~Neuenfeld, \emph{{Soft photon theorems from CFT Ward identites
  in the flat limit of AdS/CFT}},
  \href{https://doi.org/10.1007/JHEP11(2020)009}{\emph{JHEP} {\bfseries 11}
  (2020) 009} [\href{https://arxiv.org/abs/2005.03667}{{\ttfamily
  2005.03667}}].

\bibitem{Catani:2000pi}
S.~Catani and M.~Grazzini, \emph{{The soft gluon current at one loop order}},
  \href{https://doi.org/10.1016/S0550-3213(00)00572-1}{\emph{Nucl. Phys. B}
  {\bfseries 591} (2000) 435}
  [\href{https://arxiv.org/abs/hep-ph/0007142}{{\ttfamily hep-ph/0007142}}].

\bibitem{Mao:2017tey}
P.~Mao and J.-B. Wu, \emph{{Note on asymptotic symmetries and soft gluon
  theorems}}, \href{https://doi.org/10.1103/PhysRevD.96.065023}{\emph{Phys.
  Rev. D} {\bfseries 96} (2017) 065023}
  [\href{https://arxiv.org/abs/1704.05740}{{\ttfamily 1704.05740}}].

\bibitem{Strominger:2017zoo}
A.~Strominger, \emph{{Lectures on the Infrared Structure of Gravity and Gauge
  Theory}},  \href{https://arxiv.org/abs/1703.05448}{{\ttfamily 1703.05448}}.

\bibitem{Bondi:1962px}
H.~Bondi, M.~G.~J. van~der Burg and A.~W.~K. Metzner, \emph{{Gravitational
  waves in general relativity. 7. Waves from axisymmetric isolated systems}},
  \href{https://doi.org/10.1098/rspa.1962.0161}{\emph{Proc. Roy. Soc. Lond. A}
  {\bfseries 269} (1962) 21}.

\bibitem{Sachs:1962wk}
R.~K. Sachs, \emph{{Gravitational waves in general relativity. 8. Waves in
  asymptotically flat space-times}},
  \href{https://doi.org/10.1098/rspa.1962.0206}{\emph{Proc. Roy. Soc. Lond. A}
  {\bfseries 270} (1962) 103}.

\bibitem{Sachs:1962zza}
R.~Sachs, \emph{{Asymptotic symmetries in gravitational theory}},
  \href{https://doi.org/10.1103/PhysRev.128.2851}{\emph{Phys. Rev.} {\bfseries
  128} (1962) 2851}.

\bibitem{Taylor:2017sph}
T.~R. Taylor, \emph{{A Course in Amplitudes}},
  \href{https://doi.org/10.1016/j.physrep.2017.05.002}{\emph{Phys. Rept.}
  {\bfseries 691} (2017) 1} [\href{https://arxiv.org/abs/1703.05670}{{\ttfamily
  1703.05670}}].

\bibitem{Fan:2019emx}
W.~Fan, A.~Fotopoulos and T.~R. Taylor, \emph{{Soft Limits of Yang-Mills
  Amplitudes and Conformal Correlators}},
  \href{https://doi.org/10.1007/JHEP05(2019)121}{\emph{JHEP} {\bfseries 05}
  (2019) 121} [\href{https://arxiv.org/abs/1903.01676}{{\ttfamily
  1903.01676}}].

\bibitem{Schreiber:2017jsr}
A.~Schreiber, A.~Volovich and M.~Zlotnikov, \emph{{Tree-level gluon amplitudes
  on the celestial sphere}},
  \href{https://doi.org/10.1016/j.physletb.2018.04.010}{\emph{Phys. Lett. B}
  {\bfseries 781} (2018) 349}
  [\href{https://arxiv.org/abs/1711.08435}{{\ttfamily 1711.08435}}].

\bibitem{Fan:2022vbz}
W.~Fan, A.~Fotopoulos, S.~Stieberger, T.~R. Taylor and B.~Zhu, \emph{{Elements
  of celestial conformal field theory}},
  \href{https://doi.org/10.1007/JHEP08(2022)213}{\emph{JHEP} {\bfseries 08}
  (2022) 213} [\href{https://arxiv.org/abs/2202.08288}{{\ttfamily
  2202.08288}}].

\bibitem{Mizera:2022sln}
S.~Mizera and S.~Pasterski, \emph{{Celestial geometry}},
  \href{https://doi.org/10.1007/JHEP09(2022)045}{\emph{JHEP} {\bfseries 09}
  (2022) 045} [\href{https://arxiv.org/abs/2204.02505}{{\ttfamily
  2204.02505}}].

\bibitem{Pasterski:2021raf}
S.~Pasterski, M.~Pate and A.-M. Raclariu, \emph{{Celestial Holography}},  in
  \emph{{2022 Snowmass Summer Study}}, 11, 2021,
  \href{https://arxiv.org/abs/2111.11392}{{\ttfamily 2111.11392}}.

\bibitem{Pasterski:2021rjz}
S.~Pasterski, \emph{{Lectures on celestial amplitudes}},
  \href{https://doi.org/10.1140/epjc/s10052-021-09846-7}{\emph{Eur. Phys. J. C}
  {\bfseries 81} (2021) 1062}
  [\href{https://arxiv.org/abs/2108.04801}{{\ttfamily 2108.04801}}].

\bibitem{Fotopoulos:2020bqj}
A.~Fotopoulos, S.~Stieberger, T.~R. Taylor and B.~Zhu, \emph{{Extended Super
  BMS Algebra of Celestial CFT}},
  \href{https://doi.org/10.1007/JHEP09(2020)198}{\emph{JHEP} {\bfseries 09}
  (2020) 198} [\href{https://arxiv.org/abs/2007.03785}{{\ttfamily
  2007.03785}}].

\bibitem{Banerjee:2021uxe}
N.~Banerjee, T.~Rahnuma and R.~K. Singh, \emph{{Asymptotic Symmetry of Four
  Dimensional Einstein-Yang-Mills and Einstein-Maxwell Theory}},
  \href{https://arxiv.org/abs/2110.15657}{{\ttfamily 2110.15657}}.

\bibitem{Fotopoulos:2019vac}
A.~Fotopoulos, S.~Stieberger, T.~R. Taylor and B.~Zhu, \emph{{Extended BMS
  Algebra of Celestial CFT}},
  \href{https://doi.org/10.1007/JHEP03(2020)130}{\emph{JHEP} {\bfseries 03}
  (2020) 130} [\href{https://arxiv.org/abs/1912.10973}{{\ttfamily
  1912.10973}}].

\bibitem{Bern:2006kd}
Z.~Bern, L.~J. Dixon and R.~Roiban, \emph{{Is N = 8 supergravity ultraviolet
  finite?}}, \href{https://doi.org/10.1016/j.physletb.2006.11.030}{\emph{Phys.
  Lett. B} {\bfseries 644} (2007) 265}
  [\href{https://arxiv.org/abs/hep-th/0611086}{{\ttfamily hep-th/0611086}}].

\bibitem{Arkani-Hamed:2008owk}
N.~Arkani-Hamed, F.~Cachazo and J.~Kaplan, \emph{{What is the Simplest Quantum
  Field Theory?}}, \href{https://doi.org/10.1007/JHEP09(2010)016}{\emph{JHEP}
  {\bfseries 09} (2010) 016} [\href{https://arxiv.org/abs/0808.1446}{{\ttfamily
  0808.1446}}].

\bibitem{Golden:2012hi}
J.~Golden and M.~Spradlin, \emph{{Collinear and Soft Limits of Multi-Loop
  Integrands in N=4 Yang-Mills}},
  \href{https://doi.org/10.1007/JHEP05(2012)027}{\emph{JHEP} {\bfseries 05}
  (2012) 027} [\href{https://arxiv.org/abs/1203.1915}{{\ttfamily 1203.1915}}].

\bibitem{Bourjaily:2011hi}
J.~L. Bourjaily, A.~DiRe, A.~Shaikh, M.~Spradlin and A.~Volovich, \emph{{The
  Soft-Collinear Bootstrap: N=4 Yang-Mills Amplitudes at Six and Seven Loops}},
  \href{https://doi.org/10.1007/JHEP03(2012)032}{\emph{JHEP} {\bfseries 03}
  (2012) 032} [\href{https://arxiv.org/abs/1112.6432}{{\ttfamily 1112.6432}}].

\bibitem{Nandan:2012rk}
D.~Nandan and C.~Wen, \emph{{Generating All Tree Amplitudes in N=4 SYM by
  Inverse Soft Limit}},
  \href{https://doi.org/10.1007/JHEP08(2012)040}{\emph{JHEP} {\bfseries 08}
  (2012) 040} [\href{https://arxiv.org/abs/1204.4841}{{\ttfamily 1204.4841}}].

\bibitem{Jiang:2021xzy}
H.~Jiang, \emph{{Celestial superamplitude in $ \mathcal{N} $ = 4 SYM theory}},
  \href{https://doi.org/10.1007/JHEP08(2021)031}{\emph{JHEP} {\bfseries 08}
  (2021) 031} [\href{https://arxiv.org/abs/2105.10269}{{\ttfamily
  2105.10269}}].

\bibitem{Bern:2002kj}
Z.~Bern, \emph{{Perturbative quantum gravity and its relation to gauge
  theory}}, \href{https://doi.org/10.12942/lrr-2002-5}{\emph{Living Rev. Rel.}
  {\bfseries 5} (2002) 5}
  [\href{https://arxiv.org/abs/gr-qc/0206071}{{\ttfamily gr-qc/0206071}}].

\bibitem{Kawai:1985xq}
H.~Kawai, D.~C. Lewellen and S.~H.~H. Tye, \emph{{A Relation Between Tree
  Amplitudes of Closed and Open Strings}},
  \href{https://doi.org/10.1016/0550-3213(86)90362-7}{\emph{Nucl. Phys. B}
  {\bfseries 269} (1986) 1}.

\bibitem{Bern:2008qj}
Z.~Bern, J.~J.~M. Carrasco and H.~Johansson, \emph{{New Relations for
  Gauge-Theory Amplitudes}},
  \href{https://doi.org/10.1103/PhysRevD.78.085011}{\emph{Phys. Rev. D}
  {\bfseries 78} (2008) 085011}
  [\href{https://arxiv.org/abs/0805.3993}{{\ttfamily 0805.3993}}].

\bibitem{Cachazo:2013gna}
F.~Cachazo, S.~He and E.~Y. Yuan, \emph{{Scattering equations and
  Kawai-Lewellen-Tye orthogonality}},
  \href{https://doi.org/10.1103/PhysRevD.90.065001}{\emph{Phys. Rev. D}
  {\bfseries 90} (2014) 065001}
  [\href{https://arxiv.org/abs/1306.6575}{{\ttfamily 1306.6575}}].

\bibitem{Cachazo:2013hca}
F.~Cachazo, S.~He and E.~Y. Yuan, \emph{{Scattering of Massless Particles in
  Arbitrary Dimensions}},
  \href{https://doi.org/10.1103/PhysRevLett.113.171601}{\emph{Phys. Rev. Lett.}
  {\bfseries 113} (2014) 171601}
  [\href{https://arxiv.org/abs/1307.2199}{{\ttfamily 1307.2199}}].

\bibitem{Liu:2014vva}
Z.-W. Liu, \emph{{Soft theorems in maximally supersymmetric theories}},
  \href{https://doi.org/10.1140/epjc/s10052-015-3304-1}{\emph{Eur. Phys. J. C}
  {\bfseries 75} (2015) 105} [\href{https://arxiv.org/abs/1410.1616}{{\ttfamily
  1410.1616}}].

\bibitem{Bianchi:2008pu}
M.~Bianchi, H.~Elvang and D.~Z. Freedman, \emph{{Generating Tree Amplitudes in
  N=4 SYM and N = 8 SG}},
  \href{https://doi.org/10.1088/1126-6708/2008/09/063}{\emph{JHEP} {\bfseries
  09} (2008) 063} [\href{https://arxiv.org/abs/0805.0757}{{\ttfamily
  0805.0757}}].

\bibitem{Bern:1998sv}
Z.~Bern, L.~J. Dixon, M.~Perelstein and J.~S. Rozowsky, \emph{{Multileg one
  loop gravity amplitudes from gauge theory}},
  \href{https://doi.org/10.1016/S0550-3213(99)00029-2}{\emph{Nucl. Phys. B}
  {\bfseries 546} (1999) 423}
  [\href{https://arxiv.org/abs/hep-th/9811140}{{\ttfamily hep-th/9811140}}].

\bibitem{Bern:1998xc}
Z.~Bern, L.~J. Dixon, M.~Perelstein and J.~S. Rozowsky, \emph{{One loop n point
  helicity amplitudes in (selfdual) gravity}},
  \href{https://doi.org/10.1016/S0370-2693(98)01397-5}{\emph{Phys. Lett. B}
  {\bfseries 444} (1998) 273}
  [\href{https://arxiv.org/abs/hep-th/9809160}{{\ttfamily hep-th/9809160}}].

\bibitem{Adamo:2022dcm}
T.~Adamo, J.~J.~M. Carrasco, M.~Carrillo-Gonz\'alez, M.~Chiodaroli, H.~Elvang,
  H.~Johansson et~al., \emph{{Snowmass White Paper: the Double Copy and its
  Applications}},  in \emph{{2022 Snowmass Summer Study}}, 4, 2022,
  \href{https://arxiv.org/abs/2204.06547}{{\ttfamily 2204.06547}}.

\bibitem{He:2014bga}
S.~He, Y.-t. Huang and C.~Wen, \emph{{Loop Corrections to Soft Theorems in
  Gauge Theories and Gravity}},
  \href{https://doi.org/10.1007/JHEP12(2014)115}{\emph{JHEP} {\bfseries 12}
  (2014) 115} [\href{https://arxiv.org/abs/1405.1410}{{\ttfamily 1405.1410}}].

\bibitem{Ferro:2020lgp}
L.~Ferro, T.~\L{}ukowski and R.~Moerman, \emph{{From momentum amplituhedron
  boundaries toamplitude singularities and back}},
  \href{https://doi.org/10.1007/JHEP07(2020)201}{\emph{JHEP} {\bfseries 07}
  (2020) 201} [\href{https://arxiv.org/abs/2003.13704}{{\ttfamily
  2003.13704}}].

\bibitem{tab}
N.~Banerjee, T.~Rahnuma and R.~K. Singh, \emph{{Asymptotic Symmetry algebra of
  $\mathcal{N}=8$ Supergravity}},
  \href{https://arxiv.org/abs/2212.12133}{{\ttfamily 2212.12133}}.

\bibitem{Banerjee:2018hbl}
N.~Banerjee, A.~Bhattacharjee, I.~Lodato and T.~Neogi, \emph{{Maximally $
  \mathcal{N} $ -extended super-BMS$_{3}$ algebras and generalized 3D gravity
  solutions}}, \href{https://doi.org/10.1007/JHEP01(2019)115}{\emph{JHEP}
  {\bfseries 01} (2019) 115}
  [\href{https://arxiv.org/abs/1807.06768}{{\ttfamily 1807.06768}}].

\bibitem{Banerjee:2022abf}
N.~Banerjee, A.~Mitra, D.~Mukherjee and H.~R. Safari,
  \emph{{Supersymmetrization of deformed BMS algebras}},
  \href{https://arxiv.org/abs/2201.09853}{{\ttfamily 2201.09853}}.

\end{thebibliography}\endgroup
\end{document}